\documentclass{emulateapj}
\usepackage{amssymb,amsmath}
\def\bs{\boldsymbol}

\def\gsim{\;\rlap{\lower 2.5pt
\hbox{$\sim$}}\raise 1.5pt\hbox{$>$}\;}
\def\lsim{\;\rlap{\lower 2.5pt
\hbox{$\sim$}}\raise 1.5pt\hbox{$<$}\;}

\makeatletter
\newcommand{\vast}{\bBigg@{3}}
\newcommand{\Vast}{\bBigg@{5}}
\makeatother

\newcommand{\Rmnum}[1]{\expandafter\@slowromancap\romannumeral #1@}

\begin{document}
\title{Turbulence-Induced Relative Velocity of Dust Particles I: Identical Particles}


\author{Liubin Pan}
\affil{Harvard-Smithsonian Center for Astrophysics,
60 Garden St., Cambridge, MA 02138; lpan@cfa.harvard.edu}
\and
\author{Paolo Padoan}
\affil{ICREA \& ICC, University of Barcelona, Marti i Franqu\`{e}s 1, E-08028 Barcelona, Spain; ppadoan@icc.ub.edu}

\begin{abstract}

We study the relative velocity of inertial particles suspended in turbulent flows and discuss implications for dust particle collisions in protoplanetary disks. 
We simulate a weakly compressible turbulent flow, evolving 14 particle species with friction timescale, $\tau_{\rm p}$, covering the entire range of scales in 
the flow. The particle Stokes numbers, $St$, measuring the ratio of $\tau_{\rm p}$ to the Kolmogorov timescale, are in the range $0.1 \lsim St \lsim 800$.  Using simulation results, 
we show that the model by Pan \& Padoan (PP10) gives satisfactory predictions for the rms relative velocity between identical particles. The probability 
distribution function (PDF) of the relative velocity is found to be highly non-Gaussian. The PDF tails are well described by a 4/3 stretched exponential function for particles with $\tau_{\rm p} \simeq 1-2 T_{\rm L}$, where $T_{\rm L}$ is the Lagrangian correlation timescale, consistent with a prediction based on PP10. 
The PDF approaches Gaussian only for very large particles with $\tau_{\rm p} \gsim 54 T_{\rm L}$. We split particle pairs at given distances into two types 
with low and high relative speeds, referred to as continuous and caustic types, respectively, and compute their contributions to the collision kernel. 
Although amplified by the effect of clustering, the continuous contribution vanishes in the limit of infinitesimal particle distance, where the caustic contribution 
dominates. The caustic kernel per unit cross section rises rapidly as $St$ increases toward $\simeq 1$, reaches a maximum at $\tau_{\rm p} \simeq 2 T_{\rm L}$, 
and decreases as $\tau_{\rm p}^{-1/2}$ for $\tau_{\rm p} \gg T_{\rm L}$.


\end{abstract}
\section{Introduction}

The dynamics of particles of finite inertia suspended in turbulent flows is a fundamental problem with 
applications ranging from industrial processes (e.g. spray combustion engines) to geophysical flows 
(e.g., atmospheric clouds). The interaction between turbulence and particles has been studied to 
understand rain initiation in warm terrestrial clouds (e.g., Pinsky \& Khain 1997; Falkovich, Fouxon,\& Stepanov 2002; Shaw 2003), 
cloud evolution in the atmospheres of planets, cool stars and brown dwarfs (e.g., Rossow 1978; Pruppacher \& Klett 1997; 
Freytag et al.\ 2010; Helling et al.\ 2011), collisions and growth of dust particles in protoplanetory disks 
(e.g., Dullemond \& Dominik 2005; Zsom et al.\ 2010, 2011; Birnstiel et al.\ 2011) and in the interstellar medium (e.g., Ormel et al.\ 2009). 

The evolution of the particle size depends on the particle collision rate which may be significantly 
enhanced by turbulent motions in the carrier flow, as illustrated by recent numerical and theoretical 
advances in this field (e.g., Wang et al.\ 2000; Zhou et al.\ 2001; Falkovich et al.\ 2002; 
Zaichik \& Alipchenkov 2003, 2009; Zaichik et al.\  2003, 2006; Wilkinson et al.\ 2005, 2006; Falkovich \& Pumir 2007; 
Gustavsson \& Mehlig 2011; Gustavsson et al.\ 2012). An accurate evaluation of the collision rates 
requires understanding the effects of two interesting phenomena: the preferential concentration 
or clustering of inertial particles (e.g., Maxey 1987 and  Squires \& Eaton 1991) and the turbulence-induced 
collision velocity. In this work, we will focus on the statistics of turbulence-induced relative velocities, 
and briefly discuss the role of turbulent clustering on the collision rate 
(see Pan et al.\ 2011 for a detailed discussion of turbulent clustering in the context of planetesimal formation). 
We restrict our discussion to the relative velocity between same-size particles, 
usually referred to as the monodisperse case, and will address the general bidisperse 
case (collisions between particles of different sizes) in a follow-up paper.   

The main motivation of our study is to improve the modeling of the evolution of dust 
particles in protoplanetary disks, which sets the stage for the formation of planetesimals, 
the likely precursors to fully-fledged planets. For example, the planetesimal formation 
model by Johansen et al.\ (2007, 2009, 2011) requires particle growth up to decimeter 
to meter size, in order to achieve good frictional coupling to the disk rotation 
and hence the maximum clustering effect by the streaming instability. Cuzzi et al.\ (2008, 2010) 
and Chambers (2010) proposed an alternative model of planetesimal formation based on the strong 
turbulent clustering of chondrule-size particles.  Other studies (e.g. Lee et al.\ 2010) focus on 
the possibility that small particles settle to the disk midplane, where gravitational instability 
can result in planetesimal formation (e.g. Goldreich \& Ward 1973; Youdin 2011), 
despite the turbulence stirring caused by the Kelvin-Helmholtz instability induced 
by the vertical settling of the particles (e.g. Weidenschilling 1980; Chiang 2008).

The evolution of the size distribution of dust particles is controlled by collisions. 
Small particles tend to stick together when colliding, and thus their size grows by coagulation. 
As the size increases, the particles become less sticky (Blum \& Wurm 2010), and, depending 
on the collision velocity, the collisions may result in bouncing or fragmentation. 
A detailed summary of experimental results for the dependence of the collision outcome on the 
particle properties (such as the particle size and porosity) and on the collision velocity can be 
found in Guttler et al.\ (2010). The coagulation, bouncing and fragmentation processes 
may lead to a quasi-equilibrium distribution of particle sizes (e.g., Birnstiel et al.\ 2011; Zsom et al.\ 2010, 2011).  
Due to the dependence of the collision outcome on the collision velocity, an accurate 
evaluation of the turbulence-induced relative velocity is important for modeling the size distribution 
of dust particles. 
  
Saffman and Turner (1956) studied the relative velocity in the limit of small particles 
with the particle friction or stopping time, $\tau_{\rm p}$, much smaller than the 
Kolmogorov timescale, $\tau_\eta$, of the turbulent flow. This limit,  known as the 
Saffman-Turner limit, is usually expressed as $St \ll 1$, where the Stokes number is 
defined as $St \equiv \tau_{\rm p}/\tau_\eta$. Saffman and Turner (1956) predicted that, 
at a given distance, $r$, the relative velocity of identical particles is independent 
of $St$, and, at a given $St$, it scales linearly with $r$ for small $r$. In the 
opposite limit of large particles with $\tau_{\rm p}$ larger than the largest timescale 
of the turbulent flow, Abrahamson (1975) showed that the relative velocity scales with 
the friction time as $\tau_{\rm p}^{-1/2}$. A variety of models have been developed 
to bridge the two limits and to predict the relative velocity for particles of any size, 
i.e., with $\tau_{\rm p}$ covering the entire scale range of the carrier flow (Williams \& Crane 1983, Yuu 1984, 
Kruis \& Kusters 1997, \& Alipchenkov 2003, Zaichik et al.\ 2006, Ayala et al.\ 2008). 
Among these models, the formulation of Zaichik and collaborators is particularly 
impressive, as it examines turbulent clustering and turbulence-induced 
relative velocity simultaneously. The model prediction for the relative velocity 
agrees well with simulation results at low resolutions. However, the model lacks a transparent 
physical picture.  

Pan \& Padoan (2010) developed a new model for the relative velocity of inertial 
particles of any size that provides an insightful physical picture of the problem. 
Their formulation illustrates that the relative velocity of identical particles is 
determined by the memory of the flow velocity difference along their 
trajectories in the past. The model also shows that the separation of inertial 
particle pairs backward in time plays an important role in their relative velocity.  
The model prediction can correctly reproduce the scaling behaviors of the relative 
speed in the extreme limits of small and large particles, and was found to 
successfully match the simulation data of Wang et al.\ (2000). 
 
Falkovich et al.\ (2002) discovered an interesting effect, named the sling effect, 
which provides an important contribution to the collision rate. The basic physical picture 
of this effect is that inertial particles may be shot out of fluid streamlines with high curvature, 
causing their trajectories to cross with those of other particles (see Fig.\ 1 of Falkovich \& Pumir 2007). 
In particular, in flow regions with large negative velocity gradients, fast 
particles can catch up with the slower ones from behind. The trajectory crossing 
causes the particle velocity to be multi-valued at a given point. This gives rise to folds, usually 
referred to as caustics, in the momentum-position phase space of the particles (Wilkinson et al.\ 2006; see Fig.\ 1 of Gustavsson \& Mehlig 2011 for a 
clear illustration).  For small particles with $St \ll 1$, the sling events correspond 
to high-order statistics of the flow velocity gradient, and the effect is not reflected 
in the prediction of Saffman and Turner (1956).
The formulations of Falkovich et al.\ (2002) and Gustavsson \& Mehlig (2011) 
for the collision kernel of $St\lsim 1$ particles consist of two contributions. Following 
Wilkinson et al.\ (2006), we name them as continuous and caustic contributions, 
corresponding to two types of particle pairs with low and high relative velocities, respectively. 
In the continuous contribution, the relative speed follows the Saffman-Turner prediction 
and decreases linearly with the particle distance, $r$. The contribution is amplified by turbulent clustering.  
However, the scaling exponents of the relative speed and the degree of clustering suggest that 
the continuous contribution approaches zero in the limit $r \to 0$, as pointed out by Hubbard (2012). 
The caustic contribution to the collision kernel per unit cross section was predicted to 
be independent of the particle size or distance, $r$, and is thus expected to dominate at sufficiently small $r$.  
The effect of slings or caustics causes a rapid rise in the collision rate as $St$ 
approaches 1, which has been proposed to be responsible for the initiation of rain shower in terrestrial 
clouds (Wilkinson et al.\ 2006). Applying this effect to dust particle collisions in 
protoplanetary disks, one may expect that the collision rate greatly accelerates as the particle 
grows past sub-mm to mm size, corresponding to $St \simeq 1$ for typical protoplanetary 
turbulence conditions.  

The recent developments mentioned above have not been considered in coagulation 
models for dust particles in circumstellar disks. We will show that the general 
formulation of the collision kernel commonly used in the astrophysical literature 
for dust coagulation is inaccurate. 
In particular, the dust coagulation models usually adopt collision velocities from the work of Volk et al.\ (1980) and 
its later extensions (e.g., Markiewicz, Mizuno \& Volk 1991, Cuzzi and Hogan 2003, and Ormel \& Cuzzi 2007), 
which have a number of limitations.  Pan \& Padoan (2010) pointed out a weakness in the physical 
picture of these models. Roughly speaking, these models assume that the velocities of two particles 
induced by turbulent eddies with turnover time significantly smaller (larger) than $\tau_{\rm p}$ 
are independent (correlated).  As shown by Pan \& Padoan (2010), whether 
the particle velocities contributed by turbulent eddies of a given size are correlated 
or not also depends on how the eddy size compares to the separation of the particles at the time the eddies were encountered. 
Therefore, the eddy turnover time is not the only factor that determines the degree of correlation.
The role of the particle separation relative to the eddy size is not captured by the approach of Volk et al.
We also find that the model of Volk et al. overestimates the relative velocity by 
a factor of 2 for particles with $\tau_{\rm p}$ on the order of the large eddy turnover time of the turbulent flow.



In this paper, we conduct  a $512^3$ numerical simulation to study inertial particle dynamics in 
a hydrodynamic  turbulent flow. In the simulated flow, 
we evolve inertial particles in an extended size range, with $\tau_{\rm p}$ covering the entire scale range of the turbulent flow.  
To our knowledge, such a systematic simulation of a significant resolution has not been previously conducted
in the astrophysical literature. Using the simulation data, we first test the model prediction of Pan \& Padoan (2010) for the rms 
relative velocity of inertial particles as a function of  $St$, and validate the physical picture revealed by the model. 
We then apply the Pan \& Padoan (2010) model to interpret the probability distribution function (PDF) 
of the relative velocity. The PDF study is motivated by the importance of the PDF of the 
collision speed in modeling dust particle collisions (Windmark et al.\ 2012, Garaud et al.\ 2013), which 
determines the fractions of collisions leading to sticking, bouncing or fragmentation. 
The relative velocity PDF of inertial particles has been shown to be highly non-Gaussian 
by numerical, experimental and theoretical studies (e.g., Sundaram and Collins 1997, 
Wang et al.\ 2000, Gustavsson et al.\ 2008, Bec et al.\ 2009, de Jong et al.\ 2010, Gustavsson \& 
Mehlig 2011, Hubbard 2012). Our simulation further confirms 
high non-Gaussianity, which should be incorporated into coagulation models for 
dust particles in protoplanetary disks. 
We will also investigate the particle collision kernel as a function of $St$. 

   




Due to the computational cost, the number of particles included in our simulation is limited and only 
allows to accurately measure the relative velocity statistics at significant particle distances. The distance 
range explored is $\eta/4\le r \le \eta$ where $\eta$ is the Kolmogorov scale of the simulated flow. 
This raises the question concerning the direct applicability of our measured statistics 
to dust particle collisions. The size of dust particles is many orders of magnitudes smaller 
than the Kolmogorov scale ($\eta \sim 1$ km) in protoplanetary turbulence. Therefore, 
dust particles should be viewed as nearly point particles, and one is required
to examine the $r \to 0$  limit in order to model their collisions (Hubbard 2012, 2013). 
This suggests that the relative velocity measured in our simulation 
at $r\simeq \eta$ would be distinct from the collision speed of dust particles, unless 
the statistics have already converged at $\simeq \eta$.  We find that the measured relative 
velocity statistics for particles with $St \gsim 10$ actually converge at 
$r \simeq \eta$, and are thus directly applicable for the collision velocity 
of dust particles.  On the other hand, for small to intermediate particles with $St \lsim 10$, 
the measured statistics show an $r$-dependence in the $r$ range explored in this study. 
For these particles, an appropriate extrapolation to the $r \to 0$ limit is needed for applications to dust particle collisions.  

In the current paper, we focus on understanding the fundamental 
physics of turbulence-induced relative velocity at finite distances ($\lsim 1\eta$). 
Our theoretical and numerical results provide an important step toward 
the final goal of estimating the dust particle collision velocity at $r\to 0$.
To underhand the $r \to 0$ limit,  we make an initial and preliminary 
attempt to separate particle pairs into two types, i.e., continuous and 
caustic types, which show different scalings with $r$.  
In particular, we evaluate the contributions of two types of pairs 
to the collision kernel and examine their behaviors as $r\to0$. 
A systematical study for the $r\to0$ limit is deferred to a future work. 

In this work, we will consider the particle dynamics only in statistically homogeneous 
and isotropic turbulence. This is clearly an idealized situation, considering various complexities 
in protoplanetary disks. For example, the disk rotation induces large-scale anisotropy, 
which may  have significant effects on the prediction for particles with friction time close to the 
rotation period. 
Nevertheless, the idealized problem is a very useful tool to understand the fundamental physics. 
We also neglect the vertical settling and radial drift. 
These processes do not directly affect the relative velocity between identical particles, 
although they may provide important contributions for particles of different 
sizes that we address in a follow-up work.  



The paper is organized as follows. In \S 2, we present a simple model 
for the rms velocity of a single particle, which provides an illustration for 
our formulation of the particle relative velocity. In \S 3, we introduce 
the model of Pan \& Padoan (2010) for the relative velocity of nearby particles. 
Our simulation setup and the statistical properties of the simulated turbulent 
flow are described in \S 4.  \S 5 presents simulation results for the one-particle rms 
velocity. In \S 6, we test the model prediction of Pan \& Padoan (2010) for the rms relative 
velocity, and discuss in details the probability distribution of the relative velocity as 
a function of the particle inertia. In \S 7, we evaluate the collision kernel. 
The conclusions of our study are summarized in \S 8.

\section{The Velocity of Inertial Particles}

The dynamics of inertial particles depends crucially on its friction or stopping timescale, 
$\tau_{\rm p}$. To evaluate of the friction timescale, we first need to compare 
the particle size, $a_{\rm p}$, with the mean free path of the gas particles in the carrier flow. 
If the particle size is larger than the mean free path, the friction timescale is 
given by the Stokes law $\tau_{\rm p} = \frac{2}{9} \left(\frac{\rho_{\rm d}}{\rho} \right) \left(\frac{a_{\rm p}^2}{\nu} \right)$, where $\rho_{\rm d}$ 
($\simeq 1$ g cm$^{-3}$) is the density of the dust material, $\rho$ is the gas density, and $\nu$ is the kinematic viscosity of the flow. 
On the other hand, if $a_{\rm p}$ is smaller than the gas mean free path, the particle is in the
Epstein regime and $\tau_{\rm p} = \left(\frac{\rho_{\rm d}}{\rho}\right) \left(\frac{a_{\rm p}}{C_{\rm s}}\right)$, 
where $C_{\rm s}$ is the sound speed in the flow.  For example, for a typical gas density in protoplanetary discs, $\rho \simeq 10^{-9}$ g cm$^{-3}$, 
at 1 AU, the mean free path of the gas particles is $\sim 1$ cm, and thus particles with $a_{\rm p}$ 
larger (smaller) than 1 cm are in the Stokes (Epstein) regime.   
 
The velocity, ${\bs v}(t)$, of an inertial particle suspended in a turbulent velocity 
field, $\bs u ({\bs x}, t)$, obeys the equation,
\begin{equation}
\frac {d {\bs v} } {dt} = \frac { {\bs u} \left( {\bs X} (t), t \right) - {\bs v}} {\tau_{\rm p}},       
\label{particlemomentum}     
\end{equation}
where ${\bs X} (t)$ is the position of the particle at time $t$, and ${\bs u} \left( {\bs X} (t), t \right)$ 
corresponds to the flow velocity ``seen"  by the particle. Eq.\ (\ref{particlemomentum}) 
has a formal solution, 
\begin{equation}
{\bs v}(t) = \frac {1} {\tau_{\rm p} }
\int_{t_0}^t  {\bs u} \left({\bs X} (\tau), \tau \right) \exp \left(- \frac{t-\tau}{\tau_{\rm p}}\right) d\tau,
\label{formalsolution}           
\end{equation}
where it is assumed that $t-t_0 \gg \tau_{\rm p}$ and the particle has 
already lost the memory of its initial velocity at $t_{0}$. The formal 
solution indicates that the velocity of an inertial particle is determined 
by the memory of the flow velocity along its trajectory 
within a timescale of $\simeq \tau_{\rm p}$ in the past. 

Although the aim of the present work is the relative velocity of inertial particle 
pairs, we start with a discussion of the single-particle (or ``1-particle'') velocity 
induced by turbulent motions. We provide a simple model for the 1-particle rms 
velocity as a function of  $\tau_{\rm p}$. The derivation of this model helps to 
illustrate our formulation for the relative velocity between two nearby particles.    

The 1-particle rms velocity can be calculated using the formal solution 
of eq.\ (\ref{formalsolution}).  We assume the turbulent flow is statistically 
stationary, and the particle statistics eventually relax to a steady state. 
We consider a time when the steady state is already reached and denote this 
time as time 0. Using eq.\ (\ref{formalsolution}) at $t=0$, we have, 
\begin{equation}
\langle v_i v_j \rangle = \int_{-\infty}^0 \frac {d\tau}{\tau_{\rm p}} \int_{-\infty}^0 \frac {d\tau'}{\tau_{\rm p}} B_{{\rm T} ij}
(\tau, \tau') \exp \left( \frac {\tau}{\tau_{\rm p}} \right) \exp\left(\frac {\tau'}{\tau_{\rm p}} \right),  
\label{1particlevelocity}
\end{equation} 
where $\langle \cdot \cdot \cdot \rangle$ denotes the ensemble average and $B_{{\rm T}ij}(\tau, \tau') \equiv \left\langle u_i ({\bs X}(\tau) ,\tau) u_j ({\bs X}(\tau') ,\tau') \right\rangle$ 
is the temporal correlation tensor of the flow velocity along the trajectory, ${\bs X}(\tau)$, 
of the inertial particle. The subscript ``T" stands for ``trajectory". We changed the 
lower integration limit ($t_0$) in  eq.\ (\ref{formalsolution}) to $-\infty$, based on 
the assumption that the particle dynamics is fully relaxed at time 0 (i.e., $t_0 \ll -\tau_{\rm p}$). 

With statistical stationarity and isotropy, the trajectory correlation tensor 
can be written as $B_{{\rm T}ij} (\tau, \tau') = u'^2 \Phi_{\rm 1}(\tau'-\tau) \delta_{ij}$, 
where $u'$ is the 1D rms velocity of the turbulent flow and the correlation coefficient 
$\Phi_{\rm 1}$ is a function of the time lag only. The subscript ``1" is used to indicate 
that the correlation is along the trajectory of {\it one} particle. The correlation 
coefficient, $\Phi_{\rm 1}$, is unknown, and a common assumption 
is to approximate it with the Lagrangian correlation function, $\Phi_{\rm L}$, 
of tracer particles (or fluid elements), which has been extensively studied. 
The assumption is likely valid for small particles, but cannot be justified for large particles on a theoretical basis. We will validate the assumption
{\it a posteriori} using simulation results. 
  
The simplest choice for $\Phi_{\rm L}$ is an exponential function, 
$\Phi_{\rm L} (\Delta \tau)= \exp(-|\Delta \tau|/T_{\rm L})$, where 
$\Delta \tau = \tau'-\tau$ is the time lag and $T_{\rm L}$ the Lagrangian 
correlation timescale. Setting $B_{{\rm T} ij} = u'^2 \exp(-|\tau'-\tau|/T_{\rm L}) \delta_{ij} $ 
in eq.\ (\ref{1particlevelocity}), we have $\langle v_i v_j \rangle = v'^2 \delta_{ij}$, where the 
1D rms particle velocity, $v'$, is given by,  
\begin{equation}
v' = u' \left( \frac {T_{\rm L}} {T_{\rm L}+\tau_{\rm p} } \right)^{1/2}. 
\label{1particlevelocityexp}
\end{equation} 
This result shows that the particle rms velocity approaches the flow 
velocity for $\tau_{\rm p} \ll T_{\rm L}$ and decreases as $(T_{\rm L}/\tau_{\rm p})^{1/2}$ for 
$\tau_{\rm p} \gg T_{\rm L}$ (e.g., Abrahamson 1975). In the large particle limit, $\tau_{\rm p} \gg T_{\rm L}$, 
the action of even the largest turbulent eddies on the particle would appear to be 
random kicks when viewed on a timescale of $\tau_{\rm p}$. In that case, 
eq.\ (\ref{particlemomentum}) is essentially a Langevin equation, and the particle 
motions are similar to Brownian motions. The $\tau_{\rm p}^{-1/2}$ scaling 
corresponds to an ``equilibrium" between the velocity of these particles 
and the turbulent motions of the flow. 
 
Numerical simulations have shown that the Lagrangian correlation 
function, $\Phi_{\rm L} (\Delta \tau)$, is better fit by a bi-exponential 
form (e.g., Sawford 1991). A single-exponential form does not 
reflect the smooth part of the correlation function for $\Delta \tau$ 
smaller than the Taylor micro timescale, $\tau_{\rm T}$. The Taylor 
timescale is defined as $\left(2 u'^2/a^2\right)^{1/2}$, where $a$ 
is the rms acceleration of the turbulent velocity field. The bi-exponential 
form for $\Phi_{\rm L} (\Delta \tau)$ is, 
\begin{gather}
\Phi_{\rm L}(\Delta \tau)= {\frac{1}{2 \sqrt {1-2z^2} }}
\vast\{ \big(1 + \sqrt{1-2z^2}\big)  \times \hspace{5cm}\notag\\
\hspace{1cm} \exp \Bigg[-\frac{2 |\Delta \tau|}{ \big(1+ \sqrt{1-2z^2}\big) T_{\rm L} }  \Bigg] -  
\big(1-\sqrt{1-2z^2}\big) \times \notag\\ 
 \exp \left [ {- \frac{2 |\Delta \tau|}{ \big(1 - \sqrt{1-2z^2} \big) T_{\rm L}} } \right] \vast\}, \hspace{1.8cm}  
\label{biexponential}
\end{gather}
where the parameter $z$ is defined as $z=\tau_{\rm T}/T_{\rm L}$. 
From the above equation, it is easy to show that 
$T_{\rm L}= \int \Phi_{\rm L} (\Delta \tau) d \Delta \tau$, and 
the bi-exponential function is smooth, $\simeq 1- (\Delta  \tau/\tau_{\rm T})^2$, 
at $\Delta \tau \ll \tau_{\rm T}$.  In the limit $z \to 0$,  eq.\ (\ref{biexponential}) 
reduces to the single exponential with a timescale of $T_{\rm L}$.   
 
Adopting the bi-exponential form, eq.\ (\ref{biexponential}), for the trajectory correlation 
coefficient, $\Phi_1$, we find that the 1-particle rms velocity is given by,  
\begin{equation}
v' = u'\left( \frac {\Omega +z^2/2} { \Omega +\Omega^2 +z^2/2}\right)^{1/2},  
\label{1particlevelocitybiexp}
\end{equation}
where $\Omega$ is defined as $\Omega \equiv \tau_{\rm p}/T_{\rm L}$. 
In the limits  $\Omega \ll1$ and $\Omega \gg 1$,  eq.\ (\ref{1particlevelocitybiexp})
has the same behavior as eq.\ (\ref{1particlevelocityexp}) from the 
single exponential correlation. In fact, the two predictions, eqs.\ (\ref{1particlevelocityexp}) 
and (\ref{1particlevelocitybiexp}), are close to each other at all values of $\Omega$, differing only by a few percent 
at $\Omega \simeq 1$. This suggests that, for a given correlation timescale, 
$T_{\rm L}$$(\equiv \int \Phi_{\rm L} (\Delta \tau) d \Delta \tau$), the integral in eq.\ (\ref{1particlevelocity}) 
is insensitive to the exact function form of  $\Phi_1(\Delta\tau)$. 
We will measure $z$ and $T_{\rm L}$ using Lagrangian tracer particles 
in our simulated turbulent flow, and test the predictions, eqs.\ (\ref{1particlevelocityexp}) and (\ref{1particlevelocitybiexp}), against the simulation data. 
  
\section{Turbulence-induced Relative Velocity of Inertial Particles}

We briefly review the 2-point Eulerian statistics of the velocity field in fully-developed 
turbulence, which is crucial to understand the relative velocity of two inertial particles. 
We consider the structure tensor of a turbulent flow, defined as $S_{ij} (\bs {\ell})= \left \langle  \Delta u_i   \Delta u_j \right \rangle $ 
where $\Delta u_i  = u_i({\bs x} + {\bs \ell}, t) - u_i({\bs x}, t)$ is the velocity increment 
across a separation ${\bs \ell}$. The statistics of $\Delta u_i$ is independent of ${\bs x}$ and $t$ from the assumption of 
homogeneity and stationarity. With statistical isotropy, the velocity structure tensor 
takes the form (e.g., Monin and Yaglom 1975), 
\begin{equation}
S_{ij} (\bs \ell)= S_{\rm nn}(\ell) \delta_{ij}  + \left[S_{\rm ll}(\ell)- S_{\rm nn}(\ell)\right]\frac{\ell_i \ell_j} {\ell^2}
\label{flowstructure}
\end{equation}
where  the longitudinal and transverse structure functions, $S_{\rm ll}$ and $S_{\rm nn}$, 
are functions of the amplitude, $\ell$, but not the direction (${\bs \ell}/\ell$) of ${\bs \ell}$.
From eq.\ (\ref {flowstructure}), we see  $S_{\rm ll} = S_{ij} (\bs \ell) \ell_i \ell_j/\ell^2 = \left \langle  \Delta u_{\rm r}^2 \right \rangle $, 
where $\Delta u_{\rm r} = \Delta u_i \ell_i/\ell $ is the radial component of $ {\bs \Delta u}$. Similarly, 
$S_{\rm nn}$ can be written as $S_{\rm nn} = \langle (\Delta u_{\rm t} )^2 \rangle$ 
with $\Delta u_{\rm t}$ being one of the two components of 
${\bs \Delta u}$ on the tangential/transverse plane perpendicular 
to ${\bs \ell}$. The statistical isotropy indicates that the 
probability distribution of $\Delta u_{\rm t}$ is invariant under 
any rotation about the direction ${\bs \ell}/\ell$. In incompressible 
turbulence,  which is approximately the case for gas flows 
in protoplanetary disks, we have the relation $S_{\rm nn} = S_{\rm ll} + \frac{1}{2} {\ell} dS_{\rm ll}/d\ell$, 
which can be derived from the incompressibility condition: $\partial_j S_{ij} ({ \bs \ell}) = 0$ (Monin and Yaglom 1975).  

The structure functions exhibit different scaling behaviors 
in different scale ranges. There are three subranges 
divided by two length scales, the Kolmogorov length scale, $\eta$, and 
the integral length scale $L$. The Kolmogorov scale, $\eta$, 
is defined as $\eta = (\nu^3/\bar{\epsilon})^{1/4}$, where $\nu$ and $\bar{\epsilon}$ are, 
respectively, the kinematic viscosity and the average energy dissipation rate in the turbulent flow. 
It essentially corresponds to the size of the smallest eddies. Scales below 
$\eta$ are called the viscous or dissipation range, where the velocity field 
is laminar and differentiable due to the smoothing effect of the viscosity. 
In the dissipation range, the velocity difference scales linearly with $\ell$, and 
the longitudinal structure function is $S_{\rm ll} =  \frac{\bar{\epsilon}}{15 \nu} \ell^2$. $S_{\rm nn}$ 
is twice larger, i.e., $S_{\rm nn} =\frac{2\bar{\epsilon}}{15 \nu}\ell^2$, as required 
by the incompressibility constraint.  In the inertial range, $\eta \lsim \ell \lsim L$, $S_{\rm ll}$ follows the 
Kolmogorov scaling, $S_{\rm ll}  = C_{\rm K }(\bar{\epsilon} \ell)^{2/3}$, where $C_{\rm K}$ is  the 
Kolmogorov constant. The typical value of $C_{\rm K}$ is $\simeq 2$. The incompressibility condition 
gives $S_{\rm nn} = 4 S_{\rm ll}/3$ in the inertial range.  The integral scale, $L$, is essentially the correlation 
length of the velocity field. At $\ell \gg L$, the velocity field is uncorrelated, 
and both $S_{\rm ll } (\ell)$ and $S_{\rm nn} (\ell)$ are constant and equal to $2 u'^2$ 
with $u'$ the 1D rms velocity of the flow. 

To bridge the scalings of $S_{\rm ll}$ in the three scale ranges, we adopt a 
connecting formula (Zaichik et al.\  2006),   
\begin{gather}
{\displaystyle S_{\rm ll}  = 2 u'^2 \left[ 1- \exp \left( - \frac{(\ell/ \eta)}{(15C_{\rm K})^{3/4}} \right) \right]^{4/3}   \times} \hspace{2cm} \notag \\
\hspace{3cm}
{\displaystyle \left [\frac{(\ell/ \eta)^4}{(\ell/ \eta)^4 + (2 u'^2 /C_{\rm K} u_\eta^2)^{6} } \right]^{1/6}},  
\label{sll}
\end{gather}
where $u_\eta$ is the Kolmogorov velocity scale defined as $(\nu \bar{\epsilon})^{1/4}$. 
With eq.\ (\ref{sll}) for $S_{\rm ll}$, 
we can obtain $S_{\rm nn}$ using the incompressibility condition (see above). 
Alternatively, one may adopt a separate connecting formula for $S_{\rm nn}$ (see \S 4.3).

The goal of this work is to understand the relative velocity of two nearby inertial 
particles. The relative velocity across a distance, $r$, equal to the sum of the particle 
radii corresponds to the speed at which the two particles collide 
(Saffman and Turner 1956). 
As mentioned earlier, dust particles in protoplanetary disks are nearly point-like, 
as their size is much smaller than the Kolmogorov length scale, $\eta$. The collision 
speed of dust particles is therefore the relative velocity at $r\to 0$. In this paper, we focus on 
the relative speed at finite distances, $r \lsim \eta$, and the $r\to 0$ limit will be examined systematically in a future work.
  
We label two particles coming together with superscripts $(1)$ and $(2)$. For 
example, we denote their positions as ${\bs X}^{(1)}(t)$ and ${\bs X}^{(2)}(t)$, and 
their velocities as ${\bs v}^{(1)}(t)$ and $\bs{v}^{(2)}(t)$ (see Fig.\ 1 for illustration). When the 
superscripts $(1)$ and $(2)$ are not used, the discussion is general and not referring to a specific particle.  
At a given time $t$, we consider the relative velocity ${\bs w} \equiv {\bs v}^{(2)}(t) -  {\bs v}^{(1)}(t)$, 
of particle pairs at a given separation, ${\bs r}$, which  corresponds to a constraint ${\bs X}^{(2)} (t)- {\bs X}^{(1)}(t) = {\bs r}$ for the 
particle positions. 
We first present a theoretical model for the second-order moment of ${\bs w}$, 
and then use simulations to explore its full statistics including the probability distribution function (PDF).

Similar to the structure tensor of the flow velocity,  we characterize the second-order statistics 
of the particle relative velocity ${\bs w}$ by a structure tensor,
\begin{equation}
S_{{\rm p}ij} \equiv \left \langle w_i w_j \right \rangle = \left \langle \left(v^{(2)}_i -v^{(1)}_i\right) \left(v^{(2)}_j -v^{(1)}_j\right) \right \rangle,             
\label{particlestructure}
\end{equation}  
which was referred to as the particle velocity structure tensor by Pan and Padoan (2010). 
Here $\langle \cdot\cdot\cdot\rangle$ denotes the average over all particle pairs at a separation of 
${\bs r}$. 

Once the particle dynamics is fully relaxed, the particle velocity is 
expected to possess the same statistical symmetries as the flow, including 
stationarity, homogeneity and isotropy. With these symmetries, 
$S_{{\rm p} ij}$ can be written in a similar form as the structure tensor of the flow (eq.\ \ref{flowstructure}),
\begin{equation}
S_{{\rm p}ij} ({\bs r})  = \langle w_{\rm t}^2 \rangle \delta_{ij} + \left( \langle w_{\rm r}^2 \rangle - \langle w_{\rm t}^2 \rangle \right) \frac{r_i r_j}{r^2}, 
\label{particlestructure2}
\end{equation} 
where $\langle w_{\rm r}^2 \rangle$ and $\langle w_{\rm t}^2 \rangle$ are the variances of the radial/longitudinal 
component, $w_{\rm r}$ ($\equiv w_i r_i/r$), and a tangential/transverse component, $w_{\rm t}$, of the relative 
velocity, respectively. For particle collisions, we are interested in $S_{{\rm p}ij} $ at small distances 
only, with $r$ in the dissipation range of the flow. 
Under the assumption of isotropy, the tangential component, $w_{\rm t}$, is expected to be 
statistically invariant for any rotations about the axis ${\bs r}$. 
We can thus measure the statistics of the tangential relative velocity by projecting 
${\bs w}$ into an arbitrary direction on the plane perpendicular to ${\bs r}$.

In the rest of this section, we consider theoretical models for the variances, $\langle w_{\rm r}^2 \rangle$ and $\langle w_{\rm t}^2 \rangle$, 
of the relative velocity, which can be computed from the particle structure tensor $S_{{\rm p}ij}$. 
For example, $\langle w_{\rm r}^2 \rangle  = S_{{\rm p}ij} {r_i r_j}/{r^2}$. The 3D variance, $\langle w^2 \rangle$, 
of the relative velocity, is given by the contraction of the tensor, i.e., 
$S_{{\rm p}ii} = \langle w_{\rm r}^2 \rangle + 2\langle w_{\rm t}^2 \rangle$. We point out that the relative 
velocity variances cannot be directly applied to estimate the collision kernel, which depends 
on $\langle |w_{\rm r}|\rangle$ or $\langle |{\bs w}|\rangle$ (see \S 7). One may use $\langle w_{\rm r}^2 \rangle$ to approximately 
estimate the collision rate by a conversion to $\langle |w_{\rm r}|\rangle$ under an assumption for the PDF shape of $w_{\rm r}$ 
(e.g., Wang et al.\ 2000). 

Furthermore, the 3D variance $\langle w^2 \rangle$ does not accurately reflect the average collisional energy for each 
collision. As pointed out by Hubbard (2012), a collision-rate weighting is needed to evaluate the average collisional energy per collision. 
In particular, $\langle w^2 \rangle$ is defined as the variance over all particle pairs at a distance, $r$. But 
not all the pairs may give a significant contribution to the collision rate in the $r\to 0$ limit (see \S 7.2), and in that case 
$\langle w^2 \rangle$ does not provide a reliable estimate for the average collision energy for those pairs that 
dominate the collision rate at $r \to 0$.  Despite these limitations in the practical use of the overall  rms (or variance) 
of the relative speed, 
its theoretical modeling is an important step toward understanding the fundamental physics. 
As mentioned earlier, we focus on the monodisperse case with equal-size particles.




\subsection{The Limits of Small and Large Particles}

We first consider small particles in the Saffman-Turner limit (hereafter the S-T limit). 
In this limit, the friction timescale, $\tau_{\rm p}$, is much smaller than the Kolmogorov 
timescale, $\tau_{\eta}$, of the carried flow, which is defined as $\tau_\eta \equiv (\nu/\bar{\epsilon})^{1/2}$. 
The Kolmogorov timescale is the smallest timescale in a turbulent flow, 
corresponding to the turnover time of the smallest eddies. 
Therefore, the velocity of particles with $\tau_{\rm p} \ll \tau_{\eta}$ 
can be approximated by a Taylor expansion of eq.\ (\ref{particlemomentum}), 
${\bs v} (t) \simeq {\bs u} ({\bs X}, t) + \tau_{\rm p} {\bs a} ({\bs X}, t)$, 
where ${\bs a} = D{\bs u}/Dt$ is the acceleration of the local fluid 
element. Applying the approximation to both particles $(1)$ and $(2)$, 
we have ${\bs w} = \left( {\bs  u}^{(2)} - {\bs  u}^{(1)} \right)+  \left( {\bs  a}^{(2)}  - {\bs  a}^{(1)} \right) \tau_{\rm p}$, 
where ${\bs u}^{(1,2)}$ ($\equiv  {\bs  u}({\bs X}^{(1,2)}, t)$) and ${\bs  a}^{(1,2)}$ ($\equiv  {\bs  a}({\bs X}^{(1,2)}, t)$) 
are the flow velocity and acceleration at the positions of particles (1) and (2), 
respectively. Saffman and Turner (1956) assumed that the correlation 
coefficient of the flow accelerations, ${\bs  a}^{(1)}$, and ${\bs  a}^{(2)}$, across a 
small distance, $r$, is unity, which is equivalent to assuming 
${\bs  a}^{(1)} \simeq {\bs  a}^{(2)}$. The acceleration terms then cancel 
out for identical particles, and the particle structure tensor, $S_{{\rm p} ij}$, 
is simply equal to the flow structure tensor, $S_{ij}$, defined in 
eq.\ (\ref{flowstructure}). Using the flow structure functions $S_{\rm ll}$ and 
$S_{\rm nn}$ at $r \ll \eta$ in incompressible turbulence, we have the Saffman-Turner formula,  
\begin{equation}
\langle w_{\rm r}^2 \rangle = \frac{1}{15} \frac {\bar{\epsilon}}{\nu} r^2,  \hspace{0.5cm}   
\langle w_{\rm t}^2 \rangle = \frac{2}{15} \frac {\bar{\epsilon}}{\nu} r^2, 
\label {saffmanturner}
\end{equation}
for identical particles with $St \ll 1$. The equation shows that in the S-T 
limit the relative speed is caused by the flow velocity difference across the particle separation. 
The effect is usually referred to as the shear contribution\footnote{The 
term ``shear contribution" is as opposed to 
the  ``acceleration contribution" from the acceleration terms mentioned above, which do not vanish for particles of 
different sizes. The acceleration contribution in the bidisperse case will be 
discussed in a separate paper.}.  From eq.\ (\ref{saffmanturner}), the 3D variance 
of the relative velocity is given by $\langle w^2(r) \rangle =  \frac {\bar{\epsilon}}{3\nu} r^2$. 

The S-T formula predicts that the tangential variance of the relative velocity, 
$\langle w_{\rm t}^2 \rangle$, is twice larger than that in the radial direction, $\langle w_{\rm r}^2 \rangle$. Eq.\ (\ref{saffmanturner}) 
also indicates a constant relative speed at a given separation, $r$, and a 
linear scaling with $r$ at a given $St \ll 1$. The accuracy of the Saffman-Turner 
formula for the small particle limit has been questioned, as it neglects the effect of slings 
and caustic formation (e.g., Falkovich et al.\ 2002, Wilkinson et al.\ 2006). We will test 
the S-T prediction against our simulation data. 
In the S-T limit, the particle memory is short and the relative speed is determined 
largely by the local flow velocity at small scales. The memory effect becomes 
more important for larger particles with $\tau_{\rm p} > \tau_{\eta}$ (see \S 3.2).  

We next consider the other extreme limit, i.e., large particles with $\tau_{\rm p}$ 
much larger than the Lagrangian correlation time, $T_{\rm L}$, of the flow. 
As discussed in \S 2, the motions of these particles are similar to Brownian motions, and 
the velocities of any two particles are statistically independent. This is 
because the velocity of a large particle has a significant contribution from 
its memory of the flow velocity long time ago, and the flow velocities ``seen" 
by the two particles at that time were uncorrelated because the 
particle separation was likely larger than the flow integral length scale, $L$. 
With the independence of ${\bs v}^{(1)}$ and ${\bs v}^{(2)}$, the particle structure tensor 
defined in eq. (\ref{particlestructure}) can be written as $S_{{\rm p} ij } = \left[ \left(v'^{(1)}\right)^2  +  \left(v'^{(2)}\right)^2 \right] \delta_{ij}$,
where $v'^{(1)} $ and $v'^{(2)}$ are the (1D) rms velocities 
of particles (1) and (2), respectively. As shown in \S 2, for particles 
with $\tau_{\rm p} \gg T_{\rm L}$, the rms velocity is 
given by $\simeq u' \left({T_{\rm L}}/{\tau_{\rm p}}\right)^{1/2}$.
We therefore have (e.g., Abrahamson 1975),
\begin{equation}
\langle w_{\rm r}^2 \rangle = \langle w_{\rm t}^2 \rangle =  2  u'^2 \frac{T_{\rm L}}{\tau_{\rm p} }, 
\label{largelimit}
\end{equation}
for identical particles with $\tau_{\rm p} \gg T_{\rm L}$. The equation 
suggests that the rms relative speed decreases with 
$St$ as $St^{-1/2}$. The physical picture for  the large particle limit is clear, 
and eq.\ (\ref{largelimit}) is thus robust.  


In between the two extreme limits are particles in the inertial range, i.e., particles with friction 
timescale $\tau_\eta \lsim \tau_{\rm p} \lsim T_{\rm L}$, corresponding to inertial-range scales 
in the turbulent flow. 
Unlike the two extreme limits where the velocities of two nearby particles 
are either highly correlated (small particles) or essentially independent (large particles), 
the velocity correlation of nearby inertial-range particles is at an intermediate level. We will show 
that a key physics for the relative velocity of these particles is their memory of the flow 
velocity difference in the past and the separation of the particle pair backward in time.  
 
As mentioned in the Introduction, a variety of models for the particle relative velocity covering the 
whole range of particle sizes have been developed (e.g., Volk et al.\ 1980, Ormel \& Cuzzi  2007, Zaichik \& 
Alipchenkov 2003, Zaichik et al.\ 2006, and Pan \& Padoan 2010). The models listed here all 
predict a $St^{1/2}$ scaling for inertial-range particles in turbulent flows with an extended inertial 
range. The $St^{1/2}$ scaling may be obtained by a simple scale-invariant assumption for 
inertial-range particles (e.g., Hubbard 2012), which we argue, however, does not provide 
a sufficient physical picture to understand the full statistics, e.g., the PDF shape, of the relative velocity. 
The models of Zaichik and collaborators and Pan and Padoan (2010) can reproduce both 
the S-T limit (eq.\ (\ref{saffmanturner})) and the large particle limit (eq.\ (\ref{largelimit})). We 
will focus on the model of Pan and Padoan (2010), which provides a clearer physical picture than that of 
Zaichik et al.  
The physical differences between various models have been summarized in Pan and Padoan (2010). 

\subsection{The Model of Pan and Padoan (2010)}

We review the formulation and the physical picture of the model by Pan \& Padoan 
(2010; hereafter PP10) for the relative velocity of identical particles. The PP10 model aimed at predicting the 
variance or rms of the relative velocity. As mentioned earlier, although the rms 
relative velocity does not directly enter the collision kernel or the average collisional energy per collision, 
its theoretical modeling is essential for understanding the underlying physics. 
For example, the physical picture revealed by the PP10 model is very successful 
in the interpretation of the probability distribution of the relative velocity (\S 6.2), 
which, in turn, is helpful for the evaluation of the collision kernel (\S 7).  
The main idea of the PP10 model was to compute 
the particle velocity structure tensor, $S_{{\rm p} ij}({\bs r})$, using the formal 
solution (eq.\ (\ref{formalsolution})) for the particle velocity. Applying eq.\ (\ref{formalsolution}) to 
the velocities of particles (1) and (2) at $t=0$,  we have,   
\begin{equation}
v^{(2)}_i - v^{(1)}_i  =   \frac {1}{\tau_{\rm p}} \int_{-\infty}^0 \left[u_i^{(2)} (\tau) - u_i^{(1)}(\tau)\right] \exp \left( \frac {\tau}{\tau_{\rm p}} \right)  d\tau, 
\label{velocitydifference}
\end{equation} 
where ${\bs u}^{(1, 2)}(\tau)$ ($\equiv {\bs u} ({\bs X}^{(1, 2)}(\tau), \tau)$) are the flow velocities ``seen" by the two particles 
at time $\tau$. Again we have changed the lower integration limit in the formal solution, eq.\ (\ref{formalsolution}), to $-\infty$. 

Inserting eq.\ (\ref{velocitydifference}) into the definition (eq.\ (\ref{particlestructure})) of $S_{{\rm p} ij}$, it is straightforward to find that,   
\begin{gather}            
S_{{\rm p}ij}({\bs r}) = \int_{-\infty}^0  \frac {d\tau}{\tau_{\rm p}}\int_{-\infty}^0 \frac{d\tau'}{\tau_{\rm p}}
S_{{\rm T}ij} ({\bs r}; \tau, \tau') \times \hspace{1.5cm}\notag \\
\hspace{3.5cm}\exp \left(\frac{\tau}{\tau_{\rm p}} \right) \exp \left(\frac{\tau'}{\tau_{\rm p}}\right),
\label{particlestructureintegral}
\end{gather}
where $S_{{\rm T} ij}$, named the trajectory structure tensor by PP10, 
is defined as, 
\begin{equation}
S_{{\rm T}ij} ({\bs r}, \tau, \tau') = \left \langle \left[u_i^{(2)} (\tau) - u_i^{(1)} (\tau) \right] \left[ u_j^{(2)} (\tau') - u_j^{(1)} (\tau') \right]\right \rangle. 
\end{equation} 
This tensor represents the correlation of the flow velocity differences on the trajectories of the two particles 
at two times $\tau$ and $\tau'$. $S_{{\rm T}ij}$ depends on the separation, ${\bs r}$, through the constraint that ${\bs X}^{(2)} (0) -  {\bs X}^{(1)} (0) = {\bs r}$. 
Eq.\ (\ref{particlestructureintegral}) is in close analogy with eq.\ (\ref{1particlevelocity}) for the 1-particle velocity.  
Here the trajectory structure tensor, $S_{{\rm T}ij}$, replaces the trajectory 
correlation tensor, $B_{{\rm T}ij}$, in eq.\ (\ref{1particlevelocity}).

The physical meaning of eqs.\ (\ref{velocitydifference}) and (\ref{particlestructureintegral}) 
is clear: the relative velocity of two identical inertial particles is controlled by the particles' 
memory of the flow velocity difference within a friction timescale, $\sim \tau_{\rm p}$, in the 
past. The physical picture is illustrated in Fig.\ 1. The trajectory structure tensor, $S_{{\rm T}ij}$, is 
unknown, and we model it using the approach of PP10. 

\begin{figure*}[t]
\centerline{\includegraphics[width=1.75\columnwidth]{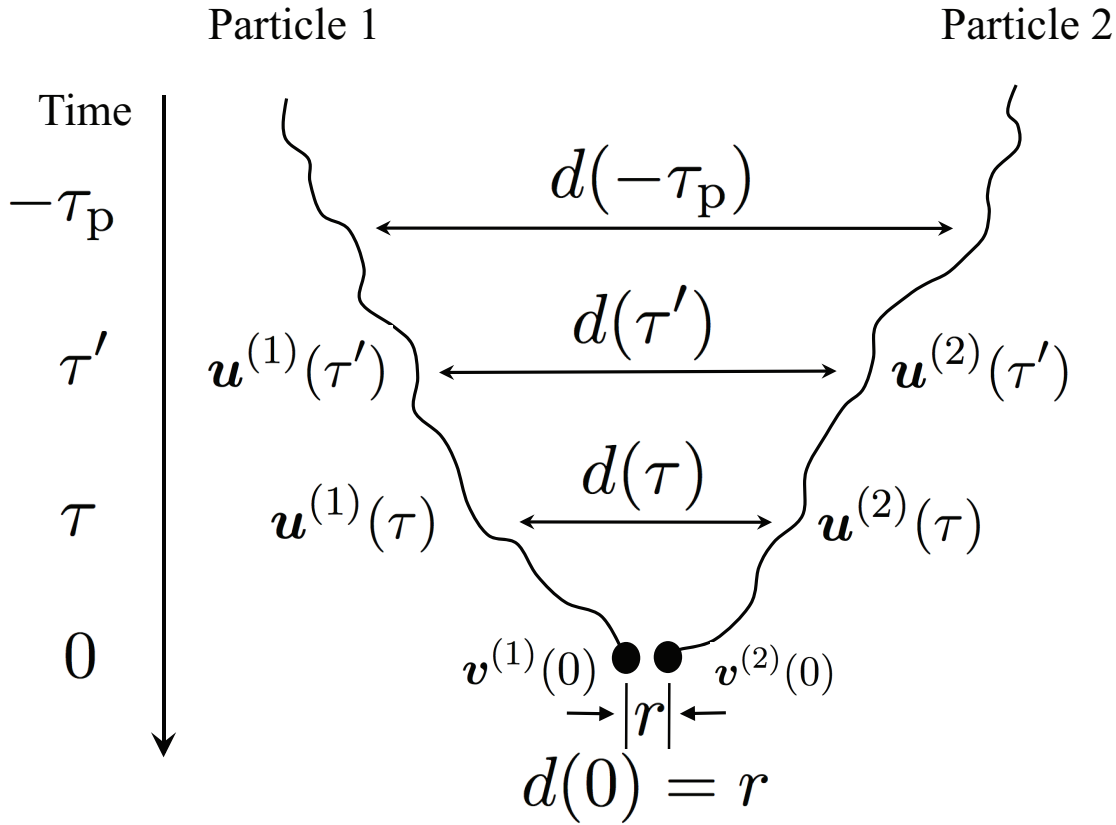}}
\caption{Schematic figure illustrating the physical picture of the PP10 model for the 
relative velocity of two equal-size particles. At time 0,  the separation of particles (1) and (2) is $r$. The velocity, ${\bs v} (0)$, of each particle 
at $t= 0$, is determined by its memory of the flow velocity, ${\bs u} (\tau)$, along the particle
trajectory in the past.  The relative velocity of the two particles mainly depends on the flow 
velocity difference, ${\bs u}^{(2)} (\tau) - {\bs u}^{(1)} (\tau)$, they ``saw" 
within about a friction timescale $\tau_{\rm p}$ in the past,  i.e., $ -\tau_{\rm p} \lsim \tau \le 0$. 
The flow velocity difference at a given time $\tau$ scales with 
the particle separation, $d(\tau)$.  The particle separation satisfies the ``initial" 
constraint $d (0) = r$ and increases backward in time. Due to particle inertia, 
a roughly ballistic separation is expected within a friction timescale. 
The trajectories plot here reflect a more-or-less linear separation of the two particles.  
The particle relative velocity also depends on the temporal correlation of the flow velocity differences the 
two particles ``saw" at different times, say $\tau$ and $\tau'$. The correlation timescale is associated with the turnover time of turbulent eddies encountered 
by the two particles.}
\label{cartoon} 
\end{figure*}

Since the flow velocity difference scales with the distance, $S_{{\rm T}ij}$ has an indirect dependence on the particle separation at $\tau$ and $\tau'$. 
We denote the particle separation at $\tau$ as ${\bs d} (\tau)$ ($\equiv {\bs X}^{(2)} (\tau) - {\bs X}^{(1)} (\tau) $). 
The vector ${\bs d}$ is stochastic because of the random dispersion 
of the particle pair by turbulent motions. $S_{{\rm T}ij}$ also has  a dependence on 
the time lag $(\tau'-\tau)$. This dependence is associated with the temporal correlation 
of turbulent structures or eddies encountered by the two particles between $\tau$ and $\tau'$, 
and the correlation time is essentially the turnover time of these eddies. 
To estimate $S_{{\rm T}ij}$, we consider the (indirect) spatial dependence 
on the particle separation and the temporal dependence on the time lag separately. 

We use a typical particle separation ${\bs R}(\tau, \tau')$ between $\tau$ and 
$\tau'$ to model the spatial dependence.  Like ${\bs d}(\tau)$ and ${\bs d}(\tau')$, 
${\bs R}(\tau, \tau')$ is also a random vector. 
We approximate the dependence on the separation by the 
Eulerian structure tensor of the flow velocity, $S_{ij} ({\bs R})$, defined in eq.\ (\ref{flowstructure}). 
We denote as $\Phi_2 (\tau'-\tau, R)$ the temporal correlation of the flow 
structure at the scale $R$. $\Phi_2$ is expected to be an even function of the time lag 
and is normalized to unity, $\Phi_2 (0, R) =1$, at zero time lag. To distinguish from the temporal 
correlation, $\Phi_{1} (\tau'-\tau)$, along the trajectory of a single particle 
(see \S 2), we have used a subscript ``2" here for the two-particle case.
The trajectory structure tensor is then modeled as the product of the two dependences (PP10), 
\begin{equation} 
S_{{\rm T}ij} ({\bs r}; \tau, \tau') \simeq \big\langle S_{ij} ({\bs R}) \Phi_2 \big(\tau' -\tau, R \big) \big\rangle_{\bs R}
\label{pp10assumption0}
\end{equation}
where $\langle \cdot \cdot \cdot \rangle_{\bs R}$ denotes the average over the statistics of the 
random vector, ${\bs R}$. This average is over the probability distributions 
of both the amplitude, $R$, and the direction of ${\bs R}$. Eq.\ (\ref{pp10assumption0}) 
implicitly assumes the statistical independence of the velocity difference, $\Delta {\bs u}$, 
seen by the two particles from their separation, ${\bs R}$. Rigorously, the amplitudes 
of $\Delta {\bs u}$ and ${\bs R}$ may have a correlation. If the particle pair encounters an eddy with a larger velocity, 
the particle separation tends to be larger. For example, if $R$ is in the inertial range
of the flow, $\Delta u \simeq \epsilon_{\rm R}^{1/3}R^{1/3}$ from the refined similarity 
hypothesis (Kolmogorov 1962), where $\epsilon_{\rm R}$ is the average dissipation 
rate over the scale $R$ seen by the particle pair. A positive correlation is expected 
between the fluctuations in $\epsilon_{\rm R}$ and $R$. Eq.\ (\ref{pp10assumption0}) 
neglects this correlation and may underestimate $S_{{\rm T}ij}$ and hence the particle relative velocity.  

The $\Phi_2$ term in eq.\ (\ref{pp10assumption0}) does not depend 
on the direction of ${\bs R}$, so one can first take the angular average of $S_{ij} ({\bs R})$ 
and then average the entire term over the PDF of the amplitude, $R$. The latter 
cannot be exactly performed because the PDF of $R$ is unknown. 
With some simple estimates, PP10 argued that simply 
using the rms of $R$ to evaluate $S_{{\rm T}ij}$ (instead of averaging over the PDF of $R$) 
only causes a small difference ($\simeq 10\%$) in the model prediction. 
Following PP10, we ignore the PDF of $R$ and insert the rms of $R$ to evaluate $S_{{\rm T}ij}$. 
For the simplicity of notation, we use $R$ to denote the rms particle distance in between $\tau$ and $\tau'$ in the 
rest of the paper. A similar notation is adopted for $d(\tau)$ and  $d(\tau')$, 
which will denote the rms separations at $\tau$ and $\tau'$, respectively. We approximate $R$ by the geometric 
average of  $d(\tau)$ and  $d(\tau')$,
\begin{equation}
R (\tau, \tau')= \left[d(\tau) d(\tau') \right]^{1/2}. 
\label{typicalseparation}
\end{equation}
The rms separation $d(\tau)$ as a function of time $\tau$ will be discussed in \S 3.2.3.

With the above assumptions, the trajectory structure tensor is modeled as,  
\begin{equation} 
S_{{\rm T}ij} ({\bs r}; \tau, \tau') \simeq \big\langle S_{ij} ({\bs R})\big\rangle_{\rm ang} \Phi_2 \big(\tau' -\tau, R \big). 
\label{pp10assumption}
\end{equation}
The angular average of $S_{ij}$ over the direction of ${\bs R}$ will be carried out in \S 3.2.2. 
In eq.\ (\ref{pp10assumption}), the dependence of $S_{{\rm T}ij}$ on ${\bs r}$ 
is through the dependence of $d(\tau)$, $d(\tau')$ and $R$ on ${\bs r}$.  
We refer to $\bs{r}$ as the ``initial" separation, although it actually corresponds to 
the current or final separation of the two particles.  Our formulation indicates that the separation of 
particle pairs backward in time is crucial for the prediction of the particle relative speed. 

Inserting eq.\ (\ref{pp10assumption}) into eq.\ (\ref{particlestructureintegral}) gives 
the PP10 model for the particle structure tensor, 
\begin{gather}        
S_{{\rm p}ij}({\bs r}) = \int_{-\infty}^0  \frac {d\tau}{\tau_{\rm p}}\int_{-\infty}^0 \frac{d\tau'}{\tau_{\rm p}}
\big\langle S_{ij} ({\bs R})\big\rangle_{\rm ang} \times \hspace{2cm} \notag\\
\hspace{2cm}
\Phi_2 \big(\tau' -\tau, R \big) \exp \left(\frac{\tau}{\tau_{\rm p}} \right) \exp \left(\frac{\tau'}{\tau_{\rm p}}\right).
\label{particlestructuremodel}
\end{gather}
We will numerically compute this double integral after evaluating or modeling the angular average, the temporal correlation 
and the particle separation backward in time. 


A simplification of the PP10 model is to set ${\bs R}$ to one of 
two distances, ${\bs d}(\tau)$ or ${\bs d}(\tau')$, instead of their geometric 
average. We find that replacing ${\bs R}$ in eq.\ (\ref{particlestructuremodel}) 
by either ${\bs d}(\tau)$ or ${\bs d}(\tau')$ leads to equivalent model 
prediction for the particle relative speed. This is because $\Phi_2$ in eq.\ (\ref{particlestructuremodel}) 
is an even function of $\Delta \tau$($\equiv \tau'-\tau$), and the product of the two exponential 
cutoffs are invariant under the exchange of $\tau$ and $\tau'$. 
If one sets ${\bs R} = {\bs d}(\tau)$ in eq.\ (\ref{particlestructuremodel}), 
the integral over $\tau'$ can be isolated, yielding,   
\begin{equation}            
S_{{\rm p}ij} = \frac {1} {\tau_{\rm p}}  \int_{-\infty}^0  \langle S_{ij} ({\bs d}(\tau))\rangle_{\rm ang}  F(\tau) \exp \left(\frac{\tau}{\tau_{\rm p}} \right) d\tau,
\label{separatedintegral}
\end{equation}
where the angular average is over the direction of ${\bs d}(\tau)$ and the function $F(\tau)$ is 
defined as, 
\begin{equation}
F(\tau) = \frac{1}{\tau_{\rm p}} \int_{-\infty}^0  \Phi_2 \big(\tau' -\tau, d(\tau)\big) \exp \left(\frac{\tau'}{\tau_{\rm p}} \right) d\tau'. 
\label{ffactor}
\end{equation}
The factor $F(\tau)$ may be roughly viewed as a response function of the particle pair 
to turbulent eddies at the scale $d(\tau)$. Although not indicated explicitly, the factor 
$F(\tau)$ also depends on $r$ through its dependence 
on $d(\tau)$. We will refer to eqs.\ (\ref{separatedintegral}) and (\ref{ffactor}) as the simplified model. 
In the simplified model, $F(\tau)$ can be 
integrated analytically using assumed function forms of $\Phi_2$ in \S 3.2.1, 
and one only needs to numerically solve a single integral in eq.\ (\ref{separatedintegral}). 
On the other hand, for the original PP10 model, one must numerically evaluate the 
double integral in eq.\ (\ref{particlestructuremodel}).  

\subsubsection{The temporal correlation $\Phi_2$} 

To estimate the temporal correlation, $\Phi_2$, in the trajectory structure tensor, $S_{{\rm T}ij}$, 
we first consider a special case where the particle separation, $R$, is much 
larger than the integral length scale, $L$, of the flow. In this case, the flow velocities, ${\bs u}^{(1)}$ and  ${\bs u}^{(2)}$, ``seen" by the two 
particles are independent, and $S_{{\rm T} ij}$ can be written 
as $\langle u^{(1)}_i(\tau) u^{(1)}_j(\tau') \rangle+ \langle u^{(2)}_i(\tau) u^{(2)}_j(\tau')\rangle$ 
(see eq.\ (\ref{particlestructure})). Both terms correspond to the trajectory correlation tensor 
$B_{{\rm T}ij}$ defined below eq.\ (\ref{1particlevelocity}) in \S 2, and for identical particles 
the two terms are equal. Therefore, for $R \gg L$,  
$\Phi_2(\Delta\tau, R)$ is the same as the temporal correlation coefficient, $\Phi_1(\Delta\tau)$, along the trajectory of one particle. 

In \S 2, we approximated $\Phi_1$ by the Lagrangian correlation function, $\Phi_{\rm L}$. Using the approximation 
again, we have $\Phi_2 (\Delta\tau, \ell)  = \Phi_1 (\Delta\tau) \simeq \Phi_{\rm L} (\Delta\tau)$ 
for $\ell \gg L$. 
Two function forms, single- and bi-exponential, were
adopted for $\Phi_{\rm L}$ in \S 2.  With 
the single-exponential form, we set $\Phi_2 (\Delta \tau, \ell) = \exp(-|\Delta \tau|/T_{\rm L})$ for $\ell \gg L$.   
An extension of this function to smaller scales gives,    
\begin{equation}
\Phi_2(\Delta \tau, \ell)= \exp \left( - \frac{|\Delta \tau|}{T(\ell)} \right),
\label{singleexponential}
\end{equation}
where $T(\ell)$ is essentially the correlation time or lifetime of turbulent eddies of size $\ell$. 
For $\ell \gg L$, we set $T(\ell)  = T_{\rm L}$. 

At smaller $\ell$, $T(\ell)$ can be estimated using the velocity scalings in the turbulent flow.  
For $\ell$ in the inertial range, we obtain $T(\ell)$ by dividing $\ell$ by the amplitude 
of the turbulent velocity fluctuations at this scale, 
which is $\left(S_{\rm ll}(\ell) + 2S_{\rm nn} (\ell)\right)^{1/2}$. Using the Kolmogorov scaling for structure functions, 
we have $T(\ell) = C_{\rm T} \bar{\epsilon}^{-1/3} \ell^{2/3}$, where $C_{\rm T} = (11C_{\rm K}/3)^{-1/2} = 0.52 C_{\rm K}^{-1/2} $. 
The factor, $11/3$, is from the incompressibility relation $S_{\rm nn} = 4S_{\rm ll}/3$ in the 
inertial range. Since the Kolmogorov constant $C_{\rm K}$ is $\simeq 2$,  
we set $C_{\rm T} \simeq 0.4$. A similar value of $C_{\rm T}$ was adopted 
by Zaichik \& Alipchenkov (2003). In the viscous range with $\ell \ll \eta $, 
the flow velocity difference goes linearly with $\ell$, and $T({\ell})$ is expected to 
be constant. Lundgren (1981) predicted that  $T(\ell)  = \sqrt{5} \tau_\eta$ for $\ell \ll \eta$, 
which was later confirmed by numerical simulations 
of Girimaji \& Pope (1990). We thus take $T(\ell) = \sqrt{5} \tau_\eta$ for $\ell \ll \eta $ 
in our model. We will use the bridging formula for $T(\ell)$ from Zaichik et al.\ (2006), 
\begin{gather}
T (\ell)  = T_{\rm L} \left[ 1- \exp \left( -   \left(\frac{C_{\rm T}}{\sqrt{5}}\right)^{3/2}  (\ell/ \eta) \right) \right]^{-2/3} \times \hspace{1cm} \notag\\
\hspace{2.7cm}
\left [\frac{(\ell/ \eta)^4}{(\ell/ \eta)^4 +  (T_{\rm L}/( C_{\rm T} \tau_\eta))^{6} } \right]^{1/6},   
\label{Tr}
\end{gather}
which satisfies the scalings of $T(\ell)$ in different scale ranges.  

One may also adopt a bi-exponential form for $\Phi_2(\Delta\tau, \ell)$ based on 
eq.\ (\ref{biexponential}) for the Lagrangian correlation function $\Phi_{\rm L}$ (see \S 2). 
Replacing $T_{\rm L}$ in eq.\ (\ref{biexponential}) by $T (\ell)$ gives, 
\begin{gather}
\Phi_{2}(\Delta \tau, \ell)= {\frac{1}{2 \sqrt {1-2z^2} }} 
\vast\{ \big(1 + \sqrt{1-2z^2}\big) \times \hspace{5cm}\notag\\
\hspace{0.9cm}\exp \Bigg[-\frac{2 |\Delta \tau|}{ \big(1+ \sqrt{1-2z^2}\big) T(\ell) }  \Bigg] - 
\big(1-\sqrt{1-2z^2}\big) \times\notag\\  
 \exp \left [ {- \frac{2 |\Delta \tau|}{ \big(1 - \sqrt{1-2z^2} \big) T(\ell)}}  \right] \vast\}.\hspace{1.3cm}
\label{biexponential2}
\end{gather}
This bi-exponential form for $\Phi_2(\Delta\tau, \ell)$ was used in all the calculations 
in PP10. We will compute the predictions of the PP10 model using both the single- and bi-exponential 
correlation functions. We find the results from the two cases are close to each other, 
suggesting that the double integral in eq.\ (\ref{particlestructuremodel}) is insensitive to the 
function form of $\Phi_2(\Delta \tau, \ell)$. After the integration, the dependence on 
$\Phi_2(\Delta \tau, \ell)$ is essentially condensed to a dependence on the 
timescale $T(\ell)$. This is similar to the case of the one-particle velocity, which 
is insensitive to the form of $\Phi_1(\Delta \tau)$ (see \S 2). 
PP10 also considered the possible dependence of the 
parameter $z$ on the length scale $\ell$. It was found that including a 
reasonable length scale dependence of $z$ barely changes the model prediction. 
We will set $z$ to be constant in this study.

We next consider the simplified model represented by eqs.\ (\ref{separatedintegral}) and (\ref{ffactor}). With  a
single-exponential $\Phi_2$ (eq.\ (\ref{singleexponential})), the response factor $F(\tau)$ defined in eq.\ (\ref{ffactor}) 
can be integrated analytcally, 
\begin{equation}            
F(\tau) = \frac{T(d)}{T(d) - \tau_{\rm p} } \exp \left(\frac{\tau}{T(d)}\right)  + \frac{2 \tau_{\rm p} T(d)}{\tau_{\rm p}^2 - T^2(d) }  \exp \left( \frac{\tau}{\tau_{\rm p}} \right).
\label{response}
\end{equation}
Since $\tau$ is negative, $F(\tau)$ is dominated by the 
first term if $T(d) \gg \tau_{\rm p}$, and it approaches $\exp\left(\tau/T(d)\right)$ in that limit.  On the other hand, 
for $T(d) \ll \tau_{\rm p}$, the leading term is $\frac{2T(d)}{ \tau_{\rm p}} \exp(\tau/\tau_{\rm p})$.  
Note that  eq.\ (\ref{response}) does not diverge at $T(d) = \tau_{\rm p}$. Applying the L'Hospital's rule 
shows that it converges to $(\frac{1}{2}-\frac{\tau}{\tau_{\rm p}}) \exp(\tau/\tau_{\rm p})$, as $T(d) \to 
\tau_{\rm p}$. Therefore, when numerically integrating eq.\ (\ref{separatedintegral}), 
we set $F(\tau) = (\frac{1}{2}-\frac{\tau}{\tau_{\rm p}}) \exp(\tau/\tau_{\rm p})$ for $T(d)$ around $\tau_{\rm p}$.

With the bi-exponential temporal correlation, eq.\ (\ref{biexponential2}), 
the response factor, $F(\tau)$, can also be integrated analytically.  
The integration is straightforward, but the resulting function for $F(\tau)$ 
is complicated and is thus omitted here. The predictions of the simplified model with single- and bi-exponential 
$\Phi_2(\Delta \tau, \ell)$ are also found to be close to each other.

\subsubsection{Averaging over the direction of ${\bs R}$}

We evaluate the angular average of $S_{ij} ({\bs R})$ over the direction of ${\bs R}$. It follows 
from eq.\ (\ref{flowstructure}) that $S_{ij} ({\bs R}) = S_{\rm nn} (R) \delta_{ij} + \left[S_{\rm ll} (R) - S_{\rm nn} (R)\right] R_i R_j/R^2$. 
The contraction of the tensor is $S_{ii} ({\bs R})= S_{\rm ll} (R) + 2 S_{\rm nn} (R)$, which does not have a direct 
dependence on the direction of ${\bs R}$. Therefore, to predict the 3D rms relative speed, we do not need to perform the angular 
average. 
However, for the radial and tangential components, one must make an assumption for the direction of ${\bs R}$ and 
compute the angular average for the term $\propto R_iR_j/R^2$. 

In PP10, we assumed that the direction of the separation change, 
$\Delta {\bs R} \equiv {\bs R} - {\bs r}$, caused by turbulent dispersion is 
completely random or isotropic. One can then insert ${\bs R} = \Delta {\bs R} + {\bs r}$ 
into $S_{ij} ({\bs R})$ and take the average over the direction of $\Delta {\bs R}$. 
From the assumed isotropy of $\Delta{\bs R}$, we have $\langle r_i \Delta R_j \rangle =0$ 
and $\langle \Delta R_i \Delta R_j  \rangle_{\rm ang} = \frac{1}{3} (R^2 -r^2) \delta_{ij}$, 
and hence $\langle  R_i R_j  \rangle_{\rm ang} \simeq r_i r_j + \frac{1}{3} (R^2 -r^2) \delta_{ij}$ 
(see PP10). The angular average $\langle S_{ij} ({\bs R}) \rangle_{\rm ang}$ is then given
by\footnote{Rigorously, 
the amplitude, $R$, of ${\bs R}$ and hence $S_{\rm ll}(R)$ and $S_{\rm nn} (R)$ 
have a dependence on the direction of $\Delta{\bs R}$. However, the average of these 
quantities over the direction of $\Delta{\bs R}$ is complicated and cannot be done analytically. 
For simplicity, we kept $R$, $S_{\rm ll}$ and $S_{\rm nn}$ fixed, and only 
accounted for the angular average of $R_i R_j$.}, 
\begin{gather}
\langle S_{ij} ({\bs R}) \rangle_{\rm ang} = \delta_{ij}\bigg[ \left(\frac{1}{3}-\frac{r^2}{3R^2}\right) S_{\rm ll}(R) + \left( \frac{2}{3} +\frac{r^2 }{3R^2} \right) \times \hspace{2cm}\notag \\ \hspace{3cm} S_{\rm nn} (R) \bigg]  + \Big[S_{\rm ll}(R)  - S_{\rm nn}(R)\Big]\frac{r_i r_j}{R^2}.  
\label{randomdirection1}
\end{gather} 
The equation approaches $S_{ij}({\bs r})$ in the limit $R \to r$. PP10 showed that eq.\ (\ref{randomdirection1}) 
reproduces the S-T formula for the radial and tangential relative speeds.
In the limit $R \gg r $, we have $\langle S_{ij}({\bs R}) \rangle_{\rm ang} \simeq \frac{1}{3}\left[S_{\rm ll}(R) + 2S_{\rm nn}(R)\right]$. 

Here we make a simpler assumption than PP10: we take the direction of ${\bs R}$ (rather than $\Delta{\bs R}$) 
to be isotropic. This means $\langle R_i R_j/R^2 \rangle_{\rm ang}  = \frac{1}{3} \delta_{ij}$, and we have,   
\begin{equation}
\langle S_{ij} ({\bs R}) \rangle_{\rm ang} = \frac{1}{3}\left[S_{\rm ll}(R) + 2 S_{\rm nn} (R) \right] \delta_{ij}, 
\label{randomdirection2}
\end{equation} 
which suggests that the particle structure tensor $S_{{\rm p} ij} \propto \delta_{ij}$ (see eq.\ (\ref{particlestructuremodel})), 
and hence $\langle w_{\rm r}^2 \rangle = \langle w_{\rm t}^2 \rangle$ (see eq.\ (\ref{particlestructure2})) 
for particles of any size. A comparison of the two assumptions, eqs.\ (\ref{randomdirection1}) 
and (\ref{randomdirection2}), shows that they differ only at $R \lsim r$.

As expected, the contraction $\langle S_{ii}(R) \rangle_{\rm ang}$ of both 
eq.\ (\ref{randomdirection1}) and eq.\ (\ref{randomdirection2}), is equal to 
$\left[S_{\rm ll}(R) + 2 S_{\rm nn} (R)\right]$, indicating that the two assumptions 
give the same prediction for the 3D rms relative velocity. 
The only difference between the two assumptions is the prediction for the radial and tangential 
components at $St \lsim 1$. 
In \S 3.2.4, we will compare the model predictions by the two assumptions. 
The angular average $\langle S_{ij} ({\bs d}) \rangle_{\rm ang}$ in the simplified model 
(eq.\ (\ref{separatedintegral})) can be evaluated similarly, and the resulting 
expressions are in the same form as eqs.\ (\ref{randomdirection1}) and (\ref{randomdirection2}) 
with $d(\tau)$ replacing $R(\tau, \tau')$. 

\subsubsection{The backward dispersion of particle pairs}

We finally specify the (rms) particle separation, $d(\tau)$, as a function of $\tau$. 
The separation of inertial particle pairs backward in time has not been explored in 
the literature. Fortunately, Bec et al.\ (2010) carried out a detailed numerical study of the 
forward-in-time pair dispersion of inertial particles. Following PP10, we use their results to 
guide the assumption for the backward dispersion. We first consider the separation behavior 
of inertial-range particles with $\tau_\eta \lsim \tau_{\rm p} \lsim T_{\rm L}$.

Bec et al.\ (2010) found that the separation of inertial particles shows different behaviors at early and late times. 
At early times, a clear ballistic phase is observed for particles with $St \gsim 3$. In this phase, 
the separation increases linearly with time, and the phase lasts for about a friction timescale. 
The ballistic behavior is easy to understand: The particle velocity tends to be roughly constant for a memory timescale, $\tau_{\rm p}$. 
This also applies to the dispersion backward in time. We thus assume that, for particle pairs at an ``initial" distance of 
$r$, the separation $d(\tau)$ in the time range $-\tau_{\rm p} \lsim \tau \le 0$ is 
given by,  
\begin{equation}
d^2 (\tau) = r^2 +\langle w^2 \rangle \tau^2 
\label{ballistic}
\end{equation}
where $\langle w^2 \rangle$ is the 3D variance of the particle relative velocity at time 0. 
The particle relative speed is actually what our model aims to predict. Therefore, the 
dependence of  $d(\tau)$ on $\langle w^2 \rangle$ in the ballistic phase leads to 
an implicit equation for  $\langle w^2 \rangle$(see \S 3.2.4).   

Bec et al.\ (2010) also showed that, after a friction timescale, the dispersion of 
inertial-range particles make a transition to a tracer-like phase, where the separation 
variance increases as time cubed, a behavior known as the Richardson law. 
The Richardson law was first discovered for tracer pair dispersion 
at  inertial-range scales. The transition to the Richardson phase at a friction timescale or so 
suggests the ballistic separation for a duration of $\tau_{\rm p}$ already 
brings the average particle distance into the inertial range of the flow.
The Richardson behavior was observed in the tracer pair dispersion both forward and backward in time (Berg et al.\  2006; see Appendix A). 
It is thus likely to exist also in the backward separation of inertial 
particles.  We connect the Richardson phase to the ballistic 
phase at $\tau \simeq -\tau_{\rm p}$, and use the Richardson law 
\begin{equation}
d^2 (\tau) \simeq g \bar{\epsilon} |\tau|^3
\end{equation}
at $\tau\lsim-\tau_{\rm p}$, where $g$ is called the Richardson constant and $\bar{\epsilon}$ is the 
average dissipation rate of the flow. As the backward separation is typically faster than the forward case, the transition to the 
Richardson phase might occur slightly earlier than assumed here. 
Bec et al.\ (2010) did not report the value of $g$ in the Richardson phase of inertial particle 
pair dispersion. As in PP10, we will take $g$ as a parameter. In our model, we use a combined separation behavior that connects 
a ballistic and a Richardson phase at $\tau \simeq -\tau_{\rm p}$. 

The Richardson behavior would end when the separation becomes larger than the 
integral length scale, $L$, of the turbulent flow. At such a large distance, the flow velocities ``seen" 
by the two particles is uncorrelated, and the particle separation is expected to be diffusive 
like in a random walk. It is thus appropriate to switch the Richardson behavior to a 
diffusive phase with $d^2(\tau) \propto |\tau|$ at $d \gsim L$. However, we find that the exact 
separation behavior at $d \gg L$ (or $R \gg L$) does not affect the prediction of our model. 
This is because at these scales both the structure functions, $S_{\rm ll} $ and $S_{\rm ll}$, 
and the timescale, $T(d)$ (or $T(R)$), become independent of $d$ (or $R$). 
Therefore, eq.\ (\ref{particlestructuremodel}) (or eq.\ (\ref{separatedintegral})) 
is insensitive to the behavior of the separation once it becomes much larger 
than $L$. This is confirmed by the numerical solutions of eqs.\ (\ref{particlestructuremodel}) 
and (\ref{separatedintegral}). For convenience, we keep using the Richardson's law even after $d$ exceeds $L$. 

The separation behavior discussed above is based on the simulation 
results of Bec et al.\ (2010) for particles in the inertial range. For simplicity, we will use the 
same behavior for all particles, although its validity is questionable for small ($\tau_{\rm p}\lsim \tau_\eta$) 
and large ($\tau_{\rm p} \gsim T_{\rm L}$) particles. For small particles with 
$St \lsim 3$, a ballistic phase is not clearly observed in the $d^2$ vs. time plots in 
Fig.\ 5 of Bec et al.\ (2010). We expect that a short ballistic phase is likely to 
exist if one plots $(d^2 -r^2)$ (instead of $d^2$) vs. time (see Fig.\ \ref{separation} 
in Appendix A for the $(d^2 -r^2)$ vs. time plot for tracer particle pairs). 
However, for $St \lsim 3 $ particles, the connection of the short ballistic 
phase to the Richardson behavior is more complicated than in the 
case of larger particles (Fig.\ 5 of Bec et al.\ 2010). This is because the pair 
separation of these particles does not enter the inertial range of the flow 
in a friction timescale or so. Therefore, an intermediate phase exists in 
between the ballistic and Richardson phases. Ideally, a three-phase 
behavior should be considered. Unfortunately, the separation behavior in the intermediate phase 
is completely unknown, and thus, to include it, one must adopt a 
pure parameterization. Here we take a simpler approximation: We still connect 
the Richardson behavior directly to the ballistic phase for $St \lsim 3$ particles, although it 
cannot be justified physically.  Essentially, this parameterizes 
the later two phases by a single Richardson law with a free parameter $g$.
Future numerical studies for the entire separation behavior  of small particles is needed to improve the approximation.      
For the particle distance range, $\eta/4 \lsim r \lsim \eta$, considered in our data 
analysis (see \S 6.1), 
our model with the assumed behavior does give acceptable prediction for $St\lsim3$ 
particles. 
However, in the $r \to 0$ limit, a careful study of the intermediate separation phase 
of $St\lsim3$ particles is necesary to accurately model their relative velocity.
 
The problem of using the assumed behavior for particles with $\tau_{\rm p} \gg T_{\rm L}$ is that 
the Richardson phase does not exit. The velocities of  these large particles are uncorrelated 
even at small distances (\S 3.1). Therefore, at timescales larger than $\tau_{\rm p}$, the 
separation is likely diffusive, i.e., $d(\tau) \propto |\tau|$. Realistically, one needs to connect the 
ballistic phase to a diffusive behavior rather than a Richardson law at $|\tau| \sim \tau_{\rm p}$. 
However, it turns out that, at the end of the ballistic phase of these particles, the separation is 
already $\gsim L$. As discussed earlier, once the separation exceeds $L$, the exact 
separation behavior would not significantly affect the model prediction. 
This justifies using a combined separation behavior with a 
ballistic and a Richardson phase also for $\tau_{\rm p} \gg T_{\rm L}$ particles. 

So far, the initial distance, $r$, just provides a floor value in our assumption for  the 
particle separation $d$ (eq.\ (\ref{ballistic})). It is, however, possible that the value of $r$ 
has additional effects on the separation behavior. Bec et al.\ (2010) only explored 
$r$ above the Kolmogorov scale, and it is not clear whether the separation behavior has a 
qualitative difference if $r \lsim \eta$.  
To model the particle collision speed, we are interested in the backward separation 
with $r \ll \eta$, and it would thus be helpful to systematically investigate whether and 
how the separation behavior changes as $r$ decreases below $\eta$. 
We defer such a study to a later work.  Due to the uncertainty in the separation 
behavior for $r \ll \eta$, we will focus on testing the model prediction for the relative velocity 
at significant fractions of the Kolmogorov scale ($\eta/4 \lsim r \lsim \eta$). We assume that the 
two-phase behavior discussed above applies for this range of $r$. 
Considering the existence of various uncertainties, the assumed separation behavior should be 
viewed more or less as a  parameterization. 

%

To constrain $g$ in the Richardson phase, in Appendix A we measure $g$ for the backward dispersion 
of tracer particle pairs in our simulated flow, which is used as a reference 
for inertial particles. The measured $g$ for tracers in our flow at a limited resolution shows a dependence 
on $r$, suggesting that the Richardson constant for inertial particles may also depend on $r$. 
When comparing our model prediction to the simulation  results at different $r$, we will adjust $g$ to 
obtain best fits, and examine whether the best-fit values are consistent with the 
range of $g$ measured from tracer particles. 
The Richardson constant for inertial particles may also have a dependence on $\tau_{\rm p}$ (or $St$), 
which will be ignored for simplicity. 

Finally, we point out that our model for the rms relative velocity does not directly 
account for the effect of the spatial clustering of the particles (see \S 7). Ideally, 
a theoretical model needs to consider the clustering and relative velocity statistics 
simultaneously. At a given time, the relative velocity determines the evolution 
of the spatial distribution of the particles, while the particle distribution may affect how 
the particles ``see" the flow velocity and hence the evolution of the relative 
velocity statistics. However, modeling clustering and the relative velocity together self-consistently is very 
challenging, and is out of the scope of the current work. 

\subsubsection{Qualitative Behavior of Our Model Prediction}

Our model for the particle structure tensor, $S_{{\rm p} ij}$, is now complete. Here we 
discuss the qualitative behavior of our model prediction. 
We start by considering the 3D variance, $\langle w^2 \rangle$. The contraction of 
eq.\ (\ref{particlestructuremodel}) gives, 
\begin{gather}            
{\displaystyle \langle w^2 \rangle = \int_{-\infty}^0  \frac {d\tau}{\tau_{\rm p}}\int_{-\infty}^0 \frac{d\tau'}{\tau_{\rm p}}
\big [S_{\rm ll}(R) + 2S_{\rm nn} (R) ] \times} \hspace{1cm}\notag \\
 \hspace{2cm} {\displaystyle \Phi_2 \big(\tau' -\tau, R \big)  
 \exp \left(\frac{\tau}{\tau_{\rm p}} \right) \exp \left(\frac{\tau'}{\tau_{\rm p}}\right)},
\label{w2}
\end{gather} 
which is an implicit equation of $\langle w^2 \rangle $ because $R$ 
depends on $\langle w^2 \rangle$ in the ballistic separation phase. In \S 6.1, 
we will solve the equation numerically using an iterative method. 

The qualitative behavior of the model prediction for $\langle w^2 \rangle$ 
can be obtained by analyzing the integrand in eq.\ (\ref{w2}). 
In the S-T limit ($\tau_{\rm p} \to 0$), the exponential cutoff terms, 
$\frac{1}{\tau_{\rm p}} \exp(\tau/\tau_{\rm p})$ and $\frac{1}{\tau_{\rm p}} \exp(\tau'/\tau_{\rm p})$, 
in the integrand can be viewed as delta functions at $\tau=0$ and $\tau'=0$, respectively. 
This suggests that $\langle w^2 \rangle$ is approximately given 
by $\simeq (S_{\rm ll} + 2 S_{\rm nn})$ at $R(0, 0)$. Since $R(0 ,0)= r$, 
we have $\langle w^2 \rangle =\frac{\bar{\epsilon}}{3\nu}r^2$ for $r$ in the dissipation rate, 
which is consistent with the S-T prediction (see eq.\ (\ref{saffmanturner})) for the 
3D variance of the relative velocity. 

 
The analysis of eq.\ (\ref{w2}) for larger particles is more complicated. We first 
note that $S_{\rm ll}(R)$, $S_{\rm nn}(R)$, and the timescale $T(R)$ in the correlation function $\Phi_2$ are all 
increasing functions of $R$. Since $R$ increases backward in time, the first factor in 
the integrand of eq.\ (\ref{w2}) increases with increasing $|\tau|$ 
and $|\tau'|$. A larger $T(R)$ also tends to increase the integral 
because, with increasing $T(R)$, $\Phi_2$ allows 
contributions from a broader range of time lag ($\Delta \tau$). Together with the exponential cutoffs, 
these suggest that the contribution to the integral 
peaks at $\tau, \tau' \simeq - \tau_{\rm p}$. We denote the 
particle separation at $\tau =\tau' = - \tau_{\rm p}$ as $R_{\rm p}$ 
($\equiv R(-\tau_{\rm p}, -\tau_{\rm p})$), and refer to it as the primary 
distance. 

In the extreme limit of large particles with $\tau_{\rm p} \gg T_{\rm L}$,  
$R_{\rm p}$ is expected to be much larger than the
integral scale, $L$, of the flow. At $R_{\rm p} \gg L$, we have $S_{\rm ll} = S_{\rm nn} =2u'^2$, 
and $T(R_{\rm p}) = T_{\rm L}$. 
The exponential cutoff by $\Phi_2$ indicates that only the time pairs 
($\tau$ and $\tau'$) that satisfy the constraint $|\tau'-\tau| \simeq T_{\rm L}$ give a significant contribution 
to the integral. Since $T_{\rm L} \ll \tau_{\rm p}$, $\Phi_2$ reduces
the range of $\tau$ and $\tau'$ that contributes to the double integral by 
a factor of $T_{\rm L}/\tau_{\rm p}$. Assuming the main contribution to 
the integral is from $R \simeq R_{\rm p}$ and accounting for the effect of $\Phi_2$, 
we find $\langle w^2 \rangle \simeq 6u'^2 T_{\rm L}/\tau_{\rm p}$. 
This is consistent with eq.\ (\ref{largelimit}), meaning that our model 
correctly reproduces the large particle limit. 

For inertial-range particles with $\tau_\eta \lsim \tau_{\rm p} \lsim T_{\rm L}$, the primary distance, $R_{\rm p}$, corresponds to 
inertial-range scales of the turbulent flow\footnote{Roughly speaking, the role of the 
primary distance, $R_{\rm p}$, is in analogy with the critical wavenumber, $k^*$, defined in the model of Volk et al.\ (1980). 
In the language of Volk et al., the velocity structures at scales much larger than $R_{\rm p}$ would be 
counted as Class I eddies, while structures below $R_{\rm p}$ would belong to Class III. 
However, Pan and Padoan (2010) pointed out a physical weakness in the evaluation of $k^*$ 
in the  Volk et al. model  for the two-particle relative velocity , as the role of the 
separation of the two particles is not properly accounted for. The reader is referred to Section 4.1 
of Pan and Padoan (2010) for a detailed discussion on this issue.}. Using the Kolmogorov scaling gives $S_{\rm ll}, S_{\rm nn} \propto R_{\rm p}^{2/3}$ and 
$T(R_{\rm p}) \propto R_{\rm p}^{2/3}$. 
From its definition,  $R_{\rm p}$ is roughly the particle distance 
at the time when the ballistic phase connects to the Richardson phase (see \S 3.2.3). 
We thus assume that $R_{\rm p}$ is determined by a ballistic 
separation of duration $\tau_{\rm p}$, i.e., $R_{\rm p} \simeq \langle w^2 \rangle^{1/2} \tau_{\rm p}$. 
The effect of $\Phi_2$ depends on how $T(R_{\rm p})$ compares 
to $\tau_{\rm p}$. If $T(R_{\rm p}) > \tau_{\rm p}$, $\Phi_2\simeq 1$ for all 
time pairs in the range $ -\tau_{\rm p} \lsim \tau, \tau' \lsim 0$. On the other hand, 
if $T(R_{\rm p}) < \tau_{\rm p}$, $\Phi_2$ provides a factor of $T(R_{\rm p})/\tau_{\rm p}$, 
which follows from the same argument used above for the large particle limit. 
We find that both cases lead to the same scaling of $\langle w^2 \rangle$ with $\tau_{\rm p}$. 
In the first case, eq.\ (\ref{w2}) is approximated by 
$\langle w^2 \rangle \simeq [S_{\rm ll}(R_{\rm p}) +S_{\rm nn}(R_{\rm p})] \propto R_{\rm p}^{2/3}$. 
With $R_{\rm p} \simeq \langle w^2 \rangle^{1/2} \tau_{\rm p}$, 
we obtain $\langle w^2 \rangle^{1/2} \propto \tau_{\rm p}^{1/2}$. 
In the second case with $T(R_{\rm p}) < \tau_{\rm p}$, we include a factor of $T(R_{\rm p})/\tau_{\rm p}$ 
and estimate $\langle w^2 \rangle$ as $\simeq [S_{\rm ll}(R_{\rm p}) +S_{\rm nn}(R_{\rm p})]T(R_{\rm p})/\tau_{\rm p} \propto R_{\rm p}^{4/3}/\tau_{\rm p}$. It is straightforward to see that setting $R_{\rm p} \simeq \langle w^2 \rangle^{1/2} \tau_{\rm p}$ in 
this estimate gives the same scaling, $\langle w^2 \rangle^{1/2} \propto \tau_{\rm p}^{1/2}$, as 
the first case. Therefore, whether $T(R_{\rm p})$ is larger or smaller than $\tau_{\rm p}$,  
our model predicts a $\tau_{\rm p}^{1/2}$ (or $St^{1/2}$) scaling for inertial-range particles. 

Using a similar argument, PP10 found that if the primary distance is 
determined by the Richardson's law, $R_{\rm p} \simeq (g \bar{\epsilon}\tau_p^3)^{1/2}$, 
the model also predicts a $St^{1/2}$ scaling for inertial-range particles. 
Since both the ballistic and Richardson behaviors yield a 
$St^{1/2}$ scaling, a combination of a ballistic and a Richardson phase 
produces the same scaling (PP10). 
The $St^{1/2}$ scaling has been previously predicted by models of 
Volk et al.\ (1980), Cuzzi and Hogan (2003), Ormel and Cuzzi (2007), 
and Zaichik \& Alipchenkov (2003). As mentioned earlier, the scaling 
may also be obtained from a dimensional analysis under a 
scale-invariant assumption. If the dimensional analysis exactly holds, then the departure 
from the $St^{1/2}$ scaling for inertial-range particles in a simulation of 
limited resolution is caused completely by the effects from dissipation 
or driving scales. 
The derivation of the $St^{1/2}$ scaling in all the models assumes a 
sufficiently broad inertial range. The scaling would not 
exist if the Reynolds number of the turbulent flow is low. In fact, the predicted 
$St^{1/2}$ behavior has never been confirmed by simulations due to the low 
numerical resolution. PP10 showed that, to see the $St^{1/2}$ scaling, 
the Taylor Reynolds number of the turbulent flow must be 
larger than $\simeq 300$. This is higher than in the $512^3$ 
simulation used in the present study, and thus a clear $St^{1/2}$ scaling is 
not observed.  
It appears likely that the existence of this scaling can 
be verified at a twice larger resolution. We will conduct a 
$1024^3$ simulation in a future work.

The above analysis for the scaling behavior of $\langle w^2 \rangle$ 
in different $St$ ranges can be similarly applied to the simplified 
model, eqs.\ (\ref{separatedintegral}) and (\ref{ffactor}). 
The prediction of the simplified model is qualitatively the same as the original PP10 model.

Finally, we examine the model prediction for the radial and tangential 
components of the relative velocity. 
The prediction for $\langle w_{\rm r}^2 \rangle$ and $\langle w_{\rm t}^2 \rangle$ 
depends on the angular average of $S_{{\rm T}ij}$ in eq.\ (\ref{particlestructuremodel}) 
(or eq.\ \ref{separatedintegral}). In \S 3.3.2, we made two assumptions, 
eqs.\ (\ref{randomdirection1}) and (\ref{randomdirection2}), for the angular average. 
Inserting the first assumption, eq.\ (\ref{randomdirection1}), into 
eq.\ (\ref{particlestructuremodel}) and comparing it with eq.\ (\ref{particlestructure2}), 
we find, 
\begin{gather}
\langle w_{\rm r}^2 \rangle =  \int_{-\infty}^0 \frac {d\tau} {\tau_{\rm p}}  \int_{-\infty}^0 \frac {d\tau'} {\tau_{\rm p}}  \bigg[ \left( \frac{1}{3} + \frac{2r^2}{3R^2}\right) S_{\rm ll}(R) + 
\hspace{2cm}\notag \\
\hspace{1.0cm}
\left(  \frac{2}{3}- \frac{2r^2}{3R^2} \right)  S_{\rm nn} (R) \bigg]  \Phi_2 \big(\tau' -\tau, R \big)
\exp \left(\frac{\tau+\tau'}{\tau_{\rm p}}\right) , \notag
\end{gather}
\begin{gather}
\langle w_{\rm t}^2 \rangle =\int_{-\infty}^0 \frac {d\tau} {\tau_{\rm p}}  \int_{-\infty}^0 \frac {d\tau'} {\tau_{\rm p}}
\bigg[ \left( \frac{1}{3}-  \frac{r^2}{3R^2}\right) S_{\rm ll}(R) +  \hspace{2cm} \notag\\  
\hspace{1.0cm}
\bigg(  \frac{2}{3} +  \frac{r^2}{3R^2}   \bigg) S_{\rm nn} (R) \bigg] \Phi_2 \big(\tau' -\tau, R \big)
\exp \left(\frac{\tau+\tau'}{\tau_{\rm p}} \right).
\label{wrwt}
\end{gather} 
In order to integrate these two equations, one needs to first solve eq.\ (\ref{w2}) for $\langle w^2 \rangle$ 
due to the dependence of $R$ on $\langle w^2 \rangle$ in the ballistic phase. It is 
easy to show that, in the limit $\tau_{\rm p} \to 0$, $R \to r$, and eq.\ (\ref{wrwt}) reduces to 
$\langle w_{\rm r}^2 \rangle =S_{\rm ll}(r)$ and $\langle w_{\rm t}^2 \rangle =S_{\rm nn}(r)$, 
reproducing the S-T formula, eq.\ (\ref{saffmanturner}). 
For larger particles with $\tau_{\rm p} \gg \tau_\eta$,  we have $R_{\rm p} \gg r$, and thus eq.\ (\ref{wrwt}) predicts $\langle w_{\rm r}^2 \rangle = \langle w_{\rm t}^2 \rangle = \frac{1}{3} \langle w^2 \rangle$, 
Therefore, like $\langle w^2 \rangle$, both $\langle w_{\rm r}^2 \rangle$ and $\langle w_{\rm t}^2 \rangle $ 
scale as $St^{1/2}$ for inertial-range particles and as $St^{-1/2}$ in the large particle limit.    
 
As discussed in \S 3.3.2, the second assumption, eq.\ (\ref{randomdirection2}), for the 
angular average of $S_{{\rm T}ij}$
predicts that $\langle w_{\rm r}^2 \rangle = \langle w_{\rm t}^2 \rangle = \frac{1}{3} \langle w^2 \rangle$ 
for all particles. 
In the S-T limit, eq.\ (\ref{w2}) gives $\langle w^2 \rangle = \frac{\bar{\epsilon}}{3\nu} r^2$, 
and thus $\langle w_{\rm r}^2 \rangle = \langle w_{\rm t}^2 \rangle = \frac{\bar{\epsilon}}{9\nu} r^2$. 
This means that the prediction by the second assumption for the radial and tangential relative 
speeds of $St \ll1$ particles differs from the S-T formula, although it reproduces the S-T prediction for the 3D rms.  
We will test the model predictions for the relative velocity variances measured from our simulation data.


\section{Statistics of the Simulated Flow}

In this section, we describe the numerical method used in our simulation and discuss 
the statistical properties of the simulated flow. 
Our simulation was conducted in a periodic $512^3$ box with a 
length of $2\pi$ on each side. Using the Pencil code\footnote{http://pencil-code.nordita.org} 
(Brandenburg \& Dobler 2002, Johansen, Andersen, \& Brandenburg 2004), we 
evolved the hydrodynamic equations, 
\begin{gather}
\frac{\partial \rho}{\partial t} + \frac{\partial}{\partial x_i}\left(\rho  u_i\right) = 0,  \notag \\
\frac{\partial {u}_i} {\partial t} +  {u_j} \frac{\partial {u_i}}{\partial x_j}=  \frac{1}{\rho}\frac{\partial}{\partial x_j}\left[\rho \nu\left(\frac{\partial u_i}{\partial x_j} + \frac{\partial u_j}{\partial x_i} - \frac{2}{3}\delta_{ij} \frac{\partial u_k}{\partial x_k} \right)  \right] \hspace{3.5cm} \notag \\
-\frac{1}{\rho}\frac{\partial p}{\partial x_i} + { f_i}, \hspace{1.7cm}
\end{gather}
with an isothermal equation of state, $p = \rho C_{\rm s}^2$. The sound speed 
is set to unity, i.e., $C_{\rm s} = 1$. The kinematic viscosity, $\nu$, is taken to be 
constant, $\nu = 5\times 10^{-5}$. A large-scale force, $f_i$, generated in 
Fourier space using 20 modes in the wavenumber range of 
$1\le k \le 2$ is applied to drive and maintain the turbulent flow. The driving length scale, $L_{\rm f}$, 
is  thus about 1/2 box size. The balance between the energy input by the 
driving force and the dissipation by viscosity leads to a statistical steady 
state with a 1D rms velocity, $u'$, of 0.05, or a 3D rms of $0.085$. This weakly 
compressible flow is suitable for the application to turbulence in protoplanetary disks. 
At an rms Mach number of $0.085$, the flow statistics is essentially the same 
as incompressible turbulence (Padoan et al.\ 2004, Pan \& Scannapieco 2011). 

The integral length scale, $L$, in our simulated flow is found to be 
$\simeq 1$, i.e., about 1/6 box size. It is about 3 times smaller 
than the driving scale, $L_{\rm f}$. The integral scale, $L$, 
represents the (longitudinal) correlation length of the velocity field, 
and we computed it from the energy spectrum, $E(k)$, of the flow, 
using the relation $L = \frac{\pi}{2u'^2} \int k^{-1} E(k) dk$ 
(Monin \& Yaglom 1975).  The energy spectrum, $E(k)$, is plot in the inset 
of Fig.\ (\ref{structureandpower}).  
With $L=1$,  the large-eddy turnover time is $T_{\rm eddy} = L/u' = 20$ 
in units in which the sound crossing time is $2 \pi$.  

The average energy dissipation rate per unit volume by the viscosity term 
is given by $\bar{\epsilon} = \frac{1}{2\bar{\rho}} \langle \rho\nu (\partial_i u_j + {\partial_j u_i} -\frac{2}{3}\delta_{ij} {\partial_k u_k} )^2 \rangle$, where $\bar{\rho}$ 
is the average density. In our weakly compressible flow, the density fluctuations and the velocity 
divergence can be neglected, and the dissipation rate can  
be estimated by $\bar{\epsilon} = \nu \langle \omega^2 \rangle$, where $\langle \omega^2 \rangle$ 
is the vorticity variance. We find that $\langle \omega^2 \rangle = 0.92$, 
implying that $\bar{\epsilon} \simeq 4.6 \times 10^{-5}$. We also evaluated the 
dissipation rate from the 3rd-order longitudinal structure function using Kolmogorov's 4/5 law, 
$\langle \Delta u_{\rm r}(\ell)^3 \rangle = -\frac{4}{5}\bar{\epsilon}\ell$, for $\ell$ in the inertial range. 
This latter method gives a larger dissipation rate, $\bar{\epsilon} = 5 \times 10^{-5}$, suggesting 
that a small fraction, $\simeq 8\%$, of kinetic energy is dissipated by numerical diffusion. 
The effective viscosity is thus larger than the adopted value by the 
same amount. We take the effective viscosity to be $5.4\times 10^{-5}$ and use it 
in our estimates of the Kolmogorov scales. We compute the Kolmogorov 
timescale from the vorticity variance as $\tau_\eta = \langle w^2 \rangle^{-1/2} =1.04$. The 
Kolmogorov length scale is estimated to be $\eta = (\nu^3/\bar{\epsilon})^{1/4} = 0.0075$, 
which corresponds to $\simeq 0.6$ cell size of the computation grid. The Kolmogorov velocity scale 
is $u_\eta = (\nu \bar{\epsilon})^{1/4} = 0.0072$ in units of the sound speed.   

The Reynolds number of our simulated flow is $Re \equiv u' L/\nu \simeq 1000$. 
A more commonly-used Reynolds number in turbulence studies is the
Taylor Reynolds number, $Re_{\lambda}  \equiv u' \lambda/\nu$, where 
the Taylor micro length scale is defined as $\lambda \equiv (15 u'^2/ \langle \omega^2\rangle )^{1/2}$.  
We find that $\lambda  = 0.2 $ in our simulated flow, and thus $Re_{\lambda} \simeq 200$. 
From the definitions of $u_\eta$ and $Re_{\lambda}$, we have $u'/u_\eta = (Re_{\lambda}/\sqrt{15})^{1/2}$. 

\subsection{The Lagrangian Correlation Function and the Timescales}

To study the Lagrangian statistics, we integrated the trajectories 
of $33.6$ million tracer particles with zero inertia in the simulated flow. 
The total number of tracer particles corresponds to an average 
number density of 1 particle per 4 computational cells. 
To obtain the particle velocity inside a cell, we selected 
the triangular-shaped-cloud interpolation method already implemented 
in the Pencil code (Johansen and Youdin 2007). We output the 
particle positions to a data file in each $0.1 \tau_\eta$. 
The Lagrangian correlation function, $\Phi_{\rm L} (\Delta \tau)$, 
is computed as the average of the velocity correlation, 
$\langle u_{i}({\bs X}(t), t) u_{i}({\bs X}(t+\Delta \tau), t +\Delta \tau)  \rangle/3u'^2$, 
along the trajectories, ${\bs X}(t)$, of all particles. We considered both 
positive and negative $\Delta \tau$, corresponding to Lagrangian
trajectories forward and backward in time, respectively. Our data 
confirmed that $\Phi_{\rm L}$ is an even function of $\Delta \tau$, as expected 
from statistical stationarity (see \S 2).  We find that the Lagrangian 
correlation timescale, $T_{\rm L}$ ($\equiv \int \Phi_{\rm L} (\Delta \tau) d\Delta \tau$), 
is $\simeq 15$, which is about 0.75 eddy turnover time, $T_{\rm eddy}$. 
This is consistent with the simulation result of Yeung et al.\ (2006). 
Since $\tau_{\eta} = 1.04$ in our flow, 
we have $T_{\rm L} = 14.4 \tau_{\eta}$.

\begin{figure}[t]
\centerline{\includegraphics[width=1.1\columnwidth]{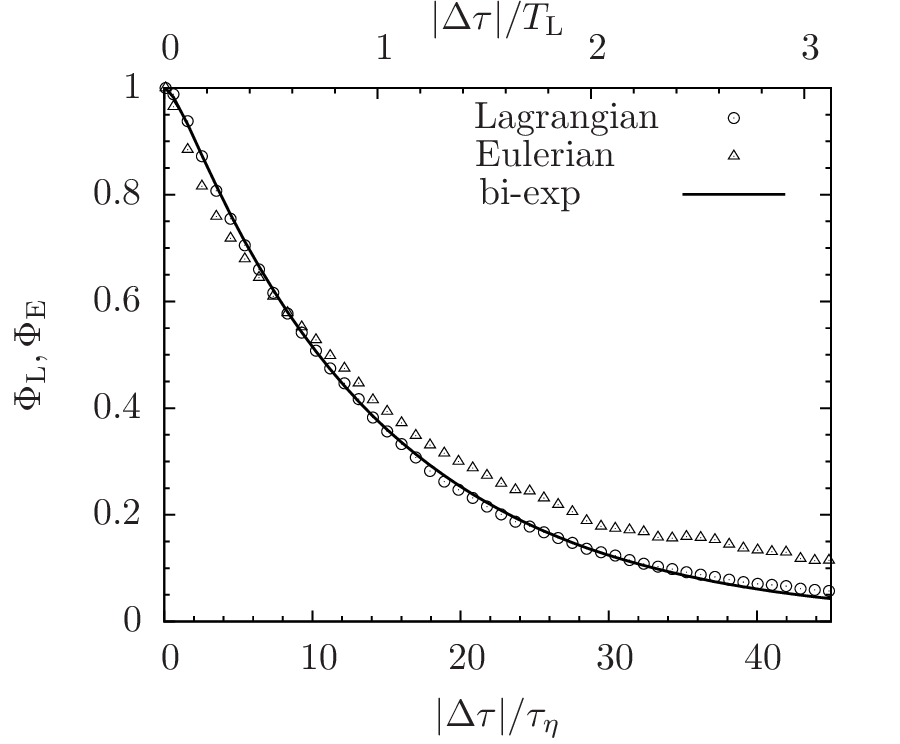}}
\caption{Lagrangian ($\Phi_{\rm L}$; circles) and Eulerian ($\Phi_{\rm E}$; triangles) temporal correlation 
functions in our simulated flow.  The time lag, $\Delta \tau$, is normalized to the Kolmogorov timescale, 
$\tau_\eta$ and the Lagrangian correlation time, $T_{\rm L}$, on the bottom and top X-axises, respectively.
The solid line shows the best fit for $\Phi_{\rm L}$  using 
the bi-exponential function, eq.\ (\ref{biexponential}).}
\label{correlation} 
\end{figure}

The Lagrangian correlation, $\Phi_{\rm L}$, in our flow is plot as circles in Fig.\  \ref{correlation}, 
where the time lag, $|\Delta \tau|$, is normalized to the Kolmogorov timescale, $\tau_{\eta}$ and 
the Lagrangian correlation time, $T_{\rm L}$, on the bottom and top X-axises, respectively. 
The solid line shows the bi-exponential function, eq.\ (\ref{biexponential}), given in \S 2. 
The parameter $z$ is set to 0.3, which suggests that the Taylor micro timescale, 
$\tau_{\rm T}$, is $\simeq 4.3 \tau_{\eta}$. This value of $\tau_{\rm T}$ corresponds 
to an acceleration variance, $a^2 \simeq 5.2 (u_{\eta}/\tau_{\eta})^2$.
The bi-exponential function matches very well the simulation data. On the other 
hand, we find that a single exponential function could not give a satisfactory fit to $\Phi_{\rm L}$.

We also considered the Eulerian temporal correlation function, $\Phi_{\rm E}(\Delta \tau)$.
It is computed as the average, $\langle u_{i}({\bs x}, t) u_{i}({\bs x}, t +\Delta \tau)  \rangle/3u'^2$, 
over all grid points ${\bs x}$. The result is plot as triangles 
in Fig.\ \ref{correlation}. 
$\Phi_{\rm E}$ is smaller than the Lagrangian correlation $\Phi_{\rm L}$ at small time lags, 
and then becomes larger at $|\Delta \tau| \gsim 8 \tau_{\eta} \simeq 0.55 T_{\rm L}$. 
Due to the slower decrease of $\Phi_{\rm E}$ at large time 
lags, the Eulerian correlation time, $T_{\rm E} \equiv \int \Phi_{\rm E} (\Delta \tau) d\Delta \tau$, 
is slightly (10\%) larger than $T_{\rm L}$. We find that $T_{\rm E} = 15.9 \tau_\eta$. 
The Eulerian correlation function is of interest for large inertial particles 
with $\tau_{\rm p} \gg T_{\rm L}$. Due to their large inertia, these particles 
have small velocities and thus may stay around as the flow sweeps by. 
Therefore, unlike small particles, the temporal series of the flow velocity ``seen" 
by the large particles may be better described by the Eulerian 
velocity. This suggests that, for $\tau_{\rm p} \gsim T_{\rm L}$, it may 
be appropriate to replace the Lagrangian correlation used in our 
model by the Eulerian correlation. However, the Eulerian correlation 
function and timescale are quite close to the Lagrangian ones, and 
using the Lagrangian correlation for all particles in our model gives satisfactory 
predictions for both the 1-particle velocity and the 2-particle relative velocity 
at any $St$ (\S 5 and \S 6.1). 

We summarize the relevant timescales in the simulated flow and list them in an increasing order. 
The smallest timescale is Kolmogorov time $\tau_{\eta}$, and we use it as a reference timescale. 
The Taylor micro scale, $\tau_{\rm T}$, was found to be $4.3 \tau_{\eta}$ from 
the bi-exponential fit to the Lagrangian correlation function. The next timescale is the 
Lagrangian correlation time, $T_{\rm L}$, which is $14.4 \tau_\eta$. The
Eulerian correlation time is slightly larger, $T_{\rm E} \simeq 15.9 \tau_\eta$. The large eddy 
turnover time, $T_{\rm eddy}$, was measured to be $\simeq 19.2 \tau_\eta$.  Another commonly-used 
timescale is the dynamical time, $\tau_{\rm dyn}$, defined as the forcing length scale, 
$L_{\rm f}$, divided by the 3D rms velocity ($\sqrt{3}u'$). We find that $\tau_{\rm dyn} = 35 \tau_\eta$. 

In this work, we will  express the particle friction time primarily by $St$ and $\Omega$.  
They correspond to normalizations to $\tau_{\eta}$ and $T_{\rm L}$, which  are convenient for small and large particles, 
respectively. One may also normalize $\tau_{\rm p}$ to the large eddy turnover time, 
and define $\Omega_{\rm eddy} = \tau_{\rm p}/T_{\rm eddy}$, which may be more convenient
for practical applications. However, we prefer using $\Omega$ than $\Omega_{\rm eddy}$, because, according to our model, 
it is $T_{\rm L}$  that directly enters the physics of turbulence-induced particle velocity. 
Using the measured values of the timescales in our simulation,  one may convert 
the normalizations by $\Omega_{\rm eddy} = 0.75 \Omega = 0.052 St$.


\subsection{The Flow Structure Functions and Energy Spectrum}

\begin{figure}[t]
\centerline{\includegraphics[width=1.05\columnwidth]{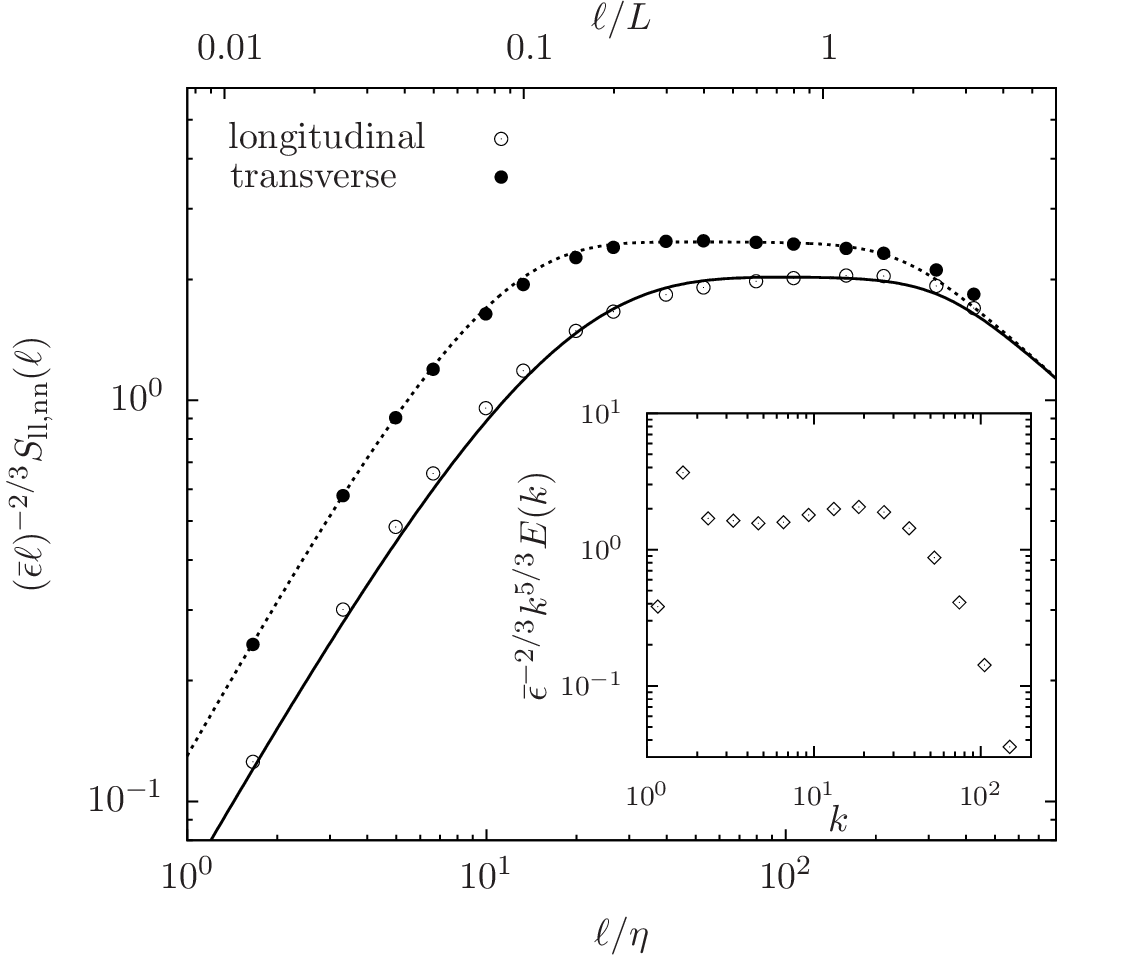}}
\caption{Longitudinal (open circles) and transverse (filled circles) 
structure functions in our simulated flow. The structure functions 
are compensated by the Kolmogorov scaling in the inertial range. 
The solid and dashed lines are fitting functions, eqs.\ (\ref{sll}) and (\ref{snn}), for $S_{\rm ll}$ and $S_{\rm nn}$, 
respectively. The bottom and top X-axises normalize $\ell$ to the Kolmogorov scale and the 
integral scale, respectively. The inset shows the energy spectrum of the flow, 
compensated by $\bar{\epsilon}^{2/3} k^{-5/3}$. 
}
\label{structureandpower} 
\end{figure}

In Fig.\ \ref{structureandpower}, we show the longitudinal ($S_{\rm ll}$; open circles) 
and transverse ($S_{\rm nn}$; filled circles) structure functions in our simulated flow. 
The structure functions are measured from the velocity differences along the 3 directions, ${\bs e}_1$, ${\bs e}_2$ and ${\bs e}_3$, 
of the simulation grid. For $S_{\rm ll}(\ell)$, we computed and averaged the variances of $\Delta u_{11} (\ell) $($\equiv u_1 ({\bs x} + \ell {\bs e}_1)- u_1({\bs x})$), 
$\Delta u_{22} (\ell)$ and $\Delta u_{33}(\ell)$ over all the points, ${\bs x}$. 
Similarly, $S_{\rm nn}(\ell)$ is obtained by averaging the variances of $\Delta u_{12} (\ell) $($\equiv u_1 ({\bs x} + \ell {\bs e}_2)- u_1({\bs x})$), $\Delta  u_{13} (\ell) $, 
$\Delta u_{21} (\ell) $, $\Delta u_{23} (\ell) $, $\Delta  u_{31} (\ell) $ and $\Delta  u_{32}(\ell)$. 

As discussed earlier, Kolmogorov's similarity theory predicts that 
$S_{\rm ll}(\ell)\simeq C_{\rm K} (\bar{\epsilon} \ell)^{2/3}$ for $\ell$ in the inertial range.
We thus compensated the structure functions by $(\bar{\epsilon} \ell)^{2/3}$ 
in Fig.\ \ref{structureandpower}. A limited inertial range is seen in both 
$S_{\rm ll}$ and $S_{\rm nn}$. The Kolmogorov constant $C_{\rm K}$ is about 2. 
In the inertial range, the scaling exponent for $S_{\rm ll}$ is found to be 
slightly larger than 2/3, while the slope of $S_{\rm nn}$ is close to 2/3. 
The ratio of the two structure functions in the inertial range 
is about 1.25, slightly smaller than the value, 4/3, expected from the 
incompressibility condition (see \S 3). This is perhaps because our flow is 
weakly compressible. Another possibililty is that the inertial range 
is too short to allow an accurate measurement of this ratio.    
Both structure functions become smooth, i.e., $\propto \ell^2$, as $\ell$ decreases toward the 
Kolmogorov scale, and approach $2u'^2$ in the limit $\ell \gg L$ (\S 3).  

The solid line in Fig.\ \ref{structureandpower} is the connecting formula, eq.\ (\ref{sll}), for $S_{\rm ll}$ (\S 3). 
We set $C_{\rm K}=2$ in the formula. The line gives a fairly good fit to the data points. As discussed in \S 3, 
with the connecting formula for $S_{\rm ll}$, one may obtain a fitting function for $S_{\rm nn}$ using the incompressibility 
relation $S_{\rm nn} = S_{\rm ll} + \frac{1}{2} \ell dS_{\rm ll}/d\ell$. However, the fitting function 
obtained this way overestimates $S_{\rm nn}$ in the inertial range, perhaps because the incompressibility 
condition does not exactly hold in our flow (see above). For a more accurate fit, we adopted a separate connecting formula for $S_{\rm nn}$,  
\begin{gather}
S_{\rm nn} = 2  u'^2 \left[ 1- \exp \left( - \frac{(\ell/ \eta)^{4/3}}{(15C_{\rm Kn}/2)} \right) \right]\times \hspace{2cm} \notag\\
\hspace{2cm}\left [\frac{(\ell/ \eta)^4}{(\ell/ \eta)^4 + (2u'^2 /C_{\rm Kn} u_\eta^2 )^{6} } \right]^{1/6},  
\label{snn}
\end{gather}
where $C_{\rm Kn}$ is the scaling coefficient for $S_{\rm nn}$ in the inertial range. 
This connecting formula correctly reproduces the scaling behaviors of $S_{\rm nn}$ 
in different scale ranges.  Its form is slightly different from  eq.\ (\ref{sll}) for $S_{\rm ll}$. 
The dotted line in Fig.\ \ref{structureandpower} corresponds to eq.\ (\ref{snn}) with 
$C_{\rm Kn} \simeq 1.25 C_{\rm K} = 2.5$. We will use eqs.\ (\ref{sll}) and (\ref{snn}) 
in the computation of our model prediction for the particle relative velocity.

The inset of Fig.\ \ref{structureandpower} show the energy spectrum, 
$E(k)$, of our flow. The Kolmogorov theory predicts $E(k) = K {\bar \epsilon}^{2/3} k^{-5/3}$ 
in the inertial range, and we compensated the spectrum by $\bar{\epsilon}^{2/3} k^{-5/3}$. 
The power-law range ($3 \le k \le 10$) in the spectrum appears to be shorter than in the structure functions. 
The constant $K$ is measured to be $\simeq 1.7$, consistent with previous studies (Ishihara et al.\ 2009). 
It is also consistent with the relation $K=0.76 C_{\rm K}$ (Monin \& Yaglon 1975), as
the Kolmogorov constant, $C_{\rm K}$, for $S_{\rm ll}$ was found to be $\simeq 2$.  

\section{One-particle Root-mean-square Velocity}

In our simulation, we included 14 species of inertial particles of different sizes. 
The friction timescale of the particles spans about four decades from 
$\simeq 0.1\tau_\eta$ to $\simeq  41 T_{\rm eddy}$ ($\simeq 54 T_{\rm L}$ or $\simeq 800 \tau_\eta$), 
covering the entire scale range of the simulated flow. The friction timescale is equally 
spaced, increasing by a factor of two in each successive species. 
The number of particles contained in each species is 33.6 million, 
corresponding to an average particle density of one per 4 computational cells. 
The same number of tracer particles was included  to study the 
Lagrangian statistics (\S 4.1). The integration of the particle trajectories is computationally very expensive. 
Using 4096 cores (512 Harpertown nodes) on the NASA/Ames Pleiades 
supercomputer, the simulation was run for 14 days, corresponding to a total CPU cost of 1.4 million hours.

To evolve the particle equation of motion (eq.\ \ref{particlemomentum}), we adopted the 
triangular-shaped-cloud (TSC) method to interpolate the flow velocity inside the 
computational cells. The TSC interpolation is a well-established method (Hockney \& Eastwood 1981, 
Johansen \& Youdin 2007) that makes use of the nearest 27 grid points in
 a 3D simulation. In 1D, the weighting factor for the nearest 3 
 grid points is set to be quadratic with the distance to the points. 
The velocity difference in our simulated flow is linear with $\ell$ around and below 
the cell size, $\Delta x$, as seen from the $\ell^2$ scaling of the structure functions 
toward $\Delta x$ (bottom data points) in Fig.\ \ref{structureandpower}. 
This implies that the subgrid velocity field can be well approximated by a linear interpolation (Pan et al.\ 2011). 
The linear scaling is also captured by the TSC method. It is straightforward to show that, 
if the flow velocity is already linear around the resolution scale 
(approximately the case in our simulated flow), the scaling of the interpolated 
velocity at subgrid scales by the TSC method would be exactly linear, 
as the quadratic terms in the weighting functions cancel out in this special 
case. In comparison to the linear interpolation, the TSC method is of higher 
order and has the advantage of smoother connections at cell boundaries.

Initially, the 33.6 million particles in each species are distributed randomly in the simulation 
box. Each component of the initial particle velocity is also random, independently drawn from a 
uniform distribution in the range [-0.01, 0.01]. Therefore, the initial (1D) rms, $v'(0)$, of 
each velocity component of all the particles is $0.01/\sqrt{3}$, 
equivalent to a 3D rms of 0.01. The numerical values given here
are in units of the gas sound speed, which was set to unity in the simulation. 
The initial particle conditions for all the 14 species are the same. We evolved 
the turbulent flow and the particle trajectories together right from the beginning of the simulation.
At time zero, the gas velocity and density are set to zero and unity, respectively. 

\begin{figure}[t]
\centerline{\includegraphics[width=1.1\columnwidth]{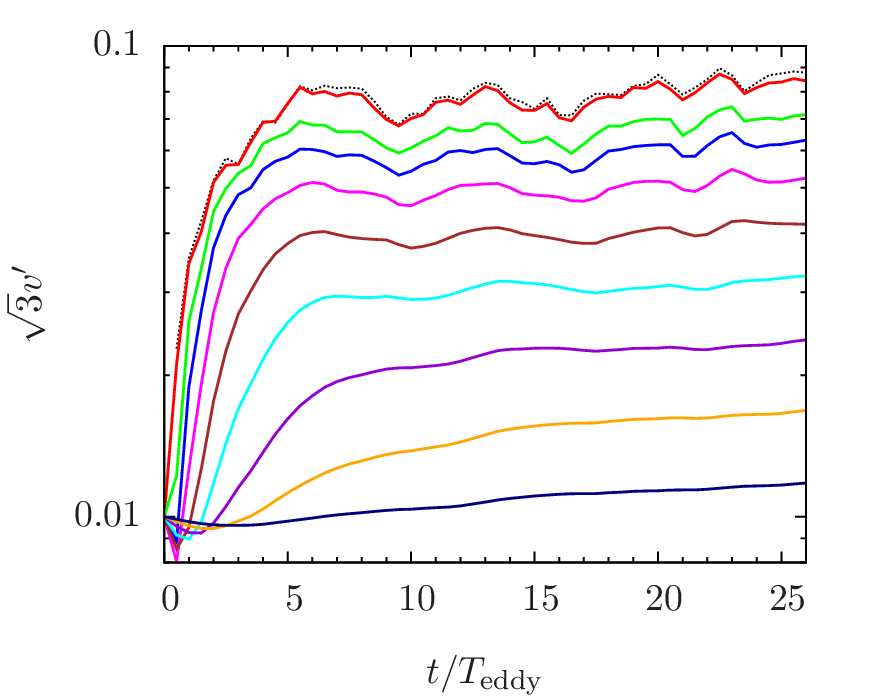}}
\caption{The temporal evolution of  the 3D rms velocity, $\sqrt{3}v'$, of inertial particles of different 
sizes. The thin dotted line shows the 3D rms flow velocity, which reaches a (quasi) steady 
state at $t \gsim 5-10 T_{\rm eddy}$.  The flow rms velocity is 0 
at time zero, and the dotted line starts from $0.5 T_{\rm eddy}$ when the first snapshot is saved. 
From top to bottom, the color lines correspond to the rms velocity for particles with 
$St = 0.39$, 6.12, 12.4, 24.9, 49.7, 99.4, 199, 398 and 795, respectively. 
The velocity is in units of the sound speed of the flow, which is set to unity in the simulation.} 
\label{rms0} 
\end{figure}

Our simulation run lasted for about $26 T_{\rm eddy}$ (or $35 T_{\rm L}$), and we saved 52 snapshots with 
an equal separation of $0.5 T_{\rm eddy}$. The black dotted  line in Fig.\ \ref{rms0}  is the 
3D rms of the flow velocity ($\sqrt{3} u'$) as a function of time, which shows that the flow is fully developed 
and reaches a (quasi) steady state at $t_{\rm dev} \simeq 5-10T_{\rm eddy}$. 
From top to bottom, the colored lines in Fig.\ \ref{rms0} plot the 3D rms 
velocities of inertial particles with $St (\Omega)[\Omega_{\rm eddy}] = 0.39(0.027)[0.02]$, 
$6.21(0.41)[0.32]$, $12.4(0.84)[0.65]$,  $24.9(1.7)[1.3]$, $49.7(3.4)[2.6]$, $99.4(6.8)[5.2]$, 
$199(13.5)[10.4]$, $398(27)[20.7]$, and 795(54)[41.4], respectively.
The numbers in the parentheses and square brackets correspond to $\Omega$ and $\Omega_{\rm eddy}$, respectively. 
We find there are dips at earlier times in the curves for relatively large particles.  When our first snapshot was saved at $0.5 T_{\rm eddy}$, 
particles with $\tau_{\rm p} \gsim 0.5 T_{\rm eddy}$ partially lost the memory of the initial 
rms velocity, and meanwhile their velocity had some contribution from the flow 
velocity, $u'$, between $t=0$ and $0.5 T_{\rm eddy}$. However, since $u'$ was 0 at $t=0$, 
this contribution turns out to be small and does not compensate the decrease due to the memory loss of the initial velocity. 
This causes a decrease in $v^{\prime}$ and leads to dips 
at $t \gsim  0.5 T_{\rm eddy}$.  
Due to their short memory time, the small particles forgot their 
initial velocity, $v'(0)$, at the first snapshot, and their velocity 
was close to the flow velocity at $0.5 T_{\rm eddy}$, which is already slightly larger than $v'(0)$. 
Therefore, no dips appear for the small particles with $\tau_{\rm p} \lsim 0.5 T_{\rm eddy}$. 
The top six color lines appear to reach a steady state at $\simeq 10 T_{\rm eddy}$. For the bottom 
three lines, $v'(t)$ keeps increasing gradually but almost monotonically. This may imply that these 
largest particles need more time to relax. It is also possible that the slow increase of $v'$ at late 
times is simply caused by the slight rise of the flow velocity (see the black dotted line).


The relaxation timescale for inertial particles in a stationary turbulent 
flow is essentially the time for the particles to forget the initial condition, and is roughly  
given by the friction timescale, if the initial velocity is not much larger than the final steady-state value 
(as is the case for our initial conditions).  The estimate for the  
relaxation time in our simulation is a little complicated because the particles are released 
to the flow before $t_{\rm dev}$. For particles with $\tau_{\rm p} \ll t_{\rm dev}$, the dynamical relaxation is expected once the flow is fully developed, 
i.e., at $\simeq t_{\rm dev}$. This is the case for the top 5-6 color lines in Fig.\ \ref{rms0}. 

For the bottom 3-4 lines, $\tau_{\rm p} \gsim t_{\rm dev}$, and we 
expect these particles would be relaxed at some time in the range ($\tau_{\rm p}$, $t_{\rm dev} + \tau_{\rm p}$). 
The lower limit is the minimum relaxation time, and the upper limit is based on 
the consideration that, if the particle evolution started at $t_{\rm dev}$ instead of time 0, 
the particles would relax at $\simeq t_{\rm dev} + \tau_{\rm p}$. From this estimate, the third largest particles ($St=199$) are relaxed by $\lsim 20 T_{\rm eddy}$,
and the second largest ones ($St=398$) are likely relaxed by the end of the simulation. 
On the other hand, the largest ($St=795$) particles may not have reached a relaxed state. 
However, the quite flat $v' (t)$ of the $St=795$ particles indicates the possibility they are actually relaxed 
toward the end of the simulation. If that is the case, a likely reason for it is that the 
chosen initial condition (e.g., the rms velocity) happens to be very similar to the expected 
relaxed state of these particles. This similarity may reduce the relaxation time. 
We assume that all particles in our simulation are relaxed in the last 5-6 $T_{\rm eddy}$.
  

In our data analysis, we average over three snapshots at $t= 21.5$, 24 and 
$26 T_{\rm eddy}$. 
For the uniformity of the data sample, we use the same snapshots for 
all particle species. Since the largest particles become relaxed around
the end of the simulation, we only select  late snapshots at $t\gsim 20 T_{\rm eddy}$. 
The purpose of averaging over a number of snapshots is to obtain better statistics by increasing the sample size. 
It is thus helpful to use well-separated snapshots with independent statistics.
A temporal separation of $\simeq 2 T_{\rm eddy}$ guarantees the 
particle velocities at the selected snapshots are independent for the first 10 particle 
species. The velocities of the largest four particles remain correlated for significantly 
longer than $\simeq 2 T_{\rm eddy}$. Therefore, unlike the case of smaller particles, using 
the selected snapshots may not effectively increase the independent
sample size or the measurement accuracy. If the computation resources allow, it would be 
ideal to run the simulation much longer and collect snapshots separated by a few friction times of the largest particles.

\begin{figure}[t]
\centerline{\includegraphics[width=1.1\columnwidth]{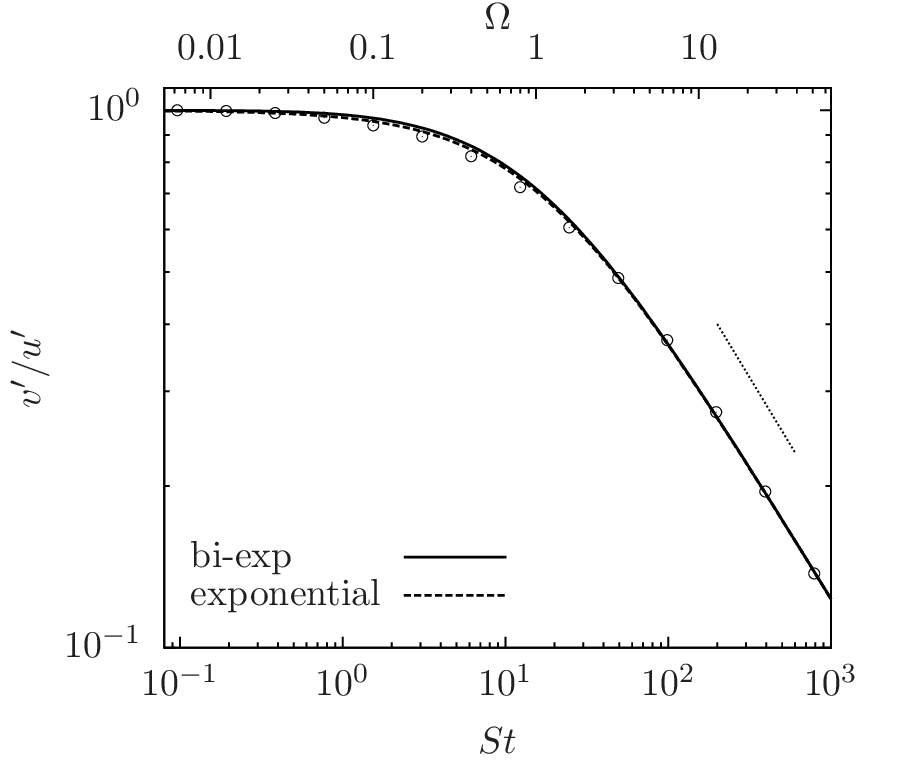}}
\caption{The 1-particle rms velocity, $v'$, as a function of the particle friction time. 
The bottom and top X-axises normalize $\tau_{\rm p}$ to $\tau_\eta$ and $T_{\rm L}$, 
respectively. 
The dotted and solid lines are the predictions, eq.\ (\ref{1particlevelocityexp}) and eq.\ (\ref{1particlevelocitybiexp}), 
of our model using single- and bi- exponential forms for the temporal correlation function, 
respectively. In both cases, $T_{\rm L}$ is set to $15.4 \tau_\eta$, and the parameter $z$ 
in the bi-exponential case is set to 0.3. The thin dotted line segment denotes a $St^{-1/2}$ scaling.} 
\label{rms} 
\end{figure}

  
In Fig.\ \ref{rms}, we show the simulation result for the 1D rms, 
$v'$, of the particle velocity as a function of $St$. We normalized $v'$ by the rms velocity, $u'$, of the flow. 
The top X-axis normalizes the friction time to the Lagrangian correlation time. 
One may convert $St$ (or $\Omega$) to $\Omega_{\rm eddy}$ by 
$\Omega_{\rm eddy}  = 0.052 St$ (or $\Omega_{\rm eddy} =0.75 \Omega)$. 
The dotted and solid lines are predictions, eq.\ (\ref{1particlevelocityexp}) and eq.\ (\ref{1particlevelocitybiexp}), 
of our model using single- and bi-exponential forms for the temporal correlation function, respectively. 
The model approximates the trajectory correlation function, $\Phi_1$, by 
the Lagrangian correlation function, $\Phi_{\rm L}$ (see \S 2). For the bi-exponential case, we set the 
parameter $z =0.3$, which best fits $\Phi_{\rm L}$ measured from the Lagrangian trajectories (see \S 4.1). The two lines almost coincide, indicating that the model prediction 
for $v'$ is insensitive to the exact form of the correlation function, 
and depends only on the correlation timescale\footnote{Unlike 
the case of the 1-particle velocity,  the choice of $\Phi_1$ is crucial 
for predicting the relative velocity between different particles in the S-T limit (PP10). 
In the bidisperse case, adopting the bi-exponential form for $\Phi_{\rm L}$ and 
$\Phi_1$ is needed to reproduce the acceleration contribution to the relative velocity, 
while a single exponential form cannot correctly capture the effect of the 
flow acceleration on $\Phi_{\rm L}$ at small time lags.}. 
In both curves, we set $T_{\rm L} = 15.4 \tau_\eta$, generally 
consistent with the directly measured value of $14.4 \tau_\eta$ (see \S 4.1). 
As expected, $v'$ follows the $St^{-1/2}$ scaling (the dotted line segment) at $\Omega \gg 1$.  
This supports our claim that the largest particles are dynamically relaxed at the end of the run. 

The flow velocity ``seen" by large particles with $\tau_{\rm p} \gsim T_{\rm L}$ may 
be closer to Eulerian than Lagrangian (\S 4.1), and thus using $\Phi_{\rm L}$ for the trajectory correlation 
$\Phi_1$ is not well justified (\S 2). However, the assumption is 
validated by our simulation result for all particles. 
This is because, first, based on our model prediction, $v'$ is 
controlled mainly by the correlation time, not the form of the correlation function,
and, second, the Eulerian correlation timescale, $T_{\rm E}$, was found to be close to $T_{\rm L}$. 
Thus, whether $\Phi_1$ is approximated by $\Phi_{\rm E}$ or $\Phi_{\rm L}$, 
the predicted 1-particle velocity would be similar, justifying the use of $\Phi_{\rm L}$ for 
all particles. The best-fit correlation timescale, $15.4 \tau_\eta$, used in Fig.\ \ref{rms} is in between 
the measured values of $T_{\rm L}$ ($14.4 \tau_\eta$) and $T_{\rm E}$ ($15.9 \tau_\eta$).  
This suggests that the temporal statistics of the flow velocity ``seen" by a large 
particle is in between that along a Lagrangian trajectory and that at a fixed Eulerian point.  
It is interesting to note that, although a single-exponential function does not 
well fit either $\Phi_{\rm E}$ or $\Phi_{\rm L}$ (\S 4.1), our model prediction for the 
rms particle velocity with an exponential correlation is in good agreement with the simulation data.


We also computed the probability distribution of the 1-particle velocity. 
The velocity of small particles is expected to be Gaussian because it simply 
samples the 1-point PDF of the flow velocity, which is close to Gaussian. 
For large particles with $\tau_{\rm p} \gg T_{\rm L}$, the velocity would also be 
Gaussian because their equation of motion is essentially a Langevin equation. 
We find the 1-particle velocity PDF is indeed nearly Gaussian at all $St$. The nearly 
Gaussian PDF for the largest particles in the three selected snapshots also
supports our assumption that these particles are relaxed at $t \gsim 20 T_{\rm eddy}$. 




\section{The Relative Velocity of Inertial Particles}

Now we explore the statistics of 2-particle relative velocity  in our simulation, focusing on the monodisperse case of 
equal-size particles. Using the simulations data, we can compute the joint probability distribution, 
$\rho({\bs r}, {\bs w}; St)$, of the particle separation and the relative velocity as a function of $St$ (see, e.g., Zaichik et al.\ 2003, Gustavsson and Mehlig 2011, 
Hubbard 2013). The joint distribution is defined such that the number of particles  located  
in a volume $dV$ at a separation ${\bs r}$ from a reference particle and moving 
at a relative velocity in the range $[{\bs w}, {\bs w} + d {\bs w}]$ is given by $\bar {n} \rho({\bs r}, {\bs w}; St) dV d {\bs w}$, where $\bar {n}$ 
is the average number density. 
Once the particle statistics become isotropic, it is convenient to study the distributions, 
$\rho(r, w_{\rm r}; St)$, $\rho(r, w_{\rm t}; St)$, and $\rho(r, |{\bs w}|; St)$, for the radial, tangential 
components and the 3D amplitude of the relative velocity. 
The normalization of $\rho(r, w_{\rm r}; St)$ with respect to $w_{\rm r}$ is given by $\int_{-\infty}^{\infty} \rho(r, w_{\rm r}; St) dw_{\rm r} =g(r, St) $, 
where $g(r, St)$ is the so-called radial distribution function (RDF). The same normalization applies
for $\rho(r, w_{\rm t}; St)$ and $\rho(r, |{\bs w}|; St)$. The RDF represents the overall probability of finding a 
neighbor with any relative speed, and is a measure of the spatial clustering of inertial particles (see \S 7). 
With the joint distributions and the RDF, we define the relative speed PDFs  as 
$P(w_{\rm r}, St) = \rho(r, w_{\rm r}; St)/g(r, St)$, $P(w_{\rm t}, St) = \rho(r, w_{\rm t}; St)/g(r, St)$, and $P(|{\bs w}|, St) = \rho(r, |{\bs w}|; St)/g(r, St)$. 
For the simplicity of notation, the dependence of the PDFs on $r$ is not explicitly indicated. A systematic study of these PDFs is given in \S 6.2. 
Our simulation results for the variances of the PDFs 
are presented in \S 6.1.


For each species ($St$), we measure the relative velocity of particle pairs mainly at three distances, $r =1\eta$, $0.5\eta$ and $0.25\eta$. These distances 
are below the resolution scale of the simulation. Measuring the statistics at subgrid scales is justified, as 
the subgrid flow velocity is reliably captured by the adopted TSC interpolation (\S 5). For each $St$ and $r$, we search the simulation box 
for all particle pairs in a distance shell from $r-\delta r/2$ to $r+\delta r/2$. For $r =1\eta$ and 
$0.5\eta$, we set the shell thickness $\delta r$ to $0.08 r$. To increase the number of particle pairs 
hence the measurement accuracy at $r=0.25 \eta$, we used a larger thickness, 
$\delta r = 0.16 r$, which is likely the largest  value one can reasonably adopt.
If $\delta r$ is  increased further by a  factor of 2, it would 
be comparable to $r$, and one may not safely attribute the measured statistics to a single 
particle distance.  For $r=0.25 \eta$ and $\delta r = 0.16 r$, 
the number of particle pairs in one snapshot is typically on the order of $\simeq 10^4$. 
This number of pairs is about enough to provide sufficient statistics, although the measured relative speed PDF  at $r=0.25 \eta$ 
already shows considerable noises at the tails (see Fig.\ \ref{pmpdf} in \S 6.2.3). 
A study of smaller values of $r$ is desirable. However, at smaller $r$, the number 
of particle pairs available becomes more limited and does not allow accurate statistical analysis. 
We take the statistical accuracy as priority, and restrict our data analysis to $r \ge \eta/4$. 

To check the relaxation of the 2-particle statistics, we examined the temporal 
evolution of the particle pair counts, or equivalently the RDF, and the rms relative velocity 
in our simulation.  The RDFs for all the 14 particle species reach a quasi steady state 
at $\simeq 10 T_{\rm eddy}$, when the turbulent flow is fully developed (Fig.\ \ref{rms0}). 
This further confirms that the dynamics of the particles in the first 10 species is well
relaxed at $10 T_{\rm eddy}$. However, the steady state of the RDF is not a perfect indicator for 
the relaxation of the largest few particles.  These particles do not show 
significantly clustering (see \S 7.1), and their RDFs are close to unity at all times. 
The rms of the relative velocity also reaches a quasi steady state at $\gsim 10 T_{\rm eddy}$ 
for essentially all particles. As discussed in \S 5, there is an uncertainty in the relaxation of the 
largest particles ($St =795$) because the expected relaxation time is larger than the end time of the 
simulation. Our data shows that the rms relative velocity of the largest particles is about equal 
to $\sqrt{2}$ times the 1-particle rms velocity, consistent with the expected relaxed state (see \S 3.1). 
The relaxation of the largest particles is also supported by our later results that the rms 
relative speed obeys the expected $St^{-1/2}$ scaling (\S 6.1) and the relative velocity 
PDF approaches Gaussian (\S 6.2). 
In summary, there is sufficient evidence that the 2-particle statistics is well relaxed toward 
the end of our simulation for all particles.  
 
Again we use the three snapshots at $21.5 T_{\rm eddy}$, $24 T_{\rm eddy}$, and $26 T_{\rm eddy}$ 
in our analysis.  For the first 10 species with $\tau_{\rm p} \lsim 2 T_{\rm eddy}$, 
the velocity of each particle is independent in the three snapshots. Averaging over these snapshots increases 
the measurement accuracy for both the 1-particle velocity and the 2-particle relative velocity.  We find that using the three snapshots also 
improves the relative velocity measurement for the four largest  particles (unlike the case of the 1-particle rms; see \S 5), 
even though the velocity of each individual particle is correlated for longer than $2$-$3 T_{\rm eddy}$.
To effectively increase the independent sample size, 
one should avoid the same two particles to appear as a pair in two 
successive snapshots selected (see Hubbard 2013). It turns out that, for the typical relative velocity of the four largest particles, 
a particle pair at $\eta/4 \lsim r \lsim \eta$ separates to a significant distance and 
no longer makes a pair in $2-3 T_{\rm eddy}$. Since the velocities of any two large particles with 
$\tau_{\rm p} \gg T_{\rm eddy}$ at any distance in a snapshot are independent (\S 3.1), 
the relative velocities of new pairs that appear in $2-3 T_{\rm eddy}$ are typically 
independent of those in the earlier snapshot. Therefore, including the three 
snapshots does increase the independent sample size and the 
measurement accuracy  for the relative velocity of the largest particles. 

To decompose the relative velocity, ${\bs w}$, into radial 
and tangential components, we set up a local coordinate system (${\bs e}'_1$, ${\bs e}'_2$ and ${\bs e}'_3$) for 
each selected particle pair. The direction ${\bs e}'_1$ is chosen to coincide with the particle separation, 
${\bs r}$. In terms of the unit base vectors, ${\bs e}_1$, ${\bs e}_2$ and ${\bs e}_3$, of the simulation grid, 
${\bs e}'_1$ is expressed as $\cos\theta\cos\phi{\bs e}_1 +  \cos\theta\sin\phi {\bs e}_2  +  \sin\theta {\bs e}_3$, 
where $\sin\theta = r_3/r$, $\cos\theta = (r_1^2 +r_2^2)^{1/2}/r$, $\cos\phi= r_1/(r_1^2 +r^2_2)^{1/2}$, 
and $\sin\phi= r_2/(r_1^2 +r_2^2)^{1/2}$.  The radial component is 
calculated as $w_{\rm r} = {\bs w} \cdot {\bs e}'_1$. For the two tangential 
directions, we set ${\bs e}'_2= -\sin\phi {\bs e}_1 + \cos\phi{\bs e}_2$ and ${\bs e}'_3= -\sin\theta \cos\phi{\bs e}_1- \sin\theta \sin\phi {\bs e}_2 + \cos\theta {\bs e}_3$, which 
are obtained by two consecutive rotations of the original coordinates. The first rotation is about ${\bs e}_3$ by $\phi$, 
which moves ${\bs e}_2$ to ${\bs e}'_2$, and the second one is about ${\bs e}'_2$ by $-\theta$, which 
further brings the original base vectors ${\bs e}_1$ and ${\bs e}_3$ to ${\bs e}'_1$ and ${\bs e}'_3$. 
We then calculate $w_{\rm t2} = {\bs w} \cdot {\bs e}'_2$ and $w_{\rm t3} = {\bs w} \cdot {\bs e}'_3$. 
The PDFs of $w_{\rm t2} $ and $w_{\rm t3}$ are found to be almost the 
same, as expected from the statistical isotropy\footnote{When 
selecting the local coordinate system, one may also perform a third rotation about ${\bs e}'_1$ 
by an arbitrary angle. This changes ${\bs e}'_2$ and ${\bs e}'_3$. However, from the statistical 
isotropy, $w_{\rm t2}$ and $w_{\rm t3}$ would be statistically invariant under this third rotation.}. 
We thus take the PDF of a tangential component, $w_{\rm t}$, to be the average PDF of $w_{\rm t2} $ and $w_{\rm t3}$. 
The variance of $w_{\rm t}$ is calculated as $\langle w_{\rm t}^2 \rangle = \frac{1}{2}\left(\langle w_{\rm t2}^2 \rangle + \langle w_{\rm t3}^2 \rangle\right)$.

\subsection{The Root-mean-square Relative Speed}

\begin{figure}[t]
\centerline{\includegraphics[width=1.1\columnwidth]{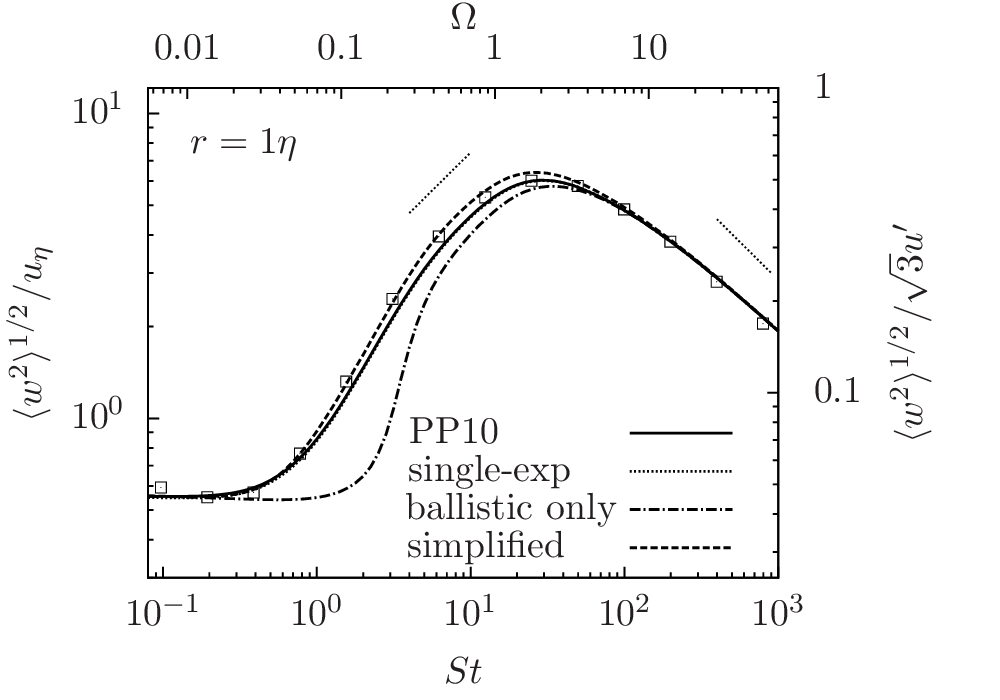}}
\caption{The 3D rms relative velocity, $\langle w^2 \rangle^{1/2}$, as a function of the Stokes number, $St$. 
The data points show the simulation result. The solid line is the prediction of the 
PP10 model (eq.\ \ref{w2}) using a bi-exponential temporal correlation $\Phi_2$ and a two-phase separation 
with $g =1.6$ in the Richardson phase. The dotted line corresponds to the same model but 
with a single-exponential $\Phi_2$. The line is barely visible because it almost 
coincides with the solid line. The dot-dashed line assumes ballistic separation 
at all times. The dashed line is the prediction of the simplified model, 
eqs.\ (\ref{separatedintegral}) and (\ref{ffactor}), using single-exponential 
$\Phi_2$ and a two-phase separation behavior with $g =1.0$. The top X-axis and 
the right Y-axis normalize the friction timescale and the relative speed to 
$T_{\rm L}$ and $\sqrt{3} u'$, respectively. The left and right dotted line segments 
show $St^{1/2}$ and $St^{-1/2}$ scalings. }
\label{3drms} 
\end{figure}

We first study the root-mean-square of the relative velocity. 
We test the prediction of the PP10 model and validate the physical picture revealed by the model.
Fig.\ \ref{3drms} shows the simulation result for the 3D rms, $\langle w^2 \rangle^{1/2}$, 
of the relative velocity as a function of the particle inertia. The data points correspond to the 
measured relative velocity at a distance of $1\eta$.  
On the bottom and top X-axises, we normalize the friction timescale to the Kolmogorov timescale ($St = \tau_{\rm p}/\tau_\eta$) and the Lagrangian 
correlation timescale($\Omega = \tau_{\rm p}/T_{\rm L}$), respectively. The left and right 
Y-axises normalize the relative speed to the Kolmogorov velocity and the 3D rms flow
velocity, respectively. As a reminder, $u' = \left(Re_\lambda/\sqrt{15} \right)^{1/2} u_\eta = 7 u_\eta$ 
in our simulated flow. Similar normalizations are adopted in most figures in the 
rest of the paper. The normalization to Kolmogorov scales is a convention commonly adopted 
in the turbulence literature, which is convenient for the study of 
small particles with $St \lsim 1$ and the related phenomena at small length scales. 
On the other hand, the normalization to large-scale quantities is more 
useful for large particles, and may be more convenient for practical applications since 
observations constrain the large-scales properties of protoplanetary turbulence\footnote{For inertial-range particles, 
one may choose to normalize the physical quantities to turbulent eddies that couple 
to the particle friction timescale. This normalization would be convenient to 
examine whether  the relative velocity of inertial-range particles  shows a 
scale-invariant behavior.}. 

At small $St$, the 3D rms relative speed is roughly constant, and its value 
is consistent with the S-T prediction, $\langle w^2 \rangle^{1/2} =u_\eta/\sqrt{3}$. 
The relative speed starts to rise at $St \simeq 1$, as the effect of the particle memory 
and the backward separation becomes important. 
For the largest particles, we find that $\langle w^2 \rangle^{1/2} \simeq \sqrt{6} u'  (T_{\rm L}/\tau_{\rm p})^{1/2}$, 
with $T_{\rm L} \simeq 14 \tau_\eta$, in agreement with eq.\ (\ref{largelimit}) 
for the large particle limit, $\tau_{\rm p} \gg T_{\rm L}$. 
Like the earlier result for the 1-particle rms velocity, this provides a validation 
for using the Lagrangian correlation function for the trajectory correlation, $\Phi_1$, 
of large particles with $\tau_{\rm p} \gg T_{\rm L}$, 
even though their trajectories may significantly deviate from 
Lagrangian tracers. The predicted $St^{1/2}$ scaling for inertial-range particles 
by various models is not observed due to the limited inertial 
range of the simulated flow. 


The solid curve in Fig.\ (\ref{3drms}) is the prediction of the PP10 model, 
and it is obtained by numerically solving eq.\ (\ref{w2}).  
In the computation, we used eq.\ (\ref{sll}) for $S_{\rm ll}$, eq.\ (\ref{snn}) 
for $S_{\rm nn}$, and eq.\ (\ref{Tr}) for $T(\ell)$, respectively.  The parameters 
in these equations were set to $C_{\rm K} = 2$, $C_{\rm Kn} = 2.5$, $C_{\rm T} =0.4$ and $T_{\rm L} = 
14.4 \tau_\eta$. 
A bi-exponential form, eq.\ (\ref{biexponential2}), is adopted for the temporal 
correlation $\Phi_2$. We used a two-phase behavior for the particle 
separation backward in time (\S 3.4.3), connecting the ballistic and Richardson phases 
at $\tau_{\rm c} =  -\tau_{\rm p}$. We set $d^2(\tau) = r^2 + \langle w^2 \rangle \tau^2$ for $\tau \ge \tau_{\rm c}$, 
and then switch to the Richardson's law, $d^2(\tau) = d^2(\tau_{\rm c}) + 
g \bar{\epsilon} (|\tau|^3 - |\tau_{\rm c}|^3)$, for $\tau < \tau_{\rm c}$, where 
$d^2(\tau_{\rm c}) = r^2 + \langle w^2 \rangle \tau_{\rm c}^2$. 
To fit the simulation data, the Richardson constant, $g$, is set to 1.6. The solid line is in good agreement 
with the data points, confirming the validity of the physical picture of our model. 
Adopting a larger  $g$ could further improve the fitting quality at intermediate $St$.

In Appendix A, we investigate the backward separation of tracer particles 
in our simulated flow, and find that $0.5 \lsim g \lsim 1.2$. Therefore, the $g$ value used 
in the solid line in Fig.\ \ref{3drms} is significantly larger than that of tracer particles. 
There exist two possibilities. First, the backward separation of inertial particles 
in the Richardson phase is indeed faster than tracers. 
Second, the accuracy of our model for the trajectory structure tensor, $S_{{\rm T}ij}$, may be inadequate.  
For example,  approximating the correlation timescale, $T(\ell)$, in $\Phi_2$ by 
the eddy turnover time at the scale of the particle separation, $R$, 
is essentially a qualitative assumption.    
Also there is an order-of-unity uncertainty in the adopted parameter $C_{\rm T}$ 
for the scaling of $T({\ell})$ in the inertia range (see eq.\ (\ref{Tr})). 
If a larger $C_{\rm T}$ were adopted, we could obtain a good fit to the data 
with a smaller $g$. Finally, in our model for $S_{{\rm T}ij}$, we neglected the 
correlation between the particle distance, $R$, and the fluctuations in the flow 
velocity difference, $\Delta u(R)$, seen by the particle pair (see \S 3.2). This tends to 
underestimate the relative velocity. 

The dotted line in Fig.\ (\ref{3drms}) is the prediction of the same PP
10 model, but with a single-exponential $\Phi_2$ (eq.\ (\ref{singleexponential})). 
The dotted line is actually not distinguishable from the solid line, confirming the earlier statement in \S 3.2.1 that 
our model prediction for the relative speed is insensitive to the 
function form of $\Phi_2$. This leaves us some freedom for the choice of 
$\Phi_2$, as long as the correlation timescale is accurately esimated. 
In particular, it provides a justification for approximating $\Phi_2$ 
by the same function form, e.g., a bi-exponential function, for all particles, 
although realistically the form of $\Phi_2$ may have a dependence on the particle inertia.  

The dot-dashed curve plots the prediction of the PP10 model assuming that the particle separation is ballistic 
with $d^2(\tau) = r^2+ \langle w^2 \rangle \tau^2$ at all times. The model is 
otherwise the same as the solid line. A pure ballistic separation is not realistic, 
and we show it here just to illustrate whether the Richardson phase provides 
important contribution to the relative velocity. At $0.5 \lsim St \lsim 5$, the dot-dashed line significantly 
underestimates $\langle w^2 \rangle$, and, from $St \simeq 5$, it becomes close to both the data points 
and the solid line using a two-phase separation behavior. 
A possible explanation for this is that, for $0.5 \lsim St \lsim 5$, 
the relative velocity receives a significant contribution from 
the Richardson phase, even though this phase occurs 
at times beyond the particle memory timescale, i.e., 
at $\tau \lsim -\tau_{\rm p}$. In that case, accounting for this phase would be necessary for particles with intermediate $St$.    
However, the validity of the above interpretation is subject to future tests. 
There is the possibility that the discrepancy between the dot-dashed 
line and the data points may be caused by various uncertainties 
in our model for the trajectory structure tensor, $S_{{\rm T}ij}$ (see above). 
 

The dashed line is the prediction of the simplified model, eqs.\ (\ref{separatedintegral}) 
and (\ref{response}), using a single-exponential $\Phi_2$. As before, the simplified 
model with a bi-exponential $\Phi_2$ gives almost the same prediction. The same 
two-phase separation as in the solid line for the original 
PP10 model is adopted. For the simplified model, the best-fit $g$ is
found to be $\simeq 1$, which is close to the $g$ values measured from tracer particles. 
The simplified model also fits the data better for intermediate $St$, although the assumption made in the model 
is physically not better than the original PP10 model. The simplified model may be a preferred choice, 
as its prediction is easier to compute.  

We find that the Stokes number, $St_{\rm m}$, at which the rms relative 
velocity peaks is $\simeq 30$, corresponding to a friction timescale of $\simeq 2T_{\rm L}$ (or $\simeq 1.5 T_{\rm eddy}$). 
The peak value of the 3D rms relative velocity is $\simeq 6.2 u_\eta$, which is about half 
the 3D rms velocity ($\sqrt{3} u'$) of the flow. We give an explanation for the behavior of the peak relative 
velocity using the qualitative analysis of our model prediction disucssed in \S 3.2.4. The 
analysis was based on the primary distance $R_{\rm p}$, 
estimated as $R_{\rm p}= \langle w^2 \rangle^{1/2} \tau_{\rm p}$. 
$R_{\rm p}$ generally increases with $\tau_{\rm p}$. Around the relative velocity peak, 
$\tau_{\rm p} \simeq 30 \tau_\eta$ and $\langle w^2 \rangle^{1/2} \simeq 6.2 u_\eta$, 
and thus  $R_{\rm p} \simeq 200 \eta$.  From eq.\ (\ref{Tr}), the correlation time, $T(\ell)$, at 
$\ell \simeq 200 \eta$ is about $14\tau_\eta$, which is close to $T_{\rm L}$. 
For $St \gsim 30$, $T(R_{\rm p})$ would be constant and $\simeq T_{\rm L}$. 
Consequently, the $\Phi_2$ term in eq.\ (\ref{w2}) provides a factor of $T_{\rm L}/\tau_{\rm p}$ 
for all particles with $St \gsim 30$ (see \S 3.2.4). Using the same analysis as 
in \S 3.2.4, one can show that this factor causes the relative speed to decrease 
with $\tau_{\rm p}$, even though the structure functions $S_{\rm ll} (R_{\rm p})$ and $S_{\rm nn} (R_{\rm p})$ 
are still increasing with $R_{\rm p}$ at $R_{\rm p}\simeq 200 \eta$ (see Fig.\ \ref{structureandpower}). 
For particles with $St \lsim 30$, both the structure functions and $T(R_{\rm p})$ 
decrease with decreasing $R_{\rm p}$, and thus the relative speed would 
decrease with decreasing $\tau_{\rm p}$. Therefore
a peak forms at $St_{\rm m} \simeq 30$.  We find that, 
for particles with $St \simeq 30$, the amplitude of the flow velocity difference at 
the primary distance ($R_{\rm p} \simeq 200 \eta$) is smaller than $\sqrt{3}u'$, 
and this is responsible for why the maximum relative velocity 
is significantly lower than the rms flow velocity. 
The discussion here shows that the relative velocity is the 
largest for the particles whose primary distance $R_{\rm p}$ corresponds to the size 
of turbulent eddies with lifetime $\simeq T_{\rm L}$. Clearly, the backward particle separation 
plays an important role in determining the peak Stokes number, $St_{\rm m}$.   
  

The model of Volk et al.\ (1980) and its later developments predict that the relative speed reaches the 
maximum when $\tau_{\rm p}$ is equal to a large eddy time, $t_{\rm L}$ (e.g., Markiewicz, Mizuno \& Volk 1991, 
Cuzzi and Hogan 2003, Ormel \& Cuzzi 2007). The definition 
of $t_{\rm L}$ in these studies is different from the timescales we 
used, and it is not clear whether, using parameters appropriate 
for our simulated flow, these models may correctly produce a peak at $St_{\rm m} =30$. 
Another issue is that, in the Volk et al.\ model,  the peak relative speed is 
predicted to be equal or close to the rms velocity of the flow. 
This overestimates the relative speed  around the peak by a factor of 2. 
A physical problem of  the Volk et al.\ model has been discussed in 
the Introduction (see PP10 for details). The performance of the Volk et al.\ (1980) 
model may improve as the Reynolds number of the flow increases, 
which will be tested by future higher-resolution simulations.

\subsubsection{Dependence on the Particle Distance}

\begin{figure}[t]
\centerline{\includegraphics[width=1.1\columnwidth]{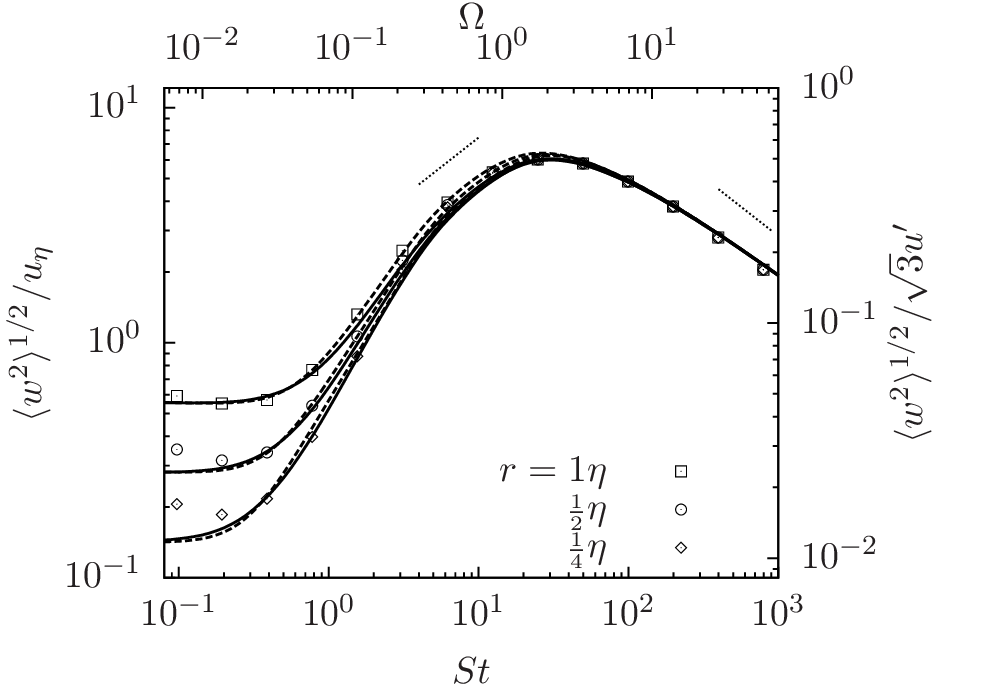}}
\caption{The 3D rms relative velocity, $\langle w^2 \rangle^{1/2}$, 
at at $r=1$ (squares), $0.5$ (circles) and $0.25\eta$ (diamonds). 
Solid and dashed lines are predictions of the PP10 model with 
bi-exponential $\Phi_2$ and the simplified model with single-exponential $\Phi_2$, 
respectively.  A two-phase separation connecting at $\tau = -\tau_{\rm p}$ 
is adopted in both models. In the PP10 model, the Richardson constant $g$ is 
set to 1.6, 1.3 and 1.0 for solid lines from top to bottom. The corresponding $g$ values 
used in the simplified model are 1, 0.7 and 0.5, respectively. The 
dotted line segments denote $St^{1/2}$ and $St^{-1/2}$ scalings.} 
\label{3drmsscale} 
\end{figure}

In Fig.\ \ref{3drmsscale}, we plot the 3D rms relative velocity at different distances, $r$. 
The squares, circles and diamonds correspond to $r=1$, $0.5$ and $0.25\eta$, respectively. 
In this distance range, the relative velocity shows a $r$-dependence at $St \lsim 6$, while 
it is independent of $r$ for $St \gsim 6$ particles.  In the context of our physical picture, 
this is because 
the friction time, $\tau_{\rm p}$, of $St \gsim 6$ particles is long enough that the backward 
particle separation after a duration of $\simeq \tau_{\rm p}$ is insensitive to the ``initial" value, $r$. 
On the other hand, the relative speed of smaller particles relies on the flow 
velocity difference they saw in the near past, when the particle separation 
was still dependent on $r$. 

The solid and dotted lines are predictions of the PP10 model with bi-exponential 
correlation function $\Phi_2$ and the simplified version with single-exponential $\Phi_2$, 
respectively. The lines for $r=1\eta$ have already been shown in Fig.\ (\ref{3drms}), 
and the Richardson constant, $g$, was set to 1.6 and 1.0, respectively, in the two 
models. At smaller $r$, the best-fit value of $g$ becomes smaller. 
For the PP10 model, we adopted $g = 1.3$ and $1.0$ for $r=0.5$ and $0.25 \eta$, 
respectively. The decrease of $g$ with decreasing $r$ is consistent with our result 
in Appendix A for the tracer particle pair dispersion. 
The backward separation of tracer pairs was found to be slower for smaller $r$. The $g$ value 
used in the simplified model also decreases with decreasing $r$. In the dashed 
lines for $r=1$, $0.5$ and $0.25 \eta$, the value of $g$ is set to 1, 0.7 and 0.5, respectively. 

The relative speed of the smallest particles ($St \simeq 0.1$) in our simulation 
appears to be larger than the second smallest ones ($St \simeq 0.2$), 
especially at smaller $r$. Slight dips are seen at $St \simeq 0.2$ (Fig.\ \ref{3drmsscale}). 
This is in contrast to the S-T formula, which predicts the relative speed 
at a given $r$ is constant at sufficiently small $St$. These dips are not 
expected from the physical picture of our model either, and their existence 
is thus questionable. 
One possibility is that the rise of the relative speed 
toward $St \simeq 0.1$ is a numerical artifact. This 
suspicion is based on the consideration that the trajectory integration 
in our simulation is likely less accurate for smaller particles. 
The accuracy of the trajectory computation 
depends on the integration time step relative to the particle
friction time. 
Since the time step is the same for all particles, the accuracy would 
be lower for smaller particles. 
This suggests the rise of the relative speed toward $St \simeq 0.1$ 
may be caused by numerical errors in the trajectory
integration. We expect it to disappear as the computation accuracy 
increases. This will be tested by future simulations with 
a better temporal resolution for the integration of small particles.  

The S-T formula predicts that $\langle w^2 \rangle^{1/2}$ scales linearly with $r$ in the 
$St \ll 1$ limit. This linear scaling is not confirmed by our simulation result 
for the smallest particles. A rough power fit for the rms relative speed 
as a function of $r$ gives $\langle w^2 \rangle^{1/2} \propto r^{0.78}$ at 
$St=0.1-0.2$. This means that, for particles with $St \simeq 0.1-0.2$, 
the S-T formula is already invalid at $r \lsim 0.5 \eta$. 
At a given $St$, a critical particle distance is expected, below which the 
linear scaling does not apply. The physical reason is that, as $r$ decreases, 
the local flow velocity difference across $r$ becomes smaller, and it is 
easier for the particle memory of the flow velocity difference in the past to 
provide a significant contribution, which tends to invalidate the S-T prediction. 
Equivalently, at a given $r$, the S-T formula is valid only below a critical 
$St$.  
In Fig.\ \ref{3drmsscale}, the lines for $r \lsim 0.5 \eta$ show that, as $St$ decreases to 
$\simeq 0.2$, the relative speed predicted by our model is not flat yet, suggesting a significant contribution from the 
particle memory.  At sufficiently small $St$, the rms relative speed at a distance 
$r \lsim 0.5 \eta$ is expected to finally become constant. To verify this, a simulation of higher accuracy for 
small particles is needed to fix the problem of the artificial rise in the relative velocity toward $St \simeq 0.1$. 
As mentioned earlier, the rise is expected to disappear as the computation accuracy for the smallest particles increases. 


Our model does not directly consider the sling effect (Falkovich et al.\ 2002, Falkovich and Pumir 2007) or 
the related caustic formation (Wilkinson \& Mehlig 2005; Wilkinson et al.\ 2006; Gustavsson \& Mehlig 2011).  
The effect of slings is usually explored for small particles with $St \lsim 1$. As mentioned in the Introduction, 
it corresponds to crossing of particle trajectories that occurs at fluid streamlines with high 
curvature or local flow regions with large velocity gradient.   
In our physical picture, the effect of slings or caustics could be viewed as a contribution 
to the backward particle separation. In the sling events, the particle pairs come together 
from a farther distance than the average. Based on the model of Wilkinson et al.\ (2006) and 
Gustavsson \& Mehlig (2011), the frequency of slings or caustic formation increases with $St$. 
The effect also becomes more important as $r$ decreases, and would finally dominate 
over the S-T contribution at a sufficiently small $r$. Falkovich and Pumir (2007) showed that the 
sling effect is already significant at $St \simeq 0.2$. In Fig.\ \ref{3drmsscale}, we see that 
our model prediction underestimates the relative velocity of $St =0.2$ particles at $r \lsim 0.5 \eta$. 
A likely reason is that the sling effect is not sufficiently reflected by the assumed backward 
separation behavior. In principle, the effect can be better incorporated into our model by directly and accurately 
evaluating the frequency of such events and their contribution to the backward separation. 
We will discuss the effect of slings or caustics on the particle collision rate in details in \S 7. 
For  $St \lsim 1$ particles, the sling events are rare.   At $St \gsim 3$, these events become 
very frequent, and essentially all particle pairs at $r\lsim \eta$ should be counted 
as sling or caustic pairs (\S 7.2).  In our physical picture, this corresponds to the fact that the 
backward separation of $St \gsim 3$ particle pairs at a friction timescale ago is 
significantly larger the initial distance, $r$.    


Concerning the $r$-dependence of  $\langle w^2 \rangle^{1/2}$ for $St \lsim 6.2$ particles seen 
in Fig.\ \ref{3drmsscale}, a fundamental question is whether the dependence disappears as $r$ 
further decreases below $\eta/4$, or, equivalently, whether $\langle w^2 \rangle^{1/2}$ 
approaches a finite constant  as $r \to 0$. 
Based on the prediction of Gustavsson \& Mehlig (2011), it is possible that, for particles that 
exhibit significant clustering (see \S 7.1), the overall rms relative velocity 
may approach 0 as $r \to 0$. In the case,  $\langle w^2 \rangle^{1/2}$ for 
$St\lsim 6.2$ particles would not converge with decreasing $r$, and thus in principle  could not be 
resolved. Intuitively, the rms relative speed, $\langle w^2 \rangle^{1/2}$, 
of intermediate particles with $1\lsim St \lsim 6.2$ may converge at sufficiently small $r$, while, 
for $St\ll 1$ particles, $\langle w^2 \rangle^{1/2}$ may decrease to zero as $r \to 0$. 
The convergence of $\langle w^2 \rangle^{1/2}$ for $St \lsim 6.2$ particles needs to checked with larger 
simulations that allow accurate measurements at $r\ll \eta/4$.

\subsubsection{The Radial and Tangential Relative Speeds}

\begin{figure}[t]
\centerline{\includegraphics[width=1.1\columnwidth]{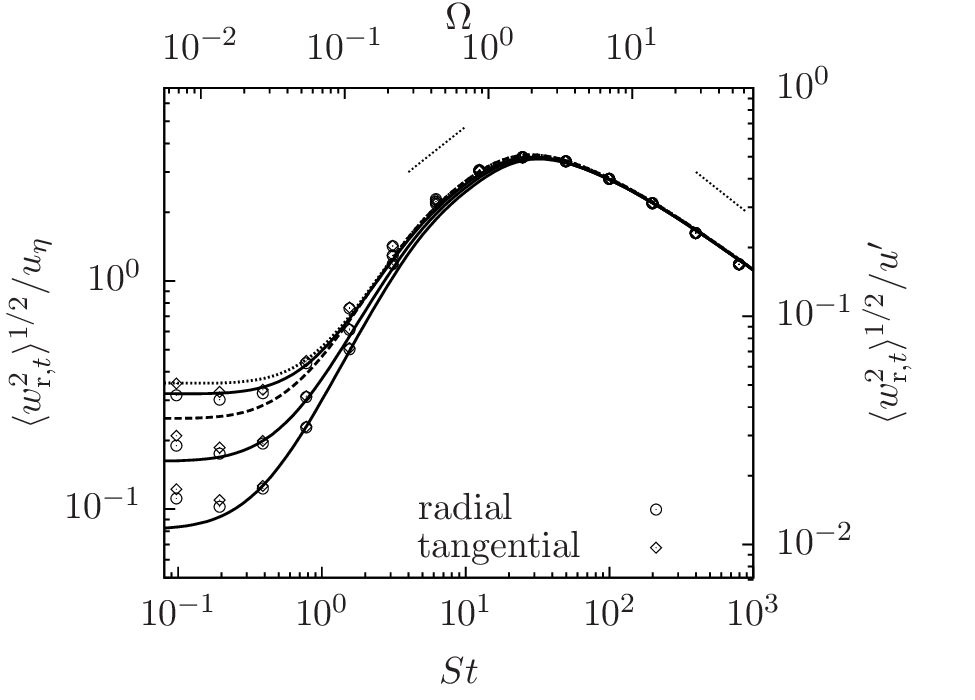}}
\caption{The rms relative speeds in the radial ($\langle w_{\rm r}^2 \rangle^{1/2}$; circles) 
and tangential ($\langle w_{\rm t}^2 \rangle^{1/2}$; diamonds) directions. From top to 
bottom, data points show simulation results at $r=1$, $0.5$, and 
$0.25\eta$, respectively. Lines are predictions of the PP10 model. 
The solid lines adopt eq.\ (\ref{randomdirection2}) for the angular average 
of $S_{{\rm T} ij}$, 
which predicts that $\langle w_{\rm r}^2 \rangle = \langle w_{\rm t}^2 \rangle = \frac{1}{3} \langle w^2 \rangle$.
The particle separation behavior assumed here is exactly the same 
as in the solid lines in Fig.\ \ref{3drmsscale} for the 3D rms. 
The Richardson constant, $g$, is set to 1.6, 1.3 and 1.0 
for $r=1$, $0.5$, and $0.25\eta$, respectively. The dashed and dotted 
lines for $r=1\eta$ are solutions of eq.\ (\ref{wrwt}) for the radial and tangential relative speeds, 
respectively. 
The two lines 
reproduce the S-T prediction for $\langle w_{\rm r}^2 \rangle^{1/2}$ and 
$\langle w_{\rm t}^2 \rangle^{1/2}$ at $St\ll1$.}    
\label{radialtangential} 
\end{figure}

Fig.\ \ref{radialtangential} shows the rms relative speeds in the radial (circles) and tangential 
(diamonds) directions for particle pairs at $r= 1$, $0.5$ and $0.25 \eta$. At $St \lsim 1$, the 
tangential rms speed, $\langle w_{\rm t}^2 \rangle^{1/2}$, is slightly larger (by $\simeq 10\%$) 
than the radial rms, $\langle w_{\rm r}^2 \rangle^{1/2}$. 
This difference is considerably smaller than the prediction of the S-T formula, 
eq.\ (\ref{saffmanturner}), which indicates that  for $St \ll 1$ particles
the tangential rms relative speed should be larger than the radial rms by 
a factor of $\sqrt{2}$. Our data implies that this prediction is not valid at least for 
particles with $St \gsim 0.1$. 
It remains to be checked whether the factor of $\sqrt{2}$ 
difference between $\langle w_{\rm r}^2 \rangle^{1/2}$ and $\langle w_{\rm t}^2 \rangle^{1/2}$ 
would be recovered at smaller Stokes numbers. 
In Fig.\ \ref{radialtangential}, we see that $\langle w_{\rm r}^2 \rangle^{1/2}$ and $\langle w_{\rm t}^2 \rangle^{1/2}$ 
become exactly equal at $St \gsim 1$. 
    
The solid lines in Fig.\ \ref{radialtangential} correspond to the prediction of 
the PP10 model using eq.\ (\ref{randomdirection2}) for the angular average, $\langle S_{{\rm T} ij} \rangle_{\rm ang}$, 
of the trajectory structure tensor. The equation 
assumes that the direction of the particle separation ${\bs R}$ at any time is completely random, 
and predicts that $\langle w_{\rm r}^2 \rangle = \langle w_{\rm t}^2 \rangle = \frac{1}{3} \langle w^2 \rangle$ 
for all particles (\S 3.2.2). This prediction is in good agreement 
with our simulation data. 
The equality of the radial and tangential rms speeds for $St \gsim 1$ is 
expected, because the separation ${\bs R}$ 
of these particles at a friction timescale ago is significantly 
larger than the initial distance, $r$, and its direction is likely 
random with respect to ${\bs r}$. On the other hand, 
the near equality of $\langle w_{\rm r}^2 \rangle^{1/2}$ and 
$\langle w_{\rm t}^2 \rangle^{1/2}$ at $St \sim 0.1-0.2$ 
is somewhat surprising. For $r=1\eta$, the backward 
separation of these particles does not contribute to make
the 3D rms, $\langle w_{\rm}^2 \rangle^{1/2}$, of the 
relative velocity significantly larger than the S-T prediction. 
This suggests that the near equality of $\langle w_{\rm r}^2 \rangle^{1/2}$ 
and $\langle w_{\rm t}^2 \rangle^{1/2}$ is due to a conversion 
of the relative velocity from the tangential to the radial direction. 
The conversion is probably caused by the deviation of the particle trajectories 
from the fluid elements. Even though the deviation does not considerably change 
the 3D amplitude $\langle w_{\rm}^2 \rangle^{1/2}$ at $r \simeq1\eta$, 
it could efficiently alter the direction of ${\bs w}$ with respect to ${\bs r}$. 
The trajectory deviation is stochastic, and thus tends to randomize the direction of 
${\bs w}$ and equalize its radial and tangential components. This reduces the 
tangential-to-radial ratio. The randomization effect is expected to be more efficient 
in the slings events, where the particles are shot out of the flow streamlines, and encounter the trajectories of other particles. 
At smaller $r$, the contribution from the backward separation to the 3D rms 
of the relative velocity is larger, and the random direction of the particle separation in the past also tends to equalize 
the radial and tangential components. In the  $r \to 0$ limit, we would expect that 
$\langle w_{\rm r}^2 \rangle^{1/2}$ 
and $\langle w_{\rm t}^2 \rangle^{1/2}$ are exactly equal at all $St$.

When computing the solid lines, we used a bi-exponential form for $\Phi_2$,
and the separation behavior adopted here is exactly the same as for the solid 
lines in Fig.\ \ref{3drmsscale} for the 3D rms.  The solid 
lines shown here correspond to those in Fig.\ \ref{3drmsscale} divided by 
$\sqrt{3}$. The Richardson constant is set to 1.6, 1.3 and 1.0 in the three lines for $r=1, 0.5$, 
and $0.25\eta$, respectively. 

  
The dashed and dotted lines for $r=1\eta$ are the solutions of eq.\ (\ref{wrwt}) 
for the radial and tangential rms relative speeds, respectively. Eq.\ (\ref{wrwt}) 
was derived from eq.\ (\ref{randomdirection1}) for $\langle S_{{\rm T} ij}\rangle_{\rm ang}$, 
which assumes that the direction of the separation change $\Delta {\bs R}$ 
(rather than ${\bs R}$ itself) is random. When solving eq.\ (\ref{wrwt}), we 
used the same two-phase separation behavior (with $g=1.6$) as in the 
corresponding solid line. At small $St$, the dashed and solid lines reproduce
the S-T prediction that $\langle w_{\rm t}^2 \rangle = 2 \langle w_{\rm r}^2 \rangle$. 
The discrepancy between the simulation data and the S-T formula 
implies that, 
for particles with $0.1 \lsim St \lsim 1$, the direction of  ${\bs R}$ is more random than assumed in 
eq.\ (\ref{randomdirection1}).  

The dependence of the radial and tangential rms relative speeds on $r$ is 
similar to that of the 3D rms (see Fig.\ \ref{3drmsscale}). In an attempt to roughly fit 
them as power-law functions of $r$, we find that $\langle w_{\rm r}^2 \rangle^{1/2}$and $\langle w_{\rm t}^2 \rangle^{1/2}$ 
scale with $r$ as $\propto r^{0.78}$ at $St =0.1-0.2$. 
Similar to the 3D rms, the slight dips at $St \simeq 0.2$ for both 
$\langle w_{\rm r}^2 \rangle^{1/2}$and $\langle w_{\rm t}^2 \rangle^{1/2}$ may be due to a numerical artifact.

\begin{figure*}[t]
\includegraphics[height=2.5in]{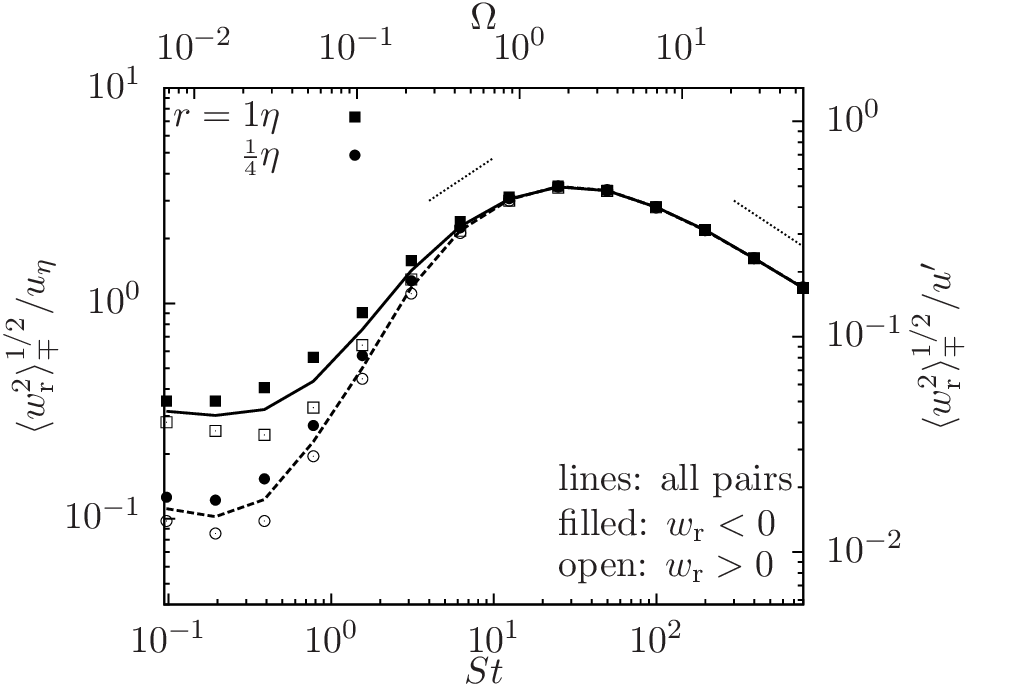}
\includegraphics[height=2.5in]{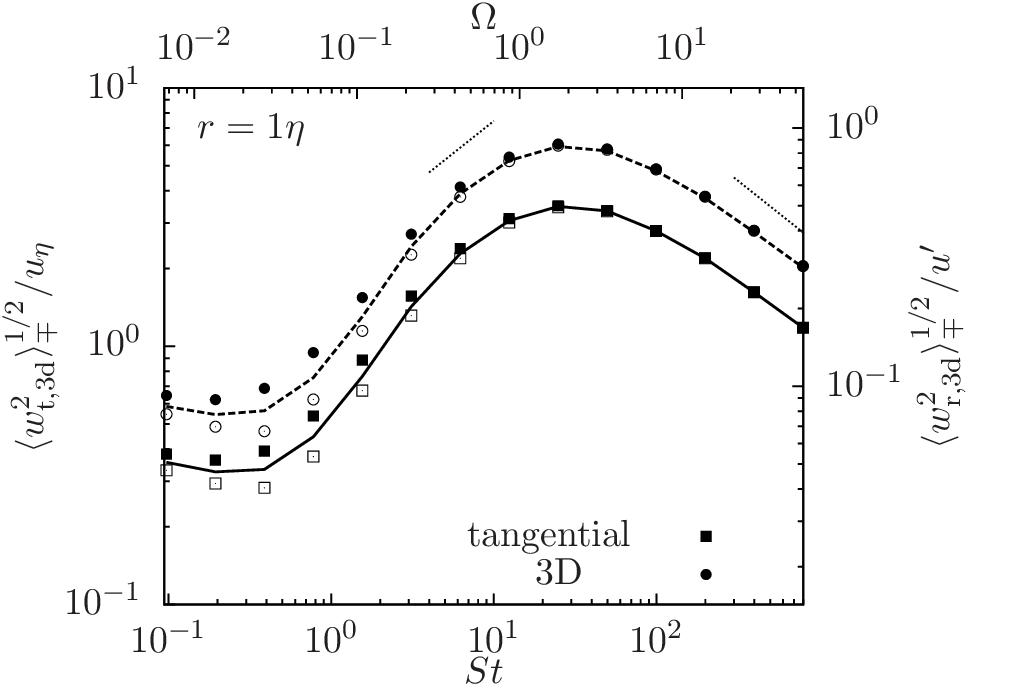}
\caption{The rms relative speeds for approaching ($w_{\rm r} <0$; filled symbols) or separating ($w_{\rm r} > 0$; 
open symbols) particle pairs. Lines correspond to the overall rms 
relative speeds counting all pairs. Left panel: The radial rms 
speed $\langle w_{\rm r}^2 \rangle_{\mp}^{1/2}$) in the minus and 
plus groups with $r=1\eta$ (squares) and $0.25\eta$ (circles).  
Right panels: the tangential (squares) and 3D (circles) rms relative 
speeds for particle pairs at $r=1\eta$. The 
dotted line segments denote $St^{1/2}$ and $St^{-1/2}$ scalings.
}  
\label{plusminus} 
\end{figure*}

The simulations of Wang et al.\ (2000) found that the tangential-to-radial variance 
ratio, $\langle w_{\rm t}^2 \rangle/\langle w_{\rm r}^2 \rangle$, is $\simeq 1.5-1.6$ 
at $St \simeq 0.1-0.2$. This is closer to the S-T prediction and larger than the 
corresponding value ($1.2-1.3$) in our simulation. This is probably because our flow has a 
much larger Reynolds number. Although Wang et al.\ (2000) claimed that 
the ratio is independent of $Re_{\lambda}$ based on several simulations with 
$Re_{\lambda} \lsim 75$, it is not clear if this is also true at 
$Re_{\lambda} \gg 75$. As speculated above, it is the deviation of the particle trajectories from 
the flow elements that tends to equalize the radial and tangential relative 
speeds of small particles. Clearly, the trajectory deviation 
would be larger in flow regions with larger velocity gradients, 
where the flow experiences a faster velocity change. 
The probability of finding large flow velocity gradients hence large trajectory 
deviations increases with $Re_{\lambda}$. Therefore, the tangential-to-radial ratio is likely 
smaller at higher $Re_{\lambda}$.
For $St \lsim 1$ particles, the sling events occur in regions 
with extreme flow velocity gradients, and the frequency of slings would increase 
with $Re_{\lambda}$ (Falkovich and Pumir 2007). This also tends to reduce 
the tangential-to-radial ratio. 
However, we cannot rule out the possibility 
that the trajectory integration of the smallest particles in our simulation 
is not sufficiently accurate to allow an accurate measurement of $\langle w_{\rm t}^2 \rangle/\langle w_{\rm r}^2 \rangle$ 
at small $St$. 



\subsubsection{Approaching and Separating Particle Pairs}

So far, our analysis for the particle relative velocity included all particle pairs at given distances. However, 
not all pairs at a small distance lead to collisions. Particles with a negative radial 
relative velocity, $w_{\rm r} < 0$, approach each other and may collide, while particle pairs with $w_{\rm r} > 0$ 
move away from each other. Since the final goal of our study is to examine the 
particle collisions, it is appropriate to split particle pairs at a given distance into two groups with $w_{\rm r} < 0$ and  
$w_{\rm r} \ge 0$, respectively.  We refer to them as the minus and plus groups. 
Although only the first group is relevant for particle collisions, it is theoretically 
interesting to compare the two groups. 

For the radial component, $w_{\rm r}$, of the relative velocity, we denote the variances in the minus 
and plus groups as $\langle w_{\rm r}^2 \rangle_{-}$ and $\langle w_{\rm r}^2 \rangle_{+}$, 
respectively. In terms of the PDF, $P(w_{\rm r}, St)$, of $w_{\rm r}$, the variances are written as $\langle w_{\rm r}^2 \rangle_{-} = \int_{-\infty}^{0} w_{\rm r}^2 P(w_{\rm r}, St) dw_{\rm r}/ \int_{-\infty}^{0} P(w_{\rm r}, St) dw_{\rm r}$ and 
$\langle w_{\rm r}^2 \rangle_{+} =  \int_{0}^{\infty} w_{\rm r}^2 P(w_{\rm r}, St) dw_{\rm r}/ \int_{0}^{\infty} P(w_{\rm r}, St) dw_{\rm r}$. 
We denote the PDFs of the tangential component in the minus and plus groups as conditional PDFs, 
$P(w_{\rm t}|w_{\rm r}< 0, St)$ and $P(w_{\rm t}|w_{\rm r}>0, St)$. The tangential variances 
in the two groups are then given by 
$\langle w_{\rm t}^2 \rangle_{\mp} = \int_{-\infty}^{\infty}  w_{\rm t}^2 P(w_{\rm t}|w_{\rm r} \lessgtr 0, St) dw_{\rm t}$. 
Similarly, for the 3D amplitude, $|{\bs w}|$, the minus and plus variances are 
expressed as $\langle w^2 \rangle_{\mp}  = \int_{-\infty}^{\infty} w^2 P(|{\bs w}||w_{\rm r} \lessgtr 0, St) d|{\bs w}|$, 
where $P(|{\bs w}||w_{\rm r} \lessgtr 0 ,St)$ are the PDFs of $|{\bs w}|$ for 
approaching and separating pairs. 
The PDFs, $P(w_{\rm r}, St)$, $P(w_{\rm t} |w_{\rm r} \lessgtr 0,St)$, and $P(|{\bs w}||w_{\rm r} \lessgtr 0, St)$, will be studied in \S 6.2.

The data points in the left panel of Fig.\ (\ref{plusminus}) show the radial rms 
relative speeds of particle pairs at $r=1\eta$ (squares) and $0.25 \eta$ (circles). 
The right panel plots the tangential (squares) and 3D (circles) rms speeds at 
$1\eta$. In both panels, the filled and open symbols correspond to particle pairs 
in the minus and plus groups, respectively, and the lines plot the overall rms 
relative velocities counting all particle pairs. If the velocity 
of $St \ll 1$ particles closely follow the flow velocity, we expect that 
$\langle w_{\rm r}^2 \rangle_{\mp}$ are determined by the variances 
of the longitudinal flow velocity increments, $\langle \Delta u_{\rm r}^2\rangle_{\mp}$, 
for negative and positive $\Delta u_{\rm r}$, respectively.  
The definition of $\langle \Delta u_{\rm r}^2\rangle_{\mp}$ is given 
in Appendix B, and they correspond to the fluctuation amplitudes in the 
left and right wings of the PDF of $\Delta u_{\rm r}$.  
In Appendix B, we find the ratio $\langle \Delta u_{\rm r}^2\rangle_{-}/\langle \Delta u_{\rm r}^2\rangle_{+}$ is 
1.47 at the size, $\Delta x$, of the computation cell. This suggests 
that, in the $St \to 0$ limit, $\langle w_{\rm r}^2 \rangle_{-}^{1/2}$ would be larger than $\langle w_{\rm r}^2 \rangle_{+}^{1/2}$ by 
$\simeq 20\%$.  The simulation result confirms  this expectation. 
At $St \simeq 0.1$,  $\langle w_{\rm r}^2 \rangle_{-}^{1/2}$ is larger than $\langle w_{\rm r}^2 \rangle_{+}^{1/2}$ 
by $\simeq 25\%$ for both $r=1$ and $0.25\eta$ (see the left panel of Fig.\ \ref{plusminus}). 
Using a similar analysis to the tangential component gives 
$\langle w_{\rm t}^2 \rangle_{\mp}^{1/2} \simeq \langle \Delta u_{\rm t}^2 \rangle_{\mp}^{1/2}$ 
for $St \ll 1$, where $\langle \Delta u_{\rm t}^2 \rangle_{\mp}$ are variances of 
the transverse flow velocity increment, $\Delta u_{\rm t}$, conditioned on the sign of the 
longitudinal increment $\Delta u_{\rm r}$ (see Appendix B). The ratio $\langle w_{\rm t}^2 \rangle_{-}^{1/2}/\langle w_{\rm t}^2 \rangle_{+}^{1/2}$
is found to be 1.16 at $St=0.1-0.2$, consistent with the ratio 
of $\langle \Delta u_{\rm t}^2 \rangle_{-}$ to $\langle \Delta u_{\rm t}^2 \rangle_{+}$ at the grid 
cell size, $\Delta x$ (see Appendix B). For the the 3D amplitude ($\langle w^2 \rangle_{\mp}^{1/2}$), the rms ratio between the minus 
and plus groups is 1.18 at $St \ll 1$.       
 
As $St$ increases, the relative speed for the plus group first decreases slightly 
and reaches a minimum at $St \simeq 0.4$ in all cases with $r=1\eta$. 
This can be explained by considering the effects of the particle memory 
and the particle separation backward in time.  
Particle pairs in the plus group with $w_{\rm r} >0$ are coming from 
smaller distances, meaning that the separation of the particles was smaller in the near past. 
As $St$ increases from 0.1 to 0.4, the contribution from the particle memory of the flow velocity 
difference becomes more important, and this contribution tends to reduce the relative speed since 
the particle distance was smaller in the immediate past. However, if we look 
back further into the past (i.e., at larger $|\tau|$), the two particles may 
pass each other, and their distance would make a transition from decreasing to 
increasing. This explains the increase of $\langle w_{\rm r}^2 \rangle_{+}$, 
$\langle w_{\rm t}^2 \rangle_{+}$, and $\langle w^2 \rangle_{+}$ at $St \gsim 0.4$. 
The minimum of $\langle w_{\rm r}^2 \rangle_{+}$ for $r=0.25\eta$
appears at $St=0.2$ instead of $St=0.4$, because, for smaller $r$, 
it takes a shorter time for the particle distance in the past to 
change from decreasing to increasing.     

For approaching particles in the minus group, the particle distance would increase monotonically 
toward the past. Therefore, the relative speed for this  group is expected to  
increase monotonically as $\tau_{\rm p} $ increases from 0  to $ \simeq T_{\rm L}$. 
This is confirmed by the filled data points in Fig.\ \ref{plusminus}, except the slight dips at $St \simeq 0.2$. 
These dips are not expected, and again may be caused by insufficient numerical accuracy 
in the trajectory integration of the smallest particles (\S 6.1.1). 
Fig.\ \ref{plusminus} shows that approaching particles tend to have a larger relative 
speed than separating ones. 
The difference between the two groups first increases with $St$ and
then decreases at $St \ge 1$. 
At $St \gsim 6.2$, the rms relative speeds in the two groups are close and coincide with 
the overall rms.  The reason is that, for these larger particles,  the separation of the 
particle pairs in the two groups at a friction timescale ago becomes insensitive to  the ``initial" condition around $\tau =0$.    

The asymmetry in the relative velocity of $St \lsim 6.2$ particles is related to the spatial 
clustering of these particles. The fact that approaching  pairs move faster than 
separating ones may imply that particles tend to cluster at small distances to 
a reference particle (see also \S 7.1). 
An interesting question is whether the asymmetry found at $\eta/4 \lsim r \lsim \eta$ would exist also 
in the $r\to 0$ limit. 
We expect the asymmetry to decrease with decreasing $r$ because the 
difference caused by the different separation behaviors in the near past for 
approaching and separating pairs would be smaller at smaller $r$. However, it is not 
clear whether it completely vanishes as $r\to 0$.  For example, the asymmetry in the flow 
velocity difference, $\Delta {\bs u}$, persists at any tiny but finite $r$, and it may leave 
an imprint on the relative velocity of small particles  (say, with $St\lsim1$).
The actual behavior of the asymmetry as $r\to 0$ will be checked by future simulations 
that allow to resolve smaller $r$. 
The prediction of the PP10 model was made only for the overall rms, and it could be 
modified to give separate predictions for approaching and separating pairs 
if the different separation behaviors of the two groups are properly specified.
We also find that $\langle w_{\rm r}^2 \rangle_{-}^{1/2}$ and $\langle w_{\rm t}^2 \rangle_{-}^{1/2}$ 
almost coincide for all $St$, suggesting that each relative velocity  component would provide equal 
amount of collision energy. On average, the radial component contributes 1/3 collision energy, while the rest 2/3 is from the two tangential components.  


\subsection{The PDF of the Particle Relative Velocity}

An accurate estimate of the PDF of the particle collision velocity is important for 
modeling the growth and evolution of dust particles in protoplanetary disks. 
As mentioned in the Introduction, the outcome of particle collisions depends on the collision velocity, 
and due to the random nature of the turbulent-induced collision velocity, collisions of particles 
with exactly the same properties may have different outcomes, and thus using a single value, e.g., the rms, for the 
collision speed of particles of a given size is insufficient.  The probability distribution of the collision velocity is 
needed to calculate the fractions of collisions resulting in sticking, bouncing or fragmentation.

In this section, we explore the probability distribution of the particle relative 
velocity. 
We will primarily show the PDFs at $r=\eta$, where the statistical 
measurements are most sufficient and accurate. 
The measured PDFs for $St \gsim 10$ particles already converge 
at $r \simeq \eta$. 
On the other hand, for smaller particles the PDFs have an $r$-dependence at $r \gsim \eta/4$, 
and an appropriate extrapolation to $r \to 0$ using larger simulations will be needed for the 
application to dust particle collisions. 

The physical picture of PP10 shows that the relative velocity of inertial particles depends on the 
flow velocity difference the particle pairs saw within a memory timescale or so. 
This suggests that the statistics of the velocity difference in the carrier flow is crucial for 
the understanding of the relative velocity PDF of inertial particles. Therefore, we analyzed the PDFs, 
$P_{\rm u} (\Delta u_{\rm r}, \ell)$ and $P_{\rm u} (\Delta u_{\rm t}, \ell)$, of 
the longitudinal and transverse velocity increments,  $\Delta u_{\rm r}$ and $\Delta u_{\rm t}$, 
as functions of the length scale, $\ell$, in our simulated flow. The results are discussed 
in details in Appendix B. The flow velocity PDFs are used in the physical explanation 
for the PDF of the particle relative velocity as a function of particle inertia in \S 6.2.2.

\begin{figure*}[t]
\includegraphics[height=2.9in]{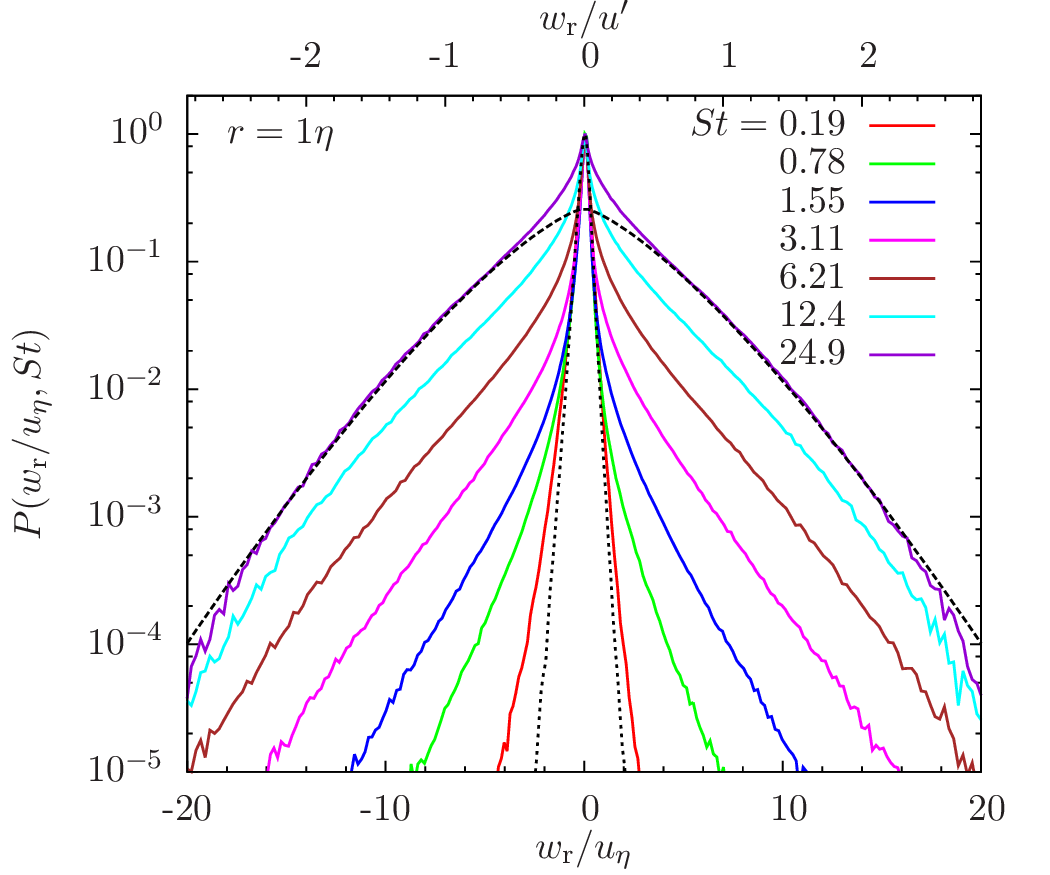}
\includegraphics[height=2.9in]{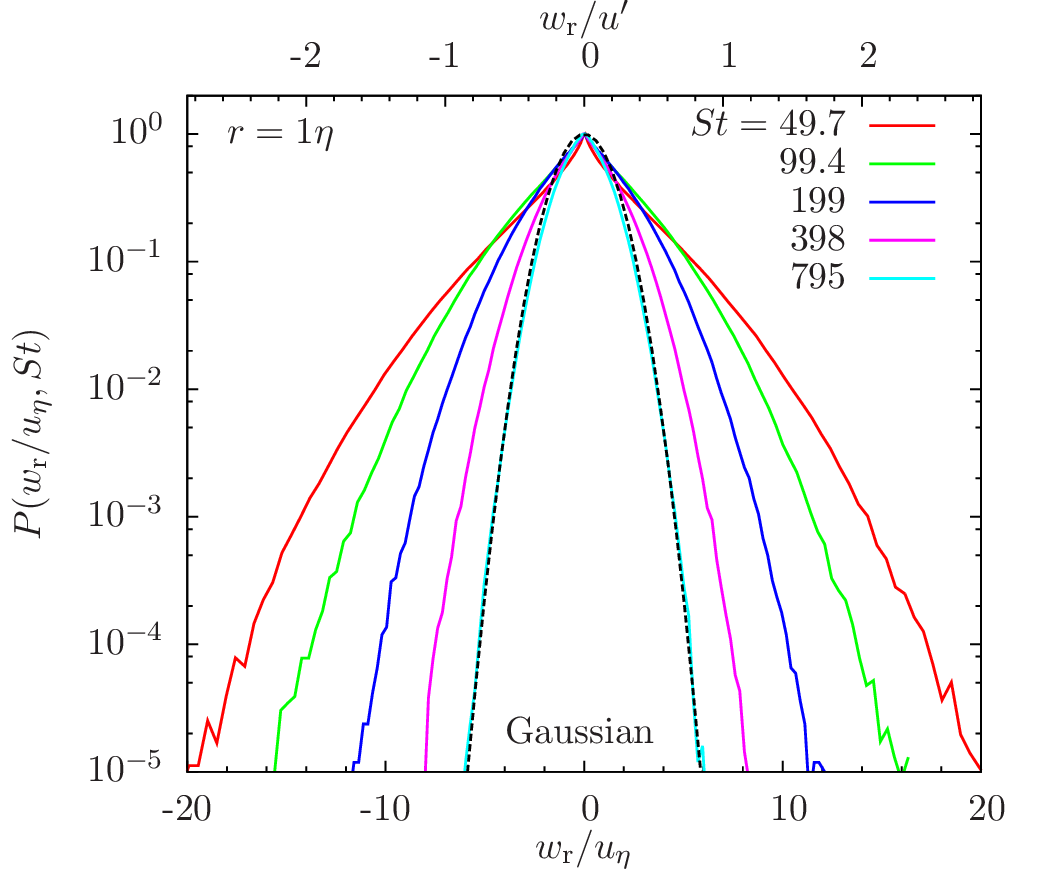}
\caption{The PDF of the radial component ($w_{\rm r}$) of the relative velocity at $r=1\eta$ as a function 
of $St$. The relative speed is normalized by the Kolmogorov velocity, $u_\eta$, and the rms flow velocity, $u'$, on bottom and top axises, respectively. 
Each PDF is normalized to its peak value at $w_{\rm r}=0$.  The left panel shows the PDFs for particles with $St \le 24.9$, while the 
right panel shows results for $St \ge 49.7$ particles. The 
dotted line in the left panel is the PDF of the radial relative 
speed of tracer particles ($St=0$). The dashed line in this panel is 
the stretched exponential function with $\alpha = 4/3$, which provides 
a good fit for the PDF tails of $St =24.9$ particles. In the right panel, the 
dashed line corresponds to the Gaussian fit  to the largest particles in our 
simulation. One can use $\Omega = St/14.4$ and $\Omega_{\rm eddy} =St/19.2$ 
to convert the normalization of the particle friction time.}
\label{rpdfs} 
\end{figure*}

\subsubsection{The PDFs of the radial and tangential components} 
 
\begin{figure*}[t]
\includegraphics[height=2.9in]{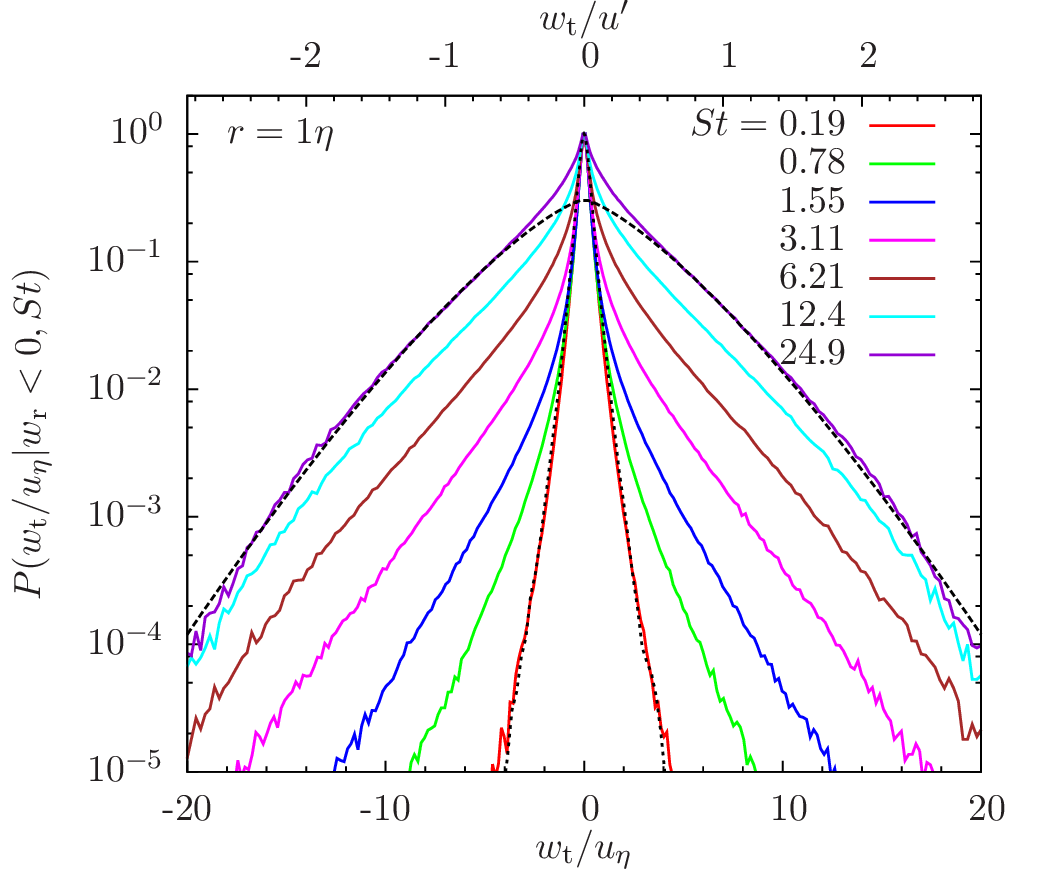}
\includegraphics[height=2.9in]{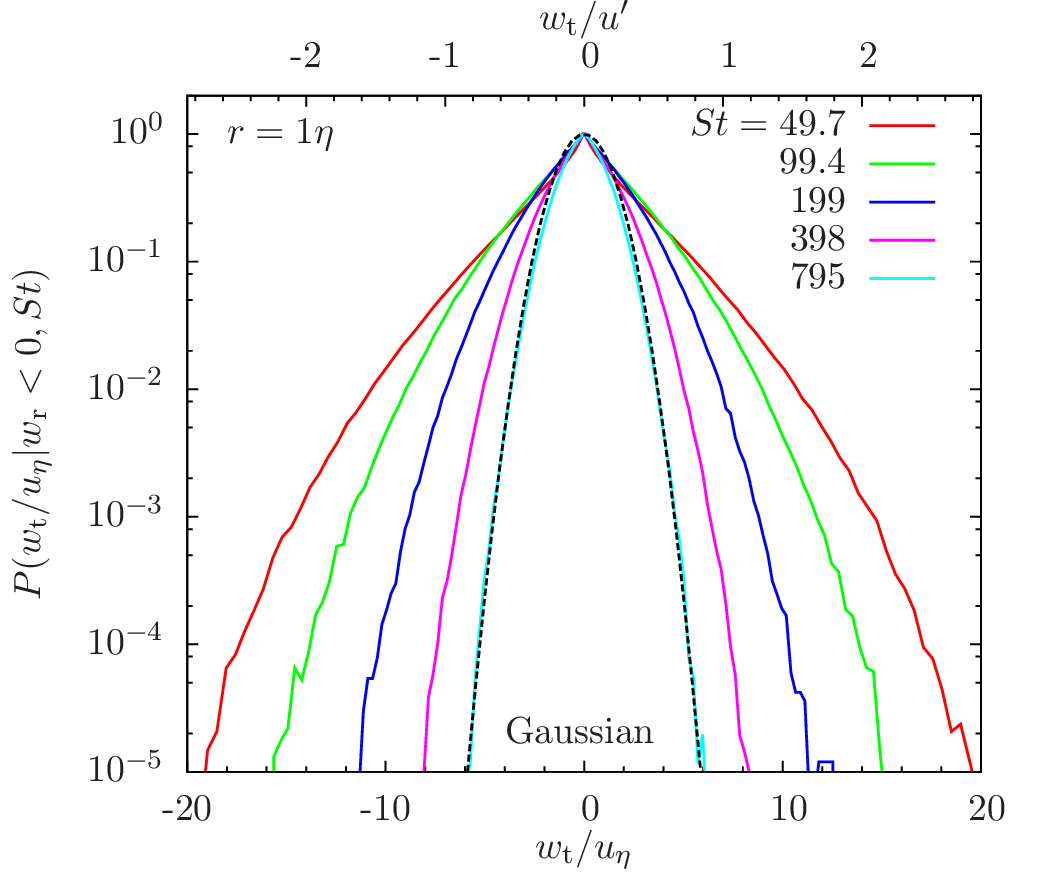}
\caption{The PDFs of the tangential relative velocity, $w_{\rm t}$,  
conditioned on $w_{\rm r} <0$.  The PDFs are measured from approaching particle pairs. 
The normalizations are the same as in Fig.\ \ref{rpdfs}. The left and right panels show results for particles with $St \le 24.9$ and $St \ge 49.7$, respectively.  
The dotted line in the left panel is the PDF of the tangential relative velocity of tracer particles ($St=0$) conditioned on $w_{\rm r} <0$.
The dashed line in this panel is a stretched exponential function with $\alpha = 4/3$. The dashed line in the right panel corresponds to 
the Gaussian fit for $St =795$ particles.}
\label{tpdfs} 
\end{figure*}
 
In Fig.\ \ref{rpdfs}, we show the PDF, $P(w_{\rm r}, St)$, of the radial component 
of the relative velocity as a function of the Stokes number. All the PDFs are 
measured at a particle distance of $1\eta$. The relative speed is 
normalized to the Kolmogorov velocity, $u_{\eta}$, and the 1D flow rms velocity, $u'$, 
on the bottom and top X-axises, respectively.  Each PDF is normalized to its 
value at the central peak. In the left panel, the thin dashed line corresponds to the PDF of 
tracer particles ($St =0$) at $1\eta$. The shape of this line is found to be identical to the PDF, $P_{\rm u}(\Delta u_{\rm r}, \ell)$, 
of the longitudinal flow velocity increment, $\Delta u_{\rm r}$, at the computational cell size ($\ell=1.7 \eta$; 
see the top line in the left panel of Fig.\ \ref{flowpdf} in Appendix B). This is expected 
as tracer particles exactly follow the flow velocity, and the shape of $P_{\rm u}(\Delta u_{\rm r}, \ell)$ is 
independent of $\ell$ in the dissipation range (Appendix B). The solid color lines of 
increasing width show the PDFs of larger particles. This corresponds to the increase
of the rms relative speed with $St$ for $St \lsim 24.9$ (see Figs. \ref{radialtangential} and \ref{plusminus}). 
For $St \le 1.55$, the PDF of $w_{\rm r}$ has a negative skewness, which is inherited from the flow 
velocity PDF $P_{\rm u}(\Delta u_{\rm r}, \ell)$. The PDF becomes symmetric at $St \gsim 3.21$. It is interesting 
to note that, as $St$ increases, the tails of the PDFs become broader, while the innermost part 
remains unaffected and the same as the PDF of the tracer particles. 
As to be shown in \S 6.2.2, the amplification of the PDF tails correspond to the effect of slings or caustic formation. 
Due to the tail amplification, the overall PDF shape becomes fatter\footnote{For definiteness, 
throughout the paper we use ``fat" or ``thin" to describe the shape of the PDF. 
The fatness can be quantified, e.g., by kurtosis. On the other hand, the extension 
or width of the PDF, corresponding to the rms, is described as ``broad"  or ``narrow".} as $St$ increases to $\simeq 1.55$.
With increasing $St$, the amplification effect proceeds towards 
the inner parts of the PDF, leading to a sharp cusp-like 
shape at the center, especially for $St \gsim 3.11$. 
For particles with $St \gsim 3.11$, the slope of the outer parts 
of the PDF tends to steepen when extending to higher tails, i.e., the PDF shape 
becomes thinner at larger $|w_{\rm r}|$. This thinning trend toward the high 
tails causes a decrease in the overall fatness of the PDF for $St$ above $3.11$. 


In the right panel, the PDF becomes narrower as  $St$ increases above $49.7$, corresponding to the decrease 
of the rms relative velocity with $St$ in the large particle limit (Figs.\ \ref{radialtangential} and \ref{plusminus}). 
For $St \ge 49.7$ ($\Omega \ge 3.4$), the friction timescale is larger than the correlation timescale 
($T_{\rm L}$) of the flow velocity at largest scales, meaning that the memory time 
of the flow velocity is shorter than the memory of the particles. This induces a factor of $T_{\rm L}/\tau_{\rm p}$ 
in the relative velocity variance (\S 3.2.4), causing a decrease of the PDF width 
at larger $St$. The dotted line in the right panel is the Gaussian fit to the PDF of the largest particles 
($St = 795$; or $\Omega =54$) in our simulation. For these particles, $\tau_{\rm p}$ is 54 times 
larger than $T_{\rm L}$, suggesting that the assumption of a Gaussian relative velocity 
PDF applies only in the extreme limit $\tau_{\rm p} \gg T_{\rm L}$.   

Fig.\ \ref{tpdfs} shows the PDF, $P(w_{\rm t}|w_{\rm r} <0, \tau_{\rm p})$, of a tangential component of the 
relative velocity conditioned on $w_{\rm r} <0$. The measurement of $P(w_{\rm t}|w_{\rm r} <0,  \tau_{\rm p})$ 
only counts particle pairs approaching each other. 
The figure is plot in the same way as Fig.\ \ref{rpdfs} for the radial component. Again, the thin dashed line in the 
left panel is for tracer particles ($St=0$).  
It corresponds to the PDF of the transverse difference, $\Delta u_{\rm t}$, 
of the flow velocity conditioned on $\Delta u_{\rm r}<0$. The shape of $P(w_{\rm t}|w_{\rm r} <0, 0)$ for tracer particles at $r= 1\eta$ is close to the 
dashed line in the right panel of Fig.\ \ref{flowpdf} for $P_{\rm u}(\widetilde{\Delta u_{\rm t}}|\Delta u_{\rm r}<0, \ell)$ at $\ell  = 1.7 \eta$ (Appendix B). 
The qualitative behavior of $P(w_{\rm t}|w_{\rm r} <0, St)$ as 
a functions of $St$ is similar to that of $P(w_{\rm r}, St)$. 

For $St=0.19$ particles,  the radial PDF tails are significantly amplified with respect to tracers (Fig.\ \ref{rpdfs}), while the 
conditional PDF of the tangential component almost coincides with the tracer PDF (Fig.\ \ref{tpdfs}). From the 
physical picture for the PDF behavior given in \S 6.2.2, the effects of the particle memory and the backward separation tend 
to amplify the PDF tails of $St=0.19$ particles. For the tangential PDF, this amplification 
effect is counteracted by the conversion of the relative velocity from the tangential to the radial direction, 
which reduces the PDF width of the tangential component. As discussed in \S 6.1.2, the 
conversion is caused by the deviation of inertial particle trajectories from the flow elements 
and the particle memory of the flow velocity in the past, which randomize the direction of $\bs {w}$ relative to the particle separation, ${\bs r}$. 
The conversion is expected to be more efficient at the PDF tails (corresponding to the sling events). 
It appears that the two opposite effects cancel out for the tangential PDF of $St \simeq 0.19$ particles, as 
it almost coincides with the dotted line for tracers. On the other hand, both effects broaden the 
PDF of the radial component, leading to significantly amplified tails with respect to tracers.   

Unlike $P( w_{\rm r}, St)$, which has a negative skewness for $St \le 1.55$ particles, $P(w_{\rm t}|w_{\rm r} <0, St)$ is 
symmetric at all $St$. As mentioned earlier, the symmetry of the  tangential PDF 
is expected from the statistical isotropy. We find that the left wing of the radial PDF $P(w_{\rm r}, St)$ 
almost coincides with that of $P(w_{\rm t}|w_{\rm r} < 0, St)$ at all $St$. This is apparently due to the 
randomization of the relative velocity direction discussed above. On the other hand, the right wing of $P(w_{\rm r}, St)$ 
is narrower than that of $P(w_{\rm r}|w_{\rm r} < 0, St)$, until it  becomes symmetric at $St \gsim 3.11$. 
To study particle collisions, we are mainly interested in the PDFs for approaching pairs, i.e., the 
left wing of $P(w_{\rm r}, St)$ and the entire tangential PDF, $P(w_{\rm t}|w_{\rm r} <0, St)$, conditioned on 
$w_{\rm r} <0$.


We give a more quantitative description for the shape of $P(w_{\rm t}|w_{\rm r} <0, St)$. 
The description also applies to the left wing of $P(w_{\rm r}, St)$, as it coincides with the left wing of 
$P(w_{\rm t}|w_{\rm r} <0,St)$. We first quantify the fatness of  $P(w_{\rm t}|w_{\rm r} <0, \tau_{\rm p})$ 
by computing the kurtosis, defined as $\langle w_{\rm t}^4 \rangle_{-}/\langle w_{\rm t}^2 \rangle_{-}^{2}$.
At $r=1\eta$, the kurtosis for $St = 0.1$ and $0.19$ particles is $\simeq 11$, which is already 
much larger than 3 for a Gaussian PDF. With increasing $St$, the kurtosis first increases due to 
the tail amplification. It reaches a maximum value of $36$ at $St=0.78$, indicating extremely 
high non-Gaussianity. The kurtosis decreases slightly to 32 at $St=1.55$, and then drops 
rapidly and approaches $\simeq 3$ for the largest particles ($St=795$). This decrease 
corresponds to the thinning trend of the high PDF tails for the large particles.  
We also measured the kurtosis for the PDFs at smaller $r$, and found that, for 
$St \lsim 6.2$ particles, it keeps increasing as $r$ decreases to $\eta/4$. The PDFs of 
these particles are fatter at smaller $r$ because the effect of the tail amplification on the PDF shape 
is relatively stronger (see more detailed discussions in \S 6.2.3). 



Following Sundaram \& Collions (1997) and Wang et al.\ (2000), we attempted to fit 
$P(w_{\rm t}|w_{\rm r} <0, \tau_{\rm p})$ with stretched exponential PDF. The generic stretched 
exponential function is given by, 
\begin{equation}
P_{\rm se} (x) =\frac{\alpha}{2 \beta \Gamma(1/\alpha)} \exp\left[- \left(\frac{|x|}{\beta} \right)^{\alpha}\right], 
\label{se}
\end{equation}
where $\Gamma$ is the Gamma function. The variance of $P_{\rm se}$ is given by $\beta^2 \Gamma(3/\alpha)/\Gamma(1/\alpha)$. Thus, to fit a given PDF by eq.\ (\ref{se}) with a chosen 
$\alpha$,  one can  fix $\beta$ by the variance of the PDF. The index $\alpha$ controls the PDF shape, 
and smaller $\alpha$ corresponds to fatter tails. The PDFs for $St = 0.1$ and $0.19$ at $r=1\eta$ almost 
have the same shape, and both can be fit by a stretched exponential with $\alpha = 0.67$. 
This value of $\alpha$ is consistent with that (0.7) used to fit the PDF tails of the flow velocity 
difference at $\ell = 1.7$ (see Appendix B and Fig.\ \ref{flowpdf}). At $St =0.39$, $0.78$ and $1.55$, the best-fit $\alpha$ is 
$0.52$, $0.48$, and $0.49$, respectively. The decrease of $\alpha$ in the $St$ range 
from 0.1 to 0.78 indicates increasing fatness of the PDF. The PDF shape at $St=1.55$ is 
very close to that at $St=0.78$.


For particles with $3.11 \le St \le 49.7$, the PDFs are more complicated, due to the existence of sharp 
cusps at the center and the steepening trend of the PDF slope toward to the far tails. 
These features cannot be captured simultaneously by a single stretched 
exponential function. It is, however, possible to fit these PDFs with a combination of two different 
stretched exponential functions for the cusp and the tails respectively. We postpone a detailed study of 
fitting functions for these intermediate particles to a future work. 
To give a quantitative idea for the PDF shape of these particles, here we 
simply list the best-fit $\alpha$ for the far tails without accounting for the central cusp. 
At $St =3.11$, 6.21, 12.4, 24.9 and $49.7$, the best-fit $\alpha$ for the PDF tails is found to be 1,  1.1,  1.3, 1.33, 
and 1.45, respectively. The stretched exponential fits for the PDF tails of $St = 24.9$ ($\Omega =1.7$) particles are 
shown as dotted lines in Figs.\  \ref{rpdfs} and \ref{tpdfs}, where the index $\alpha$ is set to $4/3$. Such a 4/3 
stretched exponential PDF was predicted by Gustavsson et al.\ (2008) assuming an exact Gaussian 
flow velocity field with Kolmogorov scaling and a rapid temporal decorrelation. An alternative 
derivation for the 4/3 stretched exponential is given in \S 6.2.2 using the physical picture 
of the PP10 model. Our derivation does not assume a short temporal correlation 
for the flow velocity, and is thus more general than that of Gustavsson et al.\ (2008). 
 
Starting from $St=99$ ($\Omega =6.8$), the central cusp becomes sufficiently small, leading to 
simpler PDF shapes. This allows the entire PDF to be satisfactorily fit by a single 
stretched exponential again. 
The measured $\alpha$ values for $St =99$, $199$, $397$ and $795$ are $1.5$, 
$1.65$, $1.75$ and $1.9$, respectively. Note that the PDF at $St =795$ is 
close to Gaussian, but the best-fit value for $\alpha$ is actually 1.9 instead of 2. 
A similar trend of the PDF fatness and the best-fit $\alpha$ as a function of $St$ 
was found in previous studies with low-resolution simulations (Sundaram \& Collions 1997; Wang et al.\ 2000).

\subsubsection{Physical picture for the PDF behavior}

We give an explanation for the behavior of  the relative velocity PDF using the 
physical picture of PP10, which was illustrated in Fig.\ \ref{cartoon}. 
We first consider particles with $\tau_{\rm p} \lsim  T_{\rm L}$. In \S 3.2.4, we showed that the 3D 
relative velocity variance of these particles may be roughly estimated by $\langle w^2 \rangle  \simeq S_{ii} \left( R_{\rm p} \right)$, 
where $S_{ij} $ is the flow structure tensor and $R_{\rm p}$ is the primary distance. For simplicity, 
we have neglected the effect of the temporal correlation function, $\Phi_2$, which may provide a 
factor, $\min(1,T(R_{\rm p})/\tau_{\rm p})$, of order of unity for particles with $\tau_{\rm p} \lsim  T_{\rm L}$. 
$R_{\rm p}$ was estimated by $R_{\rm p}^2= r^2  + \langle w^2 \rangle \tau_{\rm p}^2$, 
assuming a ballistic backward separation within a friction timescale.  

This picture for the rms relative velocity can be generalized to understand the 
behavior of the full PDF as a function of $St$. Consider a pair of particles at a distance $r$ at time 0, 
and suppose their relative velocity is $w$. Applying the above physical picture to this particular pair, 
the relative speed, $w$, is estimated as $w \simeq \Delta u (r_{\rm p})$, where $\Delta u$ is the flow 
velocity difference the two particles ``saw" at $\tau = -\tau_{\rm p}$ and $r_{\rm p}$ is the primary distance of this pair. 
We have used $w$ and $\Delta u$ to represent either the radial or the tangential component. 
The generalized picture suggests that the particle relative velocity samples the PDF of the flow velocity 
difference in a certain way. An immediate implication is that the particle relative velocity would inherit 
intermittency of the turbulent flow. Assuming a ballistic separation again, $r_{\rm p}$ is estimated 
by $(r^2 + \zeta w^2 \tau_{\rm p}^2)^{1/2}$, where $\zeta \simeq 3$ corresponds to the 
difference between the 3D separation speed of the particle pair and the 1D speed in the radial or 
tangential direction. We point out that, for particles with $ 0.8 \lsim St \lsim 6.2$, it may not be valid to assume the 
contribution to the particle relative speed is dominated by the ballistic separation phase. As discussed in 
\S 6.1, the Richardson phase may provide a crucial contribution for these particles (see Fig.\ \ref{3drms}). 
However, using the ballistic assumption to estimate $r_{\rm p}$ would be sufficient for 
a qualitative understanding of the relative velocity PDF. 

The above argument provides a satisfactory explanation for our simulation results for the relative 
speed PDF, $P(w, \tau_{\rm p})$, of particles with $\tau_{\rm p} \lsim T_{\rm L}$. At $St \ll1$, 
the primary distance $r_{\rm p}$ ($=(r^2 + \zeta w^2 \tau_{\rm p}^2)^{1/2}$) for particle pairs in the 
inner part of the PDF (i.e., at $|w| \simeq 0$) would be close to $r$. As a result, the central 
PDF follows the PDF, $P_{\rm u}(\Delta u, \ell)$, of the flow velocity difference at $\ell = r$, as 
observed in the left panels of Figs.\ \ref{rpdfs} and \ref{tpdfs}. At the tails of $P(w, \tau_{\rm p})$, $r_{\rm p}$ 
is larger, and $w$ samples the flow velocity PDF $P_{\rm u} (\Delta u, \ell)$ at larger $\ell$. This implies 
that higher tails broaden faster because $P_{\rm u}(\Delta u, \ell)$ is wider at larger $\ell$. 
The effect may be viewed as a ``self-amplification" of the PDF tails. The tail amplification makes 
the overall shape of $P(w, \tau_{\rm p}, St)$ at $St \lsim 1$ considerably fatter than the PDF of 
tracer particles. As $St$ increases, $r_{\rm p}$ becomes larger at the same 
value of $w$, and the ``amplification" proceeds deeper  into the inner part of 
the PDF,  as seen in the left panels of Figs.\ \ref{rpdfs} and \ref{tpdfs}. 
The overall PDF broadening appears to be driven by the tail amplification. 
The amplification in the far PDF tails of $St \lsim 1$ particles actually 
corresponds to the effects of slings or caustic formation. 
This is because the tail of $P(w, \tau_{\rm p})$ is associated with local flow regions 
with large velocity gradients, which are indeed where the slings or caustics are expected 
to occur. The tail amplification of $St \lsim 1$ particles thus corresponds to the caustic 
contribution to the collision kernel in the model of Wilkinson et al.\ (2006). 

As $St$ increases above 1, the range of the central PDF that follows 
$P_{\rm u}(\Delta u, \ell)$ becomes narrower, and the outer parts continue 
to get more extended.  
As discussed in \S 6.2.1, for $St  \simeq 3.11$, the PDF tails show slope changes as 
they extend to high values of $|w|$. This is because different parts of the relative velocity 
PDF samples the flow velocity difference PDF, $P_{\rm u}(\Delta u, \ell)$, at different length scales. 
As the fatness of $P_{\rm u}(\Delta u, \ell)$ decreases with 
increasing $\ell$ (see Appendix B and Fig.\ \ref{flowpdf}), the shape of $P(w, \tau_{\rm p})$ at 
higher tails becomes thinner. This thinning trend occurs at smaller values of $|w|$ for particles 
with larger $\tau_{\rm p}$. The trend explains why the overall shape of the PDF becomes less fat as $St$ 
increases above $\simeq 1$. Note that the central cusp for 
$3.11 \lsim St \lsim 24.9$ keeps a sharp shape, corresponding to $P_{\rm u}(\Delta u,\ell)$ at small $\ell$. 

The fact that the broadening of the PDF starts from the tail amplification is not captured 
by the PP10 model for the rms relative velocity (\S 3.2). The model only considers the 
2nd-order moments of the flow velocity increment, the particle separation and the 
particle relative velocity. This essentially assumes that the PDF shape does not 
considerably change with $St$, or the shape change of the PDF at the 
outer or tail parts does not have significant effects on the variance of the PDF. 
This gives rise to uncertainties in the prediction for the rms relative velocity because 
the PDF $P(w, St)$ is found to be very fat especially for $St \simeq 1$. Even the far tails 
give considerable contribution to the variance. The tail amplification also 
provides evidence for a positive correlation between the fluctuations in 
the flow velocity increment ``seen" by the particle pair 
and the particle separation. The tails of $P(w, \tau_{\rm p})$ correspond to 
the PDF tails of both the flow velocity difference, $\Delta u$, and the the 
primary distance, $r_{\rm p}$. In other words, in flow regions with $\Delta u$ larger 
than its rms value, the backward separation of particles is also faster than the rms separation rate. 
As discussed in \S 3.2 and \S 6.1, the PP10 model neglects this correlation, and thus tends 
to underestimate the rms relative speed. The effect of the PDF tail amplification on the 
variance of the relative speed may be incorporated in the PP10 model if the correlation 
between $\Delta u$ and the particle separation is properly accounted for. 
As mentioned in \S 6.1, accounting for this correlation, our model may fit the rms relative 
velocity with a smaller Richardson constant. 

In principle, if the PDF, $P_{\rm u}(\Delta u, \ell)$, of the 
flow velocity increment as a function of the scale $\ell$ is provided, 
one can derive the  particle relative velocity PDF as a function of $\tau_{\rm p}$.  For illustration, 
we consider a simplified example. We assume the flow velocity is  exactly Gaussian, i.e., 
\begin{equation} 
P_{\rm u}(\Delta u,  \ell) =  \frac{1} {\sqrt{2 \pi  S({\ell}) }} \exp \left (- \frac {\Delta u^2}{2 S(\ell)} \right),   
\label{gaussianu}
\end{equation}  
where $S(\ell)$ is the flow structure function or the variance of $\Delta u$ at $\ell$. To estimate 
the PDF, $P(w, \tau_{\rm p} )$, of the particle relative speed $w$, we ask the question what the probability 
is for two nearby particles to see a flow velocity difference of $w$ at a friction timescale ago, i.e., at $\tau \simeq -\tau_{\rm p}$. 
The probability is roughly estimated by $\propto  P_{\rm u} (w,  r_{\rm p})$. Using eq.\ (\ref{gaussianu}) for $P_{\rm u}$ 
and setting $r_{\rm p}^2= r^2 + \zeta w^2 \tau_{\rm p}^2$, we have,  
\begin{equation} 
P(w, \tau_{\rm p}) \propto \exp \left[- \frac {w^2}{2 \xi S\left(( r^2 + \zeta w^2 \tau_{\rm p}^2)^{1/2} \right)} \right],   
\label{pdfmodel}
\end{equation} 
where it is assumed all uncertainties in the rough estimate can be absorbed in a parameter $\xi$. 
If $r_{\rm p}$ is in the inertial range of the flow, we may apply the Kolmogorov scaling $S(\ell) \propto \ell^{2/3}$.  
Further assuming that $\xi$ is independent of $w$, we find that eq.\ (\ref{pdfmodel}) corresponds to a stretched 
exponential with $\alpha = 4/3$ (see eq.\ (\ref{se})) at $w \gg  r/\tau_{\rm p}$. This suggests that 
the relative speed of inertial-range particles would be non-Gaussian even if the flow statistics were exactly Gaussian. 
This non-Gaussianity originates purely from the particle dynamics, and is thus distinct from that inherited 
from the intermittency of the turbulent flow. In other words, we identified two sources, 
namely, the turbulent intermittency and the particle dynamics, that contribute to the non-Gaussianity of the particle relative velocity. 

The predicted stretched exponential with $\alpha =4/3$ was found to satisfactorily fit the PDF tails 
of $St=24.9$ ($\Omega=1.7$) particles (see dashed lines in the left panels of Figs.\ \ref{rpdfs} and \ref{tpdfs}). 
For these particles, the two assumptions made in the derivation of the stretched exponential, 
i.e., Gaussianity and Kolmogorov scaling of the flow velocity, are both satisfied. We point out 
that these assumptions are strong, and thus the validity of the 4/3 stretched exponential is quite limited. 
In fact, the prediction applies only to particles around the peak Stokes number, $St_{\rm m} \simeq 30$. As discussed in \S 6.1, 
for particles with $St \simeq St_{\rm m}$ in our simulation, the typical primary distance is around $200 \eta$. 
From Fig.\ \ref{flowpdf} (Appendix B), we see that, above this length scale, the PDF of the flow velocity increment is close to Gaussian. 
Therefore, the Gaussian assumption made in eq.\ (\ref{pdfmodel}) is valid for $St \gsim St_{\rm m}$.
 Also, Fig.\ \ref{structureandpower} shows that $200 \eta$ is toward the end of but still within the 
inertial range of the flow, meaning that, only for particles with $St \lsim St_{\rm m}$, can one apply the 
Kolmogorov scaling around 
the primary distance, $r_{\rm p}$. These suggest that the two assumptions are simultaneously met only at $St \simeq St_{\rm m}$. 
Our finding that the 4/3 stretched exponential fits the PDF tails of $St=24.9$ particles confirms the validity of 
our physical picture. The 4/3 stretched exponential can also acceptably 
fit the PDF tails of $St =12.4$ particles, but not for other particles. At the central part of the PDF 
of  $St=24.9$ (or 12.4) particles, both assumptions beak down, and the 4/3 stretched exponential does not apply.


We next consider large particles with $St \gsim St_{\rm m}$. The friction time of 
these particles is much larger than $T_{\rm L}$, and, accounting for the effect of 
the memory time of the flow velocity, the relative velocity of a given particle pair is roughly 
estimated by $w \simeq \Delta u(r_{\rm p}) (T_{\rm L}/\tau_{\rm p})^{1/2}$ (see \S 3.2.4). Due to the large friction time, 
$r_{\rm p}$ is typically comparable to or even larger than the integral scale, $L$, of the turbulent 
flow, and thus $P_{\rm u} (\Delta u, \ell)$ at $\ell \simeq r_{\rm p}$ is close to Gaussian.
Since the flow velocity ``seen" by $St \gsim St_{\rm m}$ particles 
is typically Gaussian, the PDF shape for their relative velocity is simpler than particles with intermediate 
$\tau_{\rm p}$ (see Figs.\ \ref{rpdfs} an \ref{tpdfs}).  Using eq.\ (\ref{gaussianu}) and the same analysis that led to eq.\ (\ref{pdfmodel}), 
we find $P(w, \tau_{\rm p}) \propto  \exp (- ({w^2} \tau_{\rm p})/(2 \xi S(r_{\rm p}) T_{\rm L})$ for $St \gsim St_{\rm m})$. 
The structure function, $S(\ell)$, starts to become constant at $\ell \gsim L$. Therefore, as $\tau_{\rm p} $ increases,  the typical 
$r_{\rm p}$ increases, and $S(r_{\rm p})$ becomes less dependent on $r_{\rm p}$ or $w$. As a consequence, the shape of the relative
velocity PDF becomes less fat. In the limit $\tau_{\rm p} \to \infty$, $S(r_{\rm p}) \to 2u'^2$,  and $P(w, \tau_{\rm p})$ finally approaches a Gaussian PDF 
with a variance $\propto u'^2 T_{\rm L}/\tau_{\rm p}$. As observed in Figs.\ \ref{rpdfs} and \ref{tpdfs}, 
a nearly Gaussian PDF is indeed observed for the largest particles in our simulation.  

The physical interpretation of the relative velocity PDF at the beginning of this subsection 
suggests that the PDF can be split into two parts. We use the PDF of $w_{\rm r}$ as an 
example, and, in particular, we consider the left wing of $P(w_{\rm r}, St)$, corresponding to approaching 
particle pairs. We divide the PDF into two parts using a critical value, $w_{\rm r}^{\rm c}$. 
We choose $w_{\rm r}^{\rm c}$ such that the inner PDF at $w_{\rm r}^{\rm c} \lsim w_{\rm r} \le 0$ 
follows the flow velocity difference PDF, while in the outer part ($w_{\rm r} \lsim w_{\rm r}^{\rm c}$) 
the effects of the particle memory and the backward separation dominate. 
At a given $r$, $w_{\rm r}^{\rm c}$ is roughly estimated by $-r/\tau_{\rm p}$. 
This critical value is consistent with the physical picture of Falkovich et al.\ (2002) for the sling effect. 
Falkovich et al.\ (2002) showed that the velocity gradient of small particles ($St \lsim 1$) blows up in 
a finite time once it exceeds $\tau_{\rm p}^{-1}$, leading to the sling events. At a given scale, $r$, 
this gradient criterion can be written as $|w_{\rm r}|/r \gsim \tau_{\rm p}^{-1} $.  The model of Gustavsson \& Mehlig (2011) 
suggests the same criterion for caustic formation. A similar critical value was proposed 
by Hubbard (2013) for inertial-range particles. Following  Wilkinson et al.\ (2006), 
we name the central part ($ w_{\rm r}^{c} \lsim w_{\rm r}\le 0$) and the tail part 
($ -\infty < w_{\rm r} \lsim  w_{\rm r}^{c} $) of the PDF as the continuous (or S-T) part and the caustic (sling) part, respectively. 
The terminology is based on the geometry in the position-momentum 
phase diagram of inertial particles (see Fig.\ 1 of Gustavsson and  Mehlig 2011). Because 
the two parts have different scaling behaviors with $r$,  the division is especially useful for the 
prediction of particle collisions in the $r \to 0$ limit, appropriate for applications to dust particles. 
In \S 7, we evaluate the collision kernel and show the contribution from the 
continuous part vanishes as $r \to 0$. Only the caustic particle pairs contribute to 
the collisions of nearly point-like particles. Therefore, to obtain the collision velocity PDF for the application 
to dust particles, one may exclude the contribution of the continuous part, and meanwhile push $r$ to as 
small values as possible (e.g., Hubbard 2013). A detailed study of this topic will be conducted in a future work. 
Our study for the rms relative velocity in \S 6.1 did not split the particle 
pairs into the two types, and the main purpose was to
to understand the general physics of turbulence-induced relative velocity and to validate the physical picture of PP10. 


\begin{figure*}[t]
\includegraphics[height=2.9in]{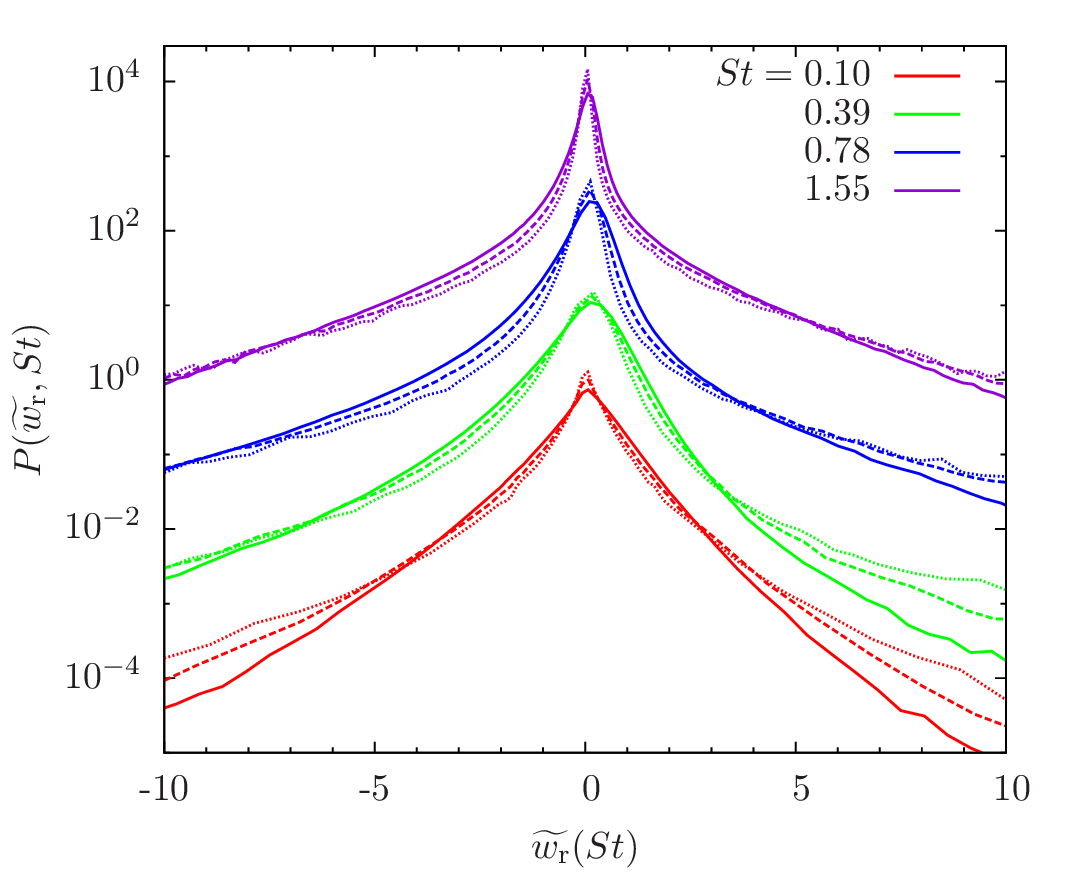}
\includegraphics[height=2.9in]{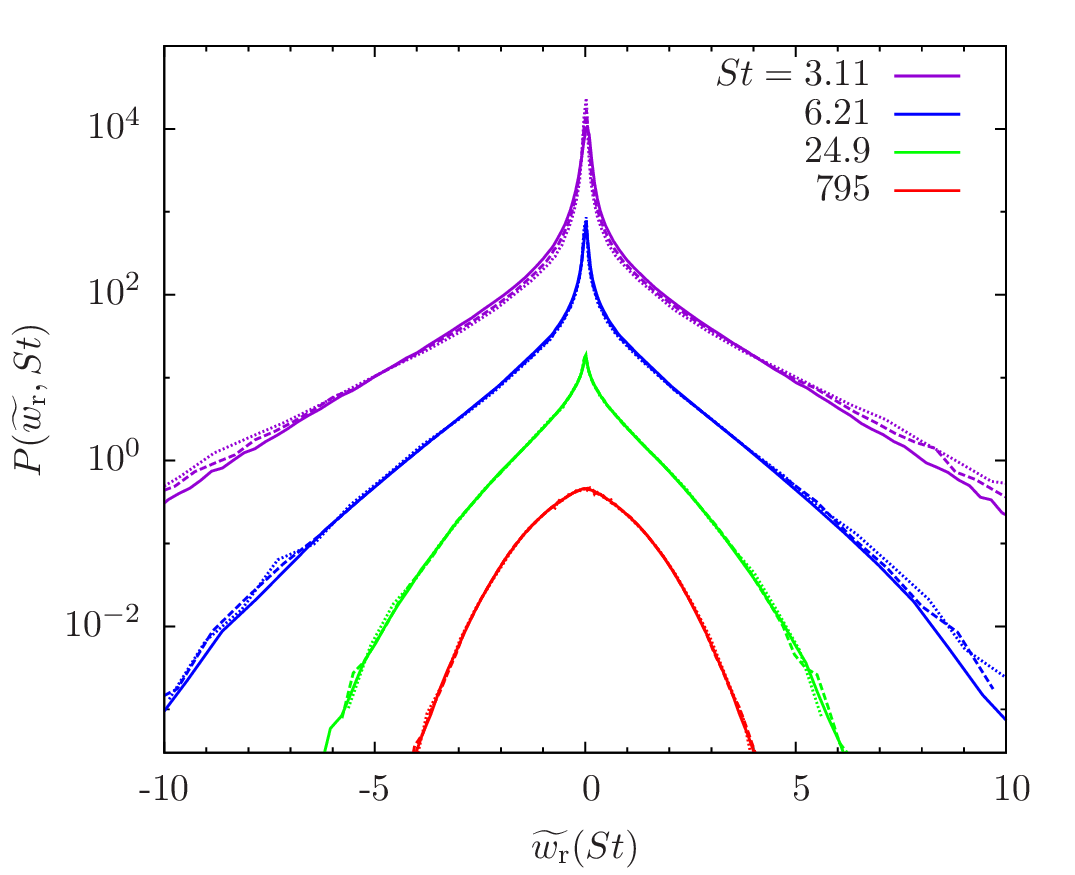}
\caption{The normalized PDF of the radial relative velocity, $w_{\rm r}$, 
as a function of the Stokes number $St$ and the particle distance, $r$. All PDFs are 
normalized to have unit variance. The normalized relative speed is defined 
as $\widetilde{w_{\rm r}} = {w_{\rm r}}/ \langle w_{\rm r}^2 \rangle^{1/2}$.  
The solid, dashed and dotted lines correspond to particle distance $r=1, 0.5$ and $0.25 \eta$, 
respectively. The left panel plots the PDFs for particles with $St \le 1.55$, 
while the right  panel shows the results for larger particles with $St \ge 3.11$. 
In each panel, the bottom lines (i.e., $St =0.1$ and $St =795$) show the actual PDF 
values, and, for clarity, the upper lines for each larger $St$ are shifted 
upward by a factor of 16.  One may change the normalization of the 
friction time using $\Omega = St/14.4$ and $\Omega_{\rm eddy} =St/19.2$.}
\label{rpdfnormalized} 
\end{figure*}

\subsubsection{The normalized PDF of the radial component}

To see the shape of the PDF more clearly, we show in Fig.\ \ref{rpdfnormalized}  the PDF 
of the radial component normalized to have unit variance. 
The radial relative speed is normalized to its rms value, $\langle w_{\rm r}^2 \rangle^{1/2}$. 
The solid, dashed and dotted lines lines are the PDFs at $r=1\eta$, $0.5$ and $0.25 \eta$, 
respectively, and the curves of different colors correspond to different $St$.  The bottom curves in the 
left and right panels plot the actual PDF values for $St=0.1$ and $St =795$($\Omega =54$), 
respectively. For clarify, the PDF curves are shifted upward by a factor of 16 for each larger $St$ 
in the left panel or each smaller $St$ in the right panel. The asymmetry of the PDFs 
at $St \lsim1.55$ is clearly seen.

As $St$ increases from 0.1 to 1.55, the central part of the normalized PDF 
$P(\widetilde{w_{\rm r}}, St)$ becomes sharper. Before normalization, the innermost part of the PDF follows the PDF 
of the flow velocity difference, and is thus essentially the same for 
particles in the range $0.1 \le St \le 1.55$ (see Fig.\ \ref{rpdfs}). Since the rms of $w_{\rm r}$ 
increases with $St$ due to the tail amplification, normalizing $w_{\rm r}$ by its rms tends to 
make the central part of the PDF sharper. 
At $St =3.11$ and $6.21$, the central cusp in the normalized PDF is very sharp.  
For $St$ above 3.11, the outer PDF parts become less fat with increasing $St$. 
The shape of the normalized PDF is helpful to understand the behavior 
of the ratio $\langle |w_{\rm r}| \rangle/\langle w_{\rm r}^2 \rangle^{1/2}$ 
with $St$ discussed in \S 7.1. 

It is interesting to note that, as $St$ increases from 0.1 to 0.39, the PDF, 
$P(\widetilde{w_{\rm r}}, St)$, at small to intermediate $\widetilde{w_{\rm r}}$ in the right wing 
decreases. This corresponds to the decrease of $\langle w_{\rm r}^2 \rangle_{+}^{1/2}$ in the $St$ 
range from 0.1 to 0.39 (see the left panel of Fig.\ \ref{plusminus}). The physical reason is that the 
right wing corresponds to separating particle pairs, and the particle distance decreases toward 
the near past. This leads to a decrease in the primary distance, $r_{\rm p}$, 
for separating pairs with small to intermediate $\widetilde{w_{\rm r}}$, as $St$ 
increases from 0.1 to 0.39. For the far right tail with large $w_{\rm r}$, 
the particle pairs may quickly move past each other, and their distance starts 
to increase within a friction time in the past. This explains why the far right tail of $St=0.39$ particles becomes slightly 
broader than that for $St=0.1$. 
As $St$ increases to $0.78$, the PDF at immediate to large $\widetilde{w_{\rm r}}$
in the right wing is significantly amplified, while the effect of the ``initial" decrease of the particle distance 
is still visible at small positive $\widetilde{w_{\rm r}}$. For $St >0.78$, the particle memory is longer, 
and the ``initial" separation phase does not cause a significant difference in the primary distances 
$r_{\rm p}$ of separating and approaching particle pairs. The two wings become almost symmetric at $St \gsim 3.11$.

For particles with $St \lsim 3.11$, the normalized PDF has a dependence on 
the distance $r$ for the range of $r$ shown here. As $r$ decreases, the central 
part of the normalized PDF becomes sharper, and the outer parts become slightly broader, leading to 
a fatter PDF shape at smaller $r$. Before the normalization, the central part of the PDF follows 
the flow velocity difference at $r$, and its width thus decreases linearly with $r$. On the other 
hand, the dependence of the tails on $r$ is weaker because the primary distance, $r_{\rm p}$, 
for particle pairs in the outer parts of the PDF has a larger contribution from backward separation. 
Also, as $r$ decreases, the contribution from the outer parts of the PDF to the variance increases. 
Consequently, normalizing the PDF to unit variance gives a sharper central 
part and broader tails at smaller $r$. The fattening of the PDF with decreasing 
$r$ can also be viewed as due to the relatively larger contribution 
of caustics at the tails. The critical value $|w_{\rm r}^{\rm c}|$ decreases with 
decreasing $r$, suggesting that, for a given $St$, there are more caustic pairs at 
smaller $r$. 
We also observe that the asymmetry in the two wings decreases with 
decreasing $r$. However, it is remains to be verified whether it exactly disappears as $r \to 0$. At $St \gsim 12.4$, 
the PDF is essentially independent of $r$, and thus directly applicable for dust 
particle collisions. For these larger particles, $r_{\rm p}$ is mainly contributed by the 
backward separation even for pairs lying around the central part of the PDF, and thus the PDF is 
independent of $r$ for $w_{\rm r}$ in any range. 


\begin{figure*}[t]
\includegraphics[height=2.9in]{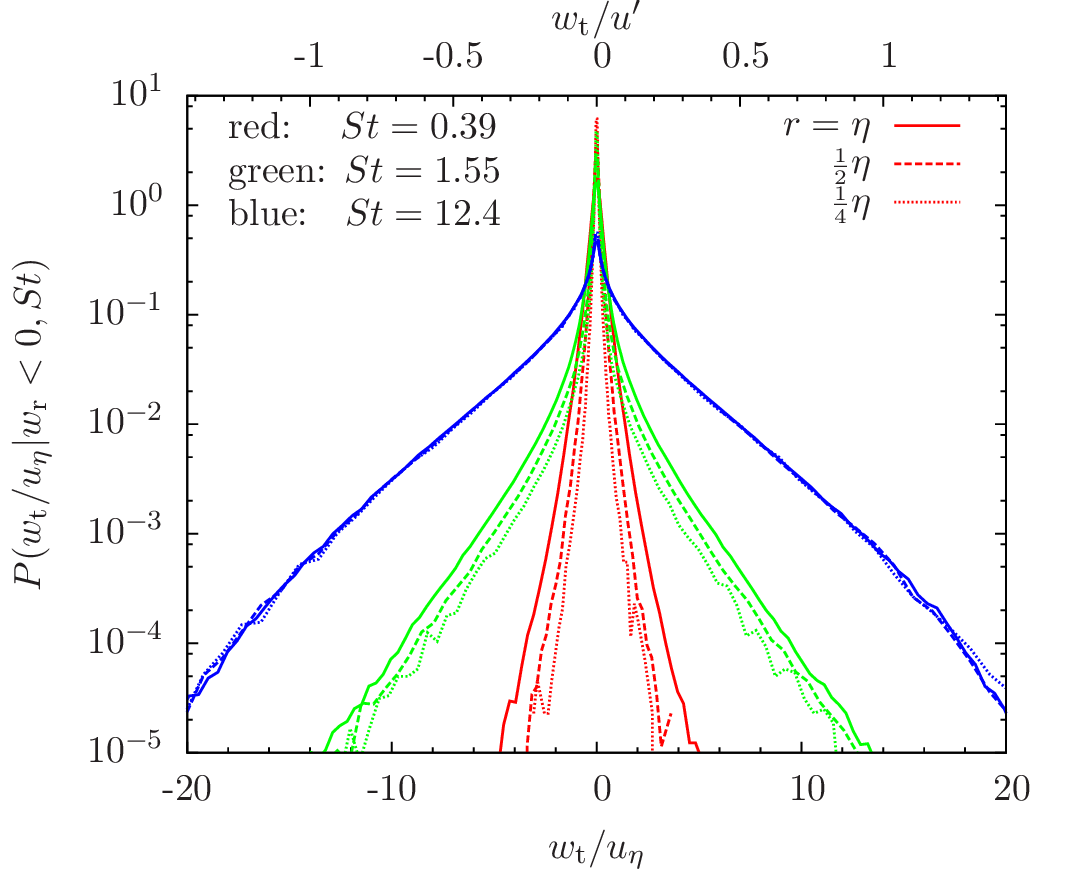}
\includegraphics[height=2.9in]{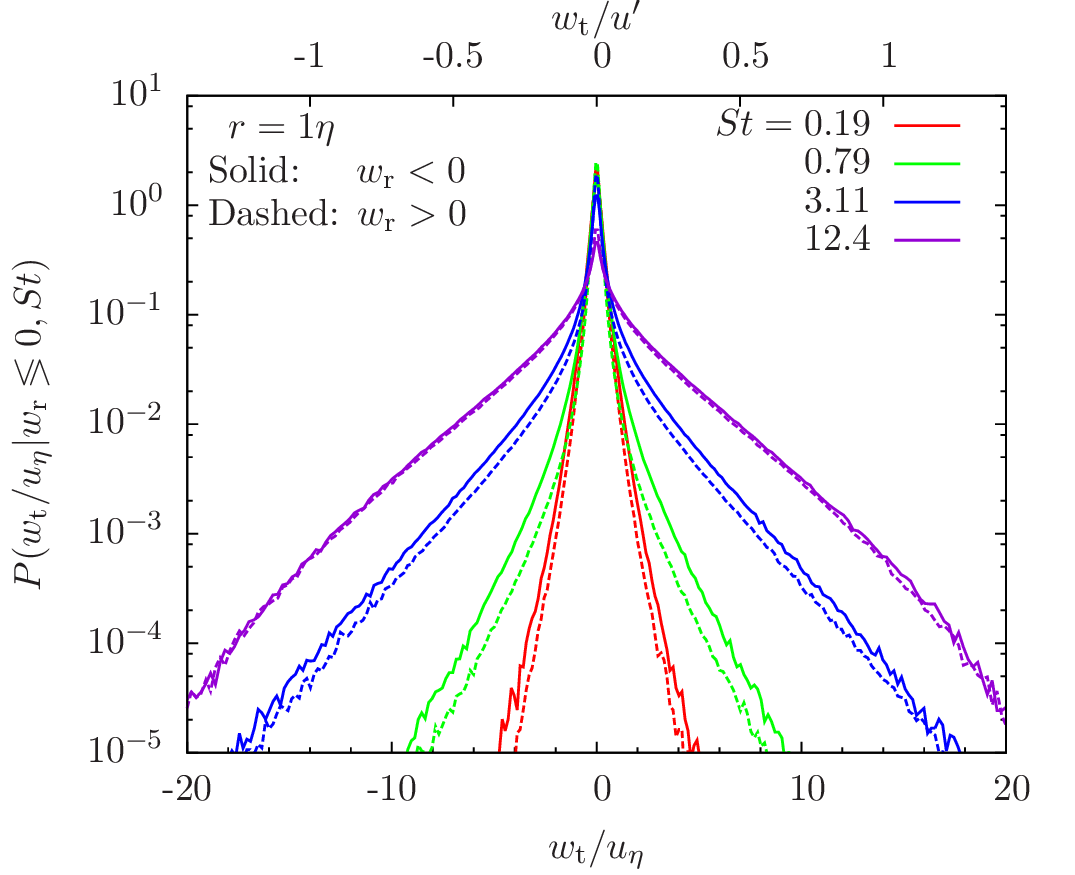}
\caption{Left panel: The PDF of tangential relative speed for approaching 
pairs at three distances for $St=0.39$, $1.55$ and $12.4$ particles. The 
PDF becomes independent of $r$ at $St \gsim 12.4$.  Right panel: the tangential PDF for 
approaching ($w_{\rm r} <0$; solid) and separating ($w_{\rm r} >0$; dashed) particle pairs at $r=1\eta$. 
For $St < 12.4$ particles, the PDF, $P(w_{\rm t}|w_{\rm r} <0$, St), of  approaching pairs is wider than 
that of separating pairs ($w_{\rm r} >0$). Above $St = 12.4$, the two conditional PDFs almost coincide.} 
\label{pmpdf} 
\end{figure*}

\subsubsection{The tangential PDFs: the $r$-dependence and approaching \& separating pairs}

The left panel of Fig.\ \ref{pmpdf} shows the PDFs of the tangential relative speed at $r=1$, 0.5, 
and $0.25\eta$ for three values of $St$. At smaller $r$, the number of particle pairs available 
is smaller, leading to a decrease in the sample size.  As a consequence, 
the PDFs become quite noisy at $r = \eta/4$, especially for small particles. For the $r$ 
range shown here, the PDF width decreases with decreasing $r$ for $St \lsim 6.21$ particles. 
The difference between the PDF tails at $\eta/4$ and $\eta/2$ appears to be smaller than that 
between $\eta/2$ and $\eta$, indicating relatively stronger contribution from caustic formation at smaller 
$r$. Like the radial PDF (Fig.\ \ref{rpdfnormalized}), the PDF becomes independent of $r$ for $St \gsim 12.4$. 

In the right panel of Fig.\ \ref{pmpdf}, we compare the PDFs of the tangential relative velocity for approaching 
($P(w_{\rm t}|w_{\rm r} <0, St)$) and separating ($P(w_{\rm t}|w_{\rm r} >0, St)$) particle pairs at $r=1\eta$. For $
St \lsim 6.2$, the PDF of approaching particles is broader than the separating ones, consistent with 
our earlier result for the rms relative speeds, $\langle w_{\rm t} ^2\rangle^{1/2}_{\mp}$ 
(see the right panel of Fig.\ \ref{plusminus}). Again, this is because, for a given ``initial" value $r$, 
the distance of approaching particles was larger in the near past than the separating ones. 
Therefore, the PDF of the relative velocity for approaching pairs samples the PDF, $P_{\rm u}(\Delta u, \ell)$, 
of the flow difference at larger $\ell$. Since the width of $P_{\rm u}(\Delta u, \ell)$ increases with 
$\ell$, $P(w_{\rm t}|w_{\rm r} <0, St)$ is expected to be broader than $P(w_{\rm t}|w_{\rm r} >0, St)$. 
At $St \gsim 12.4$, the PDFs for approaching and separating 
pairs are almost equal. For these larger particles, the primary distance at $\tau \simeq -\tau_{\rm p}$
is insensitive to the initial conditions around $\tau=0$. Although $P(w_{\rm t}|w_{\rm r} >0, St)$ for separating particle pairs is not relevant for 
particle collisions, a comparison of $P(w_{\rm t}|w_{\rm r} <0, St)$ with $P(w_{\rm t}|w_{\rm r} >0, St)$ 
provides an interesting illustration for the role of the backward separation of particle pairs in determining their relative velocity.  

\subsubsection{The normalized PDF of the 3D amplitude}

\begin{figure*}[t]
\includegraphics[height=2.9in]{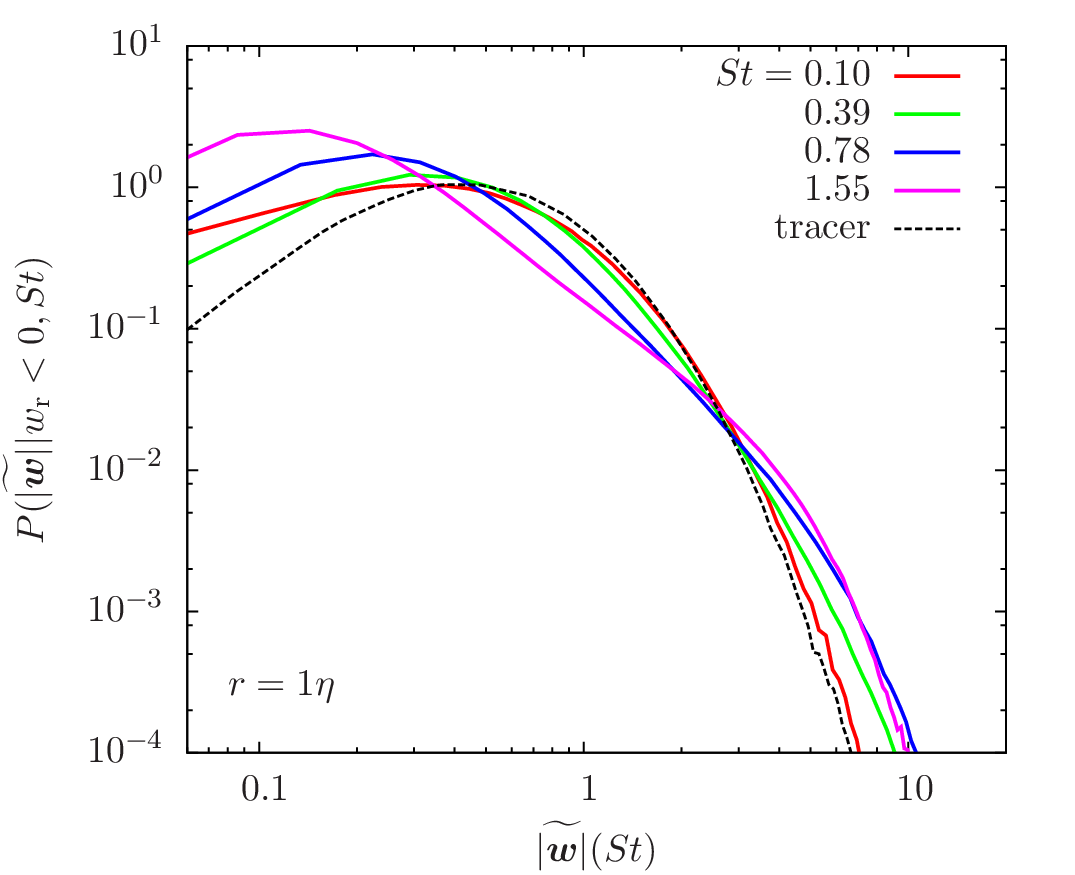}
\includegraphics[height=2.9in]{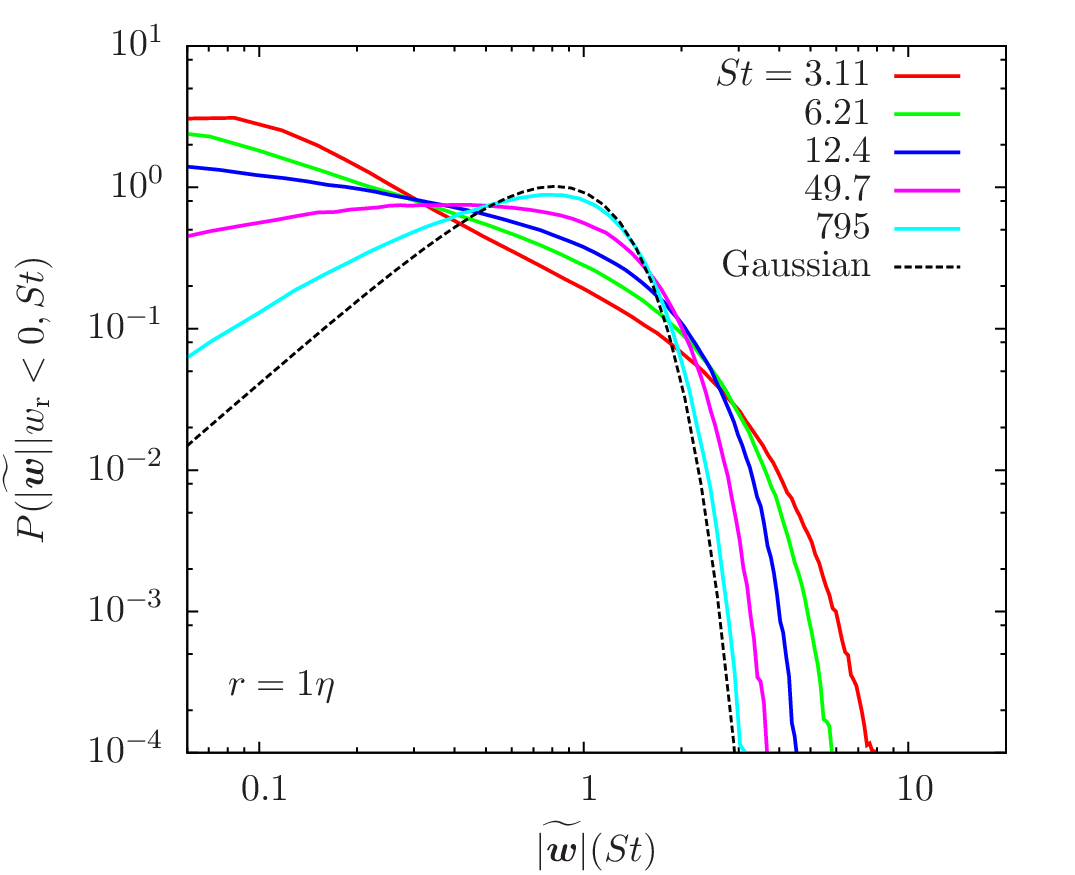}
\caption{The PDF of the 3D amplitude, $|{\bs w}|$, of the relative velocity for 
approaching particle pairs with $w_{\rm r} <0$. For each $St$, the amplitude, $|{\bs w}|$, 
is normalized to its rms value, and the PDF is normalized to have unit variance. 
In the left panel, the solid lines show results for $St  \le1.55$, and the dashed line 
corresponds to the PDF for tracer particles ($St=0$).   
The right panel plots PDFs for $St  \ge 3.11$. The dashed  line in this panel 
is the normalized PDF, $\sqrt{\frac{54}{\pi}} \widetilde{|{\bs w}|}^2\exp(-3\widetilde{|{\bs w}|}^2/2)$, 
with unit variance for the amplitude of a Gaussian vector.}
\label{3dpdf} 
\end{figure*}

Fig.\ \ref{3dpdf} plots the PDF of the 3D amplitude, $|{\bs w}|$, of the relative 
velocity for approaching particle pairs at $r=1\eta$. For each $St$, the amplitude $|{\bs w}|$ is normalized to 
the rms value $\langle w^2 \rangle_{-}^{1/2}$, i.e., $\widetilde{|{\bs w}|} = |{\bs w}|/\langle w^2 \rangle_{-}^{1/2}$, 
so that all the normalized PDFs have unit variance. The rms of the 3D amplitude, $\langle w^2 \rangle_{-}^{1/2}$, for approaching pairs 
has been shown in Fig.\ \ref{plusminus}. The left and right panels of Fig.\ \ref{3dpdf} show simulation results for 
small ($St \le1.55$) and large ($St \ge 3.11$) particles, respectively. 
The thin dashed line in the left panel corresponds to approaching tracer 
particles ($St=0$), while the dashed line in right panel is the normalized 
PDF for a Gaussian vector with three independent components of equal variance. 
The PDF for tracer particles in the left panel is already 
highly non-Gaussian, as can be seen from a comparison with the 
dashed line in the right panel. In the left panel, the degree of non-Gaussianity increases as $St$ 
increases from $0$ to $\simeq 1$. At larger $St$, the PDF peaks at smaller $\widetilde{|{\bs w}|}$. 
The PDF around the rms value (i.e., $\widetilde{|{\bs w}|} \simeq 1$) decreases 
with increasing $St$, and more probability is distributed toward smaller and larger 
values of $\widetilde{|{\bs w}|}$. This corresponds to the sharpening of $P(w_{\rm r}, St)$ 
and $P(w_{\rm t}|w_{\rm r}<0, St)$ in the central part and the broadening 
of the tail parts in this $St$ range (see Figs.\ \ref{rpdfs}, \ref{tpdfs} and \ref{rpdfnormalized}). 
The trend is reversed as $St$ increases further above $St \simeq 3.11$. The peak of the PDF 
moves back to around the rms value, $\widetilde{|{\bs w}|} \simeq 1$, 
at $St \gsim 49.7$ ($\Omega =3.5$). The PDF eventually approaches Gaussian in 
the limit $\tau_{\rm p} \gg T_{\rm  L}$.  However, note that, even for $St =795$ ($\Omega =54$) 
particles, the PDF shows a difference from the Gaussian distribution at small relative
speeds.

We find that, for $St$ in the range from 0.78 to 6.22, the PDF shows an extended power-law 
range at intermediate values of $\widetilde{|{\bs w}|}$. For example, at $St =0.78$, the PDF goes 
like $|{\bs w}|^{-2.4}$ in the range $0.5 \le \widetilde{|{\bs w}|} \le 4$. The slope of the PDF in the 
power-law range becomes shallower with increasing $St$.  For $St = 1.55$, $St = 3.11$ 
and $St =6.22$, the power-law exponent of the PDF in the intermediate $\widetilde{|{\bs w}|}$ range 
is $-1.8$, $-1.3$, and $-0.8$, respectively. The PDF of $|{\bs w}|$ can be easily computed from the 
PDFs of the radial and tangential components, if the three components are completely 
independent. In that case, one may obtain fitting functions for $P(|{\bs w}|)$ using the fitting functions discussed earlier for the 
PDFs, $P(w_{\rm r}, St)$ and $P(w_{\rm t}|w_{\rm r}<0, St)$, of approaching particle pairs.  
One could also directly fit the PDF of $|{\bs w}|$ with simple function forms or 
tabulate it as a function of $St$. 
We will provide fitting functions for  $P(|\bs w|, St) $ in a future paper.  

The strong non-Gaussianity in the amplitude of the relative velocity has interesting 
implications for the growth and evolution of dust particles in planetary disks.
We assume that the shape of the collision velocity PDF of dust particles 
at $r \to 0$ is more or less similar to the measured PDFs at $r\simeq \eta$ in our simulation, 
even though the PDFs of $St \lsim 12.4$ particles have not converged at $r \simeq \eta/4$.  
The shape of the PDFs shown in Fig.\ \ref{3dpdf} suggests that there would more collisions with 
extremely large ($\widetilde{|{\bs w}|} \gg 1$) or small ($\widetilde{|{\bs w}|}\gg 1$) relative speeds 
than estimated from a Gaussian distribution. 
The higher probability for collisions with small relative speed would favor sticking, while there are also 
more collisions that would result in fragmentations. The competition of the two opposite effects would determine whether the non-Gaussian PDF of the collision 
velocity accelerates or slows down the particle growth. 
A coagulation model incorporating the non-Gaussian statistics of the collision speed 
would give a more realistic prediction for the evolution of the size distribution of dust particles. 


\section{The Collision Kernel} 

The prediction of the collision rate is one of the main goals of our study of the particle 
relative velocity. If the mean number density of inertial particles of a given 
size is $\bar{n}_{\rm p}$, the collision rate per unit volume between these identical particles 
is estimated by $\frac{1}{2}\bar{n}_{\rm p}^2 \Gamma$, where $\Gamma$ 
is the collisional kernel and the factor 1/2 is used to avoid counting the same pair twice. 
Saffman and Turner (1956) presented two formulations for 
the collision kernel. The formulations were based on spherical and cylindrical 
geometries, respectively, and were thus named as the spherical and cylindrical 
formulations by Wang et al.\ (1998). 

For the spherical formulation, we  make use of the joint distribution, $\rho(r, w_{\rm r}; St)$, 
of the particle distance and the radial relative speed, defined at the beginning of \S 6. 
The collision kernel for identical particles is written as (e.g.,  Saffman and Turner 1956), 
\begin{equation}
\Gamma^{\rm sph} =  - 4\pi d_{\rm p}^2 \int_{-\infty}^{0} \rho(d_{\rm p}, w_{\rm r}; St) w_{\rm r} d w_{\rm r}
\label{sphericalformal}
\end{equation}
where $d_{\rm p}$ is the diameter of the particle, and the integral limits include 
only particle pairs moving toward each other. 
Using the definition of the radial PDF, $P(w_{\rm r}, St)$ as $\rho(r, w_{\rm r}; St)/g(r, St)$, eq.\ (\ref{sphericalformal}) 
can be expressed in a simpler form,  
$\Gamma^{\rm sph} = 4 \pi d_{\rm p}^2 g(r, St) F_{\rm r}^-$, 
where $F_{\rm r}$ is the inward flux of particles toward a given
reference particle defined as $F_{\rm r}^{-} = - \int_{-\infty}^{0} w_{\rm r} P(w_{\rm r}, St) dw_{\rm r}$. 


The particle statistics become stationary when the dynamics is fully relaxed. In the steady 
state, the inward flux, $F_{\rm r}^{-}$, is equal to the outward flux 
$F^{+}$($\equiv \int_{0}^{\infty} w_{\rm r} P(w_{\rm r}, St) dw_{\rm r}$) (Wang et al.\ 2000), which is confirmed by our 
simulation data\footnote{Similar 
to the variances, $\langle w_{\rm r}^2 \rangle_{\mp}$, of approaching and separating 
pairs, one can also define $\langle w_{\rm r} \rangle_{-} = F_{\rm r}^{-}/ \int_{-\infty}^{0}
P(w_{\rm r}, St) dw_{\rm r}$ and $\langle w_{\rm r} \rangle_{+} = F_{\rm r}^+/ 
\int_{0}^{\infty}P(w_{\rm r}, St) dw_{\rm r}$. Using the same reasoning in \S 6.1.3, 
we expect $\langle w_{\rm r} \rangle_{-} > \langle w_{\rm r} \rangle_{+}$ for inertial 
particles, which is confirmed by the simulation data. Together with the steady-state condition 
$F_{\rm r}^{-} = F_{\rm r}^{+}$,  we have  $\int_{-\infty}^{0}P(w_{\rm r}, St) dw_{\rm r} < \int_{0}^{\infty}P(w_{\rm r}, St) dw_{\rm r}$. This means that there tend to be more particles coming out from 
a reference particle, suggesting a larger particle density at small distances to the reference particle. 
This provides an interesting physical perspective for the origin of particle clustering.}. 
This is because the average radial velocity $\langle w_{\rm r} \rangle =0$, as expected 
from statistical isotropy, and $\langle w_{\rm r} \rangle =  F_{\rm r}^{+} -  F_{\rm r}^{-}$ by definition. 
We thus have $F_{\rm r}^{+} =  F_{\rm r}^{-} = \frac{1}{2} \langle |w_{\rm r} |\rangle$ 
where $\langle |w_{\rm r} |\rangle$ is the ensemble average of the absolute value of the 
radial relative velocity. The collision kernel can then be written as (Wang et al.\ 2000),
\begin{equation}
\Gamma^{\rm sph} = 2\pi d_{\rm p}^2 g(d_{\rm p}, St) \langle |w_{\rm r}| \rangle.
\label{spherical}
\end{equation}   
The cylindrical formulation assumes that all particles inside a cylinder of length 
$\langle |{\bs w}| \rangle dt$ located at a distance $d_{\rm p}$ from a given particle will 
collide with the particle in a time interval $dt$.   Similar to eq.\ (\ref{spherical}), 
the cylindrical collision kernel can be written as 
$\Gamma^{\rm cyl}=\pi d_{\rm p}^2 g \langle |{\bs w}| \rangle$, with $\langle |{\bs w}| \rangle$ 
being the average of the 3D amplitude of the relative velocity. 

To evaluate the collision kernel, one can start directly from eq.\ (\ref{sphericalformal}) using the 
joint distribution, $\rho(r, w_{\rm r}; St)$. However, previous studies have shown that 
considering the RDF and $\langle |w_{\rm r}| \rangle$ separately provides 
interesting insights on the estimate of the particle collision rate 
(e.g., Sundaram \& Collins 1997, Wang et al.\ 2000).  In \S 7.1, we 
compute the RDF and the absolute average of the relative speed from our simulation data. 

\subsection{The Overall RDF and Absolute Average of the Relative Speed}

The RDF represents the probability of finding a neighboring particle at a distance of $r$ 
with any relative velocity, and is a measure for the spatial clustering of the particles. 
Due to their finite inertia, inertial particles do not exactly follow the flow velocity and 
have been found to exhibit inhomogeneous distribution even in incompressible 
turbulence. Turbulent clustering of inertial particles has been extensively investigated in the 
literature (see e.g., Maxey 1987, Sundaram and Collins 1997, Wang et al.\ 2000, 
Cuzzi et al.\ 2001, Hogan \& Cuzzi 2001, Balkovsky et al.\ 2001, Zaichik et al.\ 2003, Falkovich and Pumir 2004, 
Cuzzi et al.\ 2008, Pan et al.\ 2011). The general physical interpretation 
for turbulent clustering is that vortical structures in turbulent flows tend to 
expel inertial particles. The particles are pushed out of high-vorticity regions by the 
centrifugal force, leading to the formation of clusters in strain-dominated regions. 


\begin{figure}[t]
\centerline{\includegraphics[width=1.1\columnwidth]{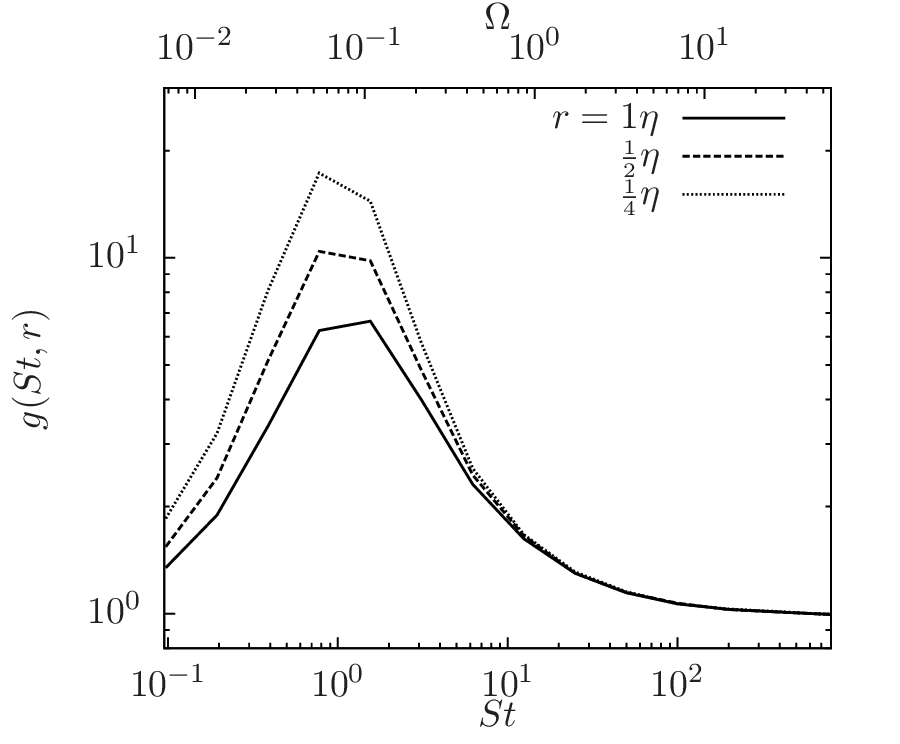}}
\caption{The radial distribution function, $g(St, r)$, as a function of the Stokes number at $r=1$ (solid), $0.5$ (dashed) and 
$0.25\eta$ (dotted).}    
\label{rdf} 
\end{figure}

Fig.\ \ref{rdf} plots the RDF as a function of $St$ at $r=1$, 0.5 and $0.25\eta$. 
Consistent with previous studies, the RDF is largest for $St \simeq 1$ particles, 
whose friction timescale couples with the smallest scale of the turbulent flow. 
For  $r$ in the dissipation rage, the RDF increases toward smaller $r$ as a power-law, 
i.e., $g(St, r) \propto r^{-\mu}$. The scaling exponent $\mu$ peaks at $St \simeq 1$, 
and approaches zero in the limits $St \ll 1$ and $St \gg1$. 
We measured $\mu$ using the values of $r$ shown in Fig.\ \ref{rdf}, and 
found that $\mu = 0.73$ for $St=0.78$, consistent with the result of Pan et al.\ (2011). 
The interested reader is referred to Pan et al.\ (2011) for the scaling exponent, $\mu$, as a 
function of $St$. 

\begin{figure*}[t]
\includegraphics[height=2.5in]{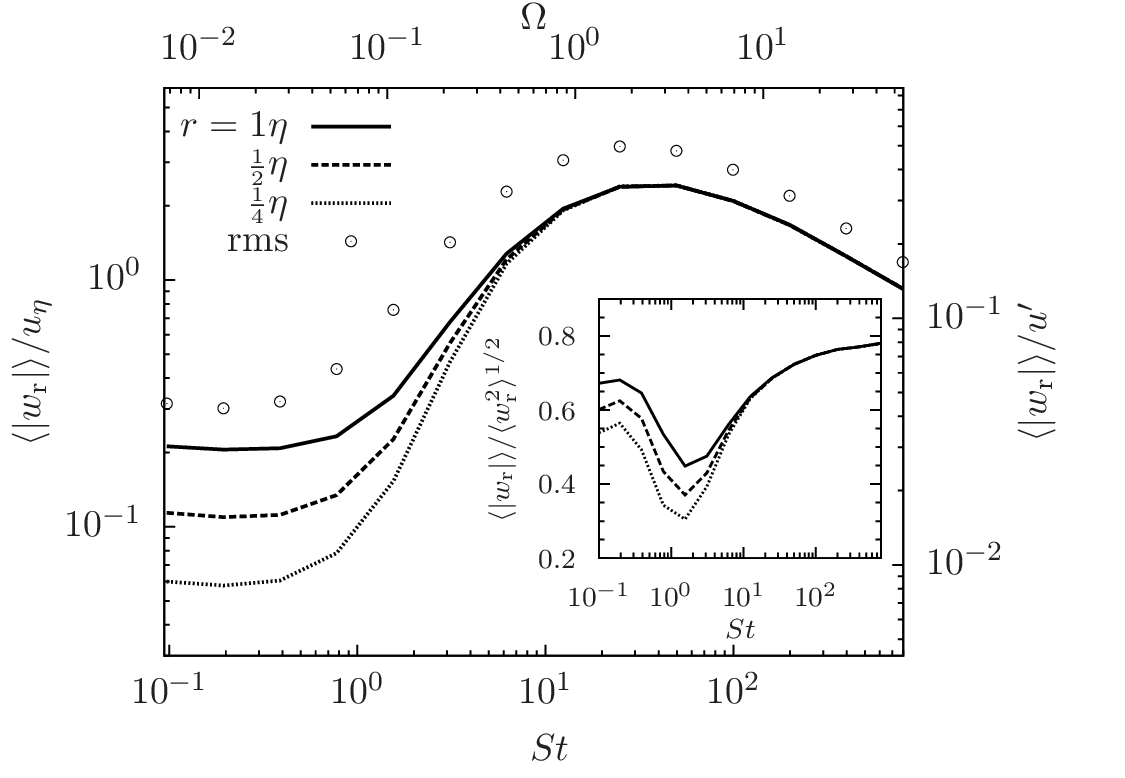}
\includegraphics[height=2.5in]{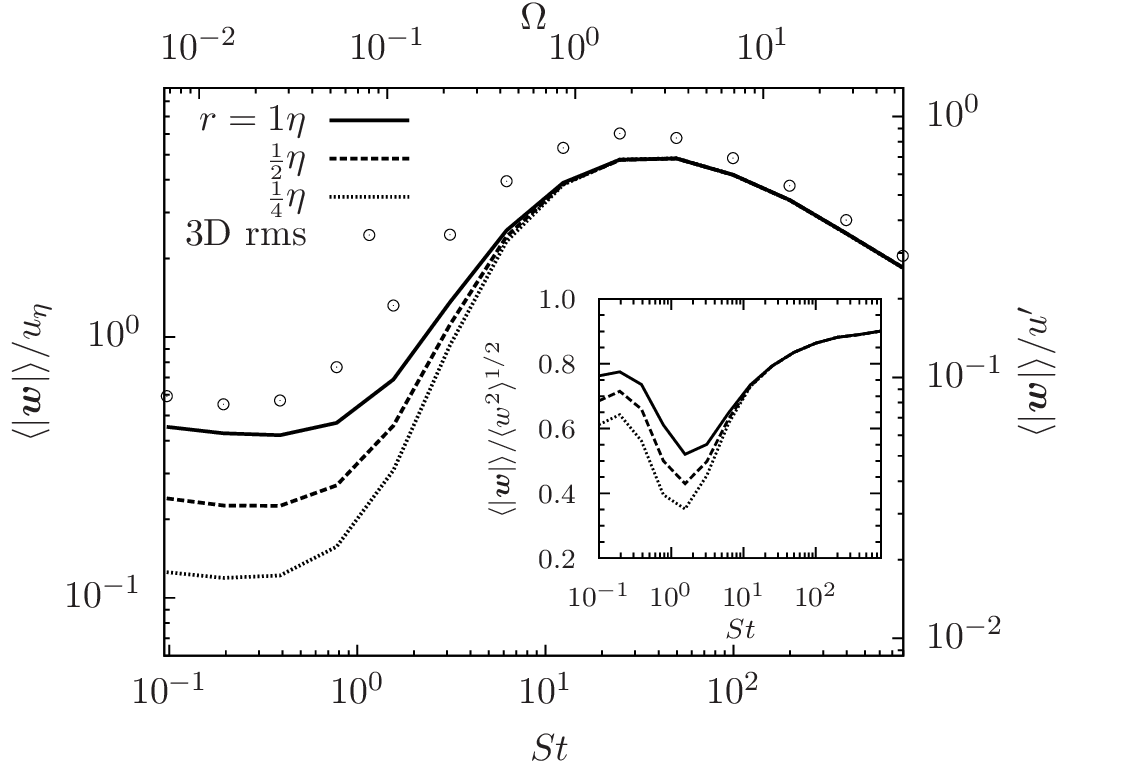}
\caption{Left panel: The average of the absolute value of the radial relative speed, $\langle |w_{\rm r} | \rangle$, 
as a function of the Stokes number. Solid, dashed and dotted lines correspond to the particle distance $r=1, 0.5$ 
and $0.25 \eta$, respectively. For comparison, the circles show the rms, $\langle w_{\rm r}^2 \rangle^{1/2}$, 
of the radial relative speed at $r=1\eta$. The inset plots the ratio, 
$\langle |w_{\rm r} | \rangle/\langle w_{\rm r}^2 \rangle^{1/2}$, at $r=1$ (solid), $0.5$ (dashed)  and 
$0.25 \eta$ (dotted). Right panel: same as the left panel, but for the 3D amplitude, $|{\bs  w}|$, 
and the 3D rms, $\langle w^2 \rangle^{1/2}$, of the relative velocity.} 
\label{absolute} 
\end{figure*}

In the left panel of Fig.\ (\ref{absolute}), we plot the simulation result for $\langle |w_{\rm r} | \rangle$ 
at $r=1, 0.5$, and $0.25 \eta$. For comparison, we also show the data (circles) for 
the rms of the radial relative speed at $r=1\eta$. Qualitatively, $\langle |w_{\rm r} | \rangle$ as 
a function of $St$ and $r$ is similar to the rms.  It is smaller than the rms, as it 
corresponds to the 1st-order moment of the PDF, $P(w_{\rm r}, St)$, of $w_{\rm r}$.
Most theoretical models, including our own, for the particle relative velocity are based on the 
computation of the variance (e.g., $\langle w_{\rm r}^2\rangle$), and cannot be directly 
applied to predict $\langle |w_{\rm r} |\rangle$. The conversion between 
$\langle w_{\rm r}^2\rangle$ and $\langle |w_{\rm r} |\rangle$ relies on the shape of the 
PDF of $w_{\rm r}$, which is difficult to predict. We thus did not attempt to fit $\langle |w_{\rm r} |\rangle$ 
with a model prediction. 

Similar to the S-T formula (eq.\ \ref{saffmanturner}) for the variances of the relative 
velocity, we would predict that $\langle |w_{\rm r} |\rangle = \langle |\Delta u_{\rm r} |\rangle$ 
in the $St \ll 1$ limit, where $\langle |\Delta u_{\rm r}| \rangle$ is the absolute average of the 
longitudinal flow velocity increment. At $\ell \lsim \eta$, $\langle |\Delta u_{\rm r}| \rangle$ 
is expected to scale linearly with $\ell$. 
We find that, for $St \ll 1$, $\langle |w_{\rm r} |\rangle \propto r^{0.9}$, which is slightly 
shallower than the linear scaling. This is likely caused by the contribution from the effect of slings or caustics. 
The scaling is steeper than $r^{0.78}$ for the radial rms velocity (see \S 6.1.2), 
suggesting that $\langle |w_{\rm r} |\rangle$ follows the flow velocity 
scaling better. The behavior of $\langle |\Delta u_{\rm r}| \rangle$ at small $St$ 
also appears to be more regular than that of $\langle w_{\rm r}^2 \rangle^{1/2}$. 
For all three values of $r$, the $\langle |w_{\rm r} |\rangle$ curves become flat at $St \lsim 0.4$.
The likely reason is that $\langle |w_{\rm r} |\rangle$ represents statistics at a 
lower order than the variance (or rms), and is thus less affected by the rare and extreme sling events, 
or by the numerical uncertainty in the particle trajectory computation. 


The inset of the left panel shows the ratio of the absolute average to the rms. The ratio 
depends on the PDF shape of $w_{\rm r}$, and particularly on the central part of the 
PDF because both $\langle |w_{\rm r} |\rangle$ and the rms are lower-order moments. As a reference, 
if the PDF $P(w_{\rm r}, St)$ is Gaussian, we have $\langle |w_{\rm r} |\rangle/\langle w_{\rm r}^2\rangle^{1/2} =(2/\pi)^{1/2} = 0.80$ (e.g., Wang et al.\ 2000), 
and for an exponential PDF it is equal to $1/\sqrt{2}$. The ratio from a Gaussian PDF was 
usually used to convert the model predictions for the rms to $\langle |w_{\rm r} |\rangle$ (Wang et al.\ 2000, Zaichik et al. 2003, 2006).  
Generally, the ratio is smaller if the central PDF is sharper and the tails are fatter. 
As seen in the inset, the ratio 
reaches a minimum at $St\simeq 1$, corresponding to a maximum fatness 
of the PDF at $St\simeq 1$ (see \S 6.2.3). At $r \le \eta$, 
the minimum is smaller than 0.45, corresponding to highly non-Gaussian 
PDF. 
At $St \simeq 800$, $\langle |w_{\rm r} |\rangle/\langle w_{\rm r}^2\rangle^{1/2}$ 
reaches 0.78, close to the expected value for a Gaussian PDF. For the range of $r$ shown here, 
the ratio also decreases with decreasing $r$ for $St \lsim 6$, which is expected from the trend of the PDF shape with $r$ for these particles (\S 6.2.3). 

Our simulation result for $\langle |{\bs w}| \rangle$ is shown in the right panel of Fig.\ (\ref{absolute}), 
which is very similar to the left panel for $\langle |w_{\rm r}| \rangle$.  
At $St\lsim 0.4$, we find $\langle |{\bs w}| \rangle$ also scales as $r^{0.9}$ with $r$. 
The ratio of $\langle |{\bs w}| \rangle$ to $\langle w^2 \rangle^{1/2}$ also shows 
a dip at $St \simeq 1$. The ratio approaches 0.9 at the largest $St$, as 
expected from a 3D Gaussian distribution.

\begin{figure}[b]
\centerline{\includegraphics[width=1.1\columnwidth]{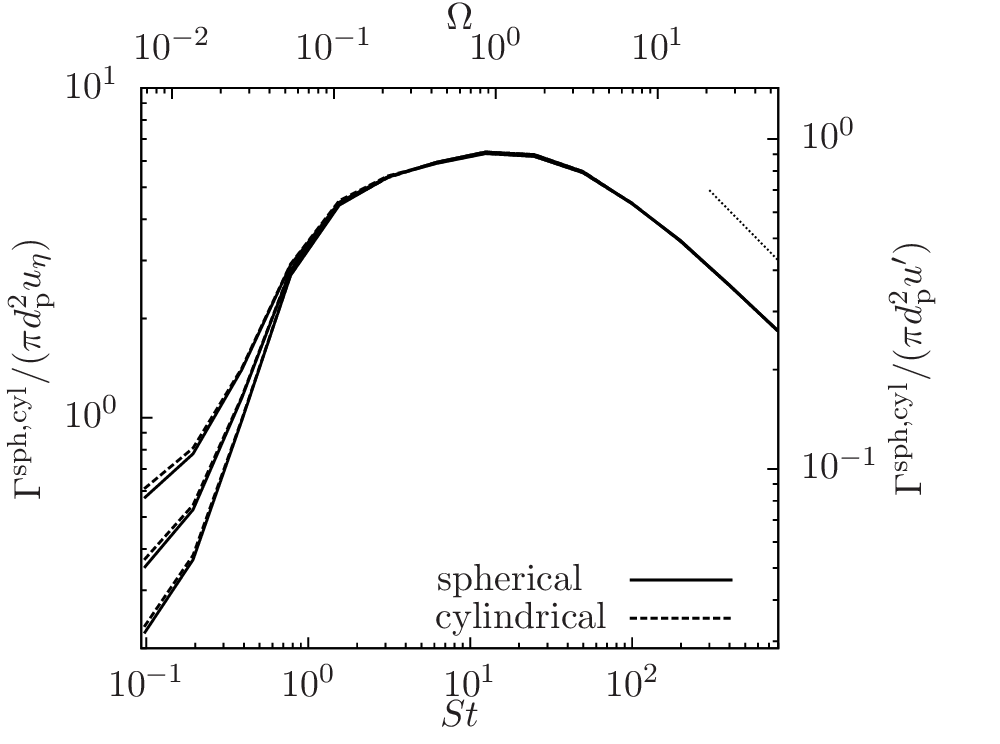}}
\caption{The collision kernels normalized to $\pi d_{\rm p}^2 u_\eta$. 
The solid lines plot $2 g(St, r) \langle |w_{\rm r}| \rangle/u_\eta$ for the spherical 
formulation, and the dashed lines show $g(St, r) \langle |{\bs w}| \rangle/u_\eta$ for the 
cylindrical formulation. The lines from bottom to top correspond to normalized kernels 
measured at $r=0.25$, $0.5$ and $1\eta$, respectively. The dotted line segment corresponds to a $St^{-1/2}$ scaling.}    
\label{product} 
\end{figure}

In Fig.\ \ref{product}, we plot $\Gamma^{\rm sph}$ and $\Gamma^{\rm cyl}$ normalized to 
$\pi d_{\rm p}^2 u_\eta$. 
The normalized kernels correspond to 
the product $2g(St, r)\langle |w_{\rm r}| \rangle/u_\eta$ (solid lines)  for the spherical formulation and
$g(St, r)\langle |{\bs w}| \rangle/u_\eta$ (dashed lines) for the cylindrical formulation. 
We will also refer to these products as the collision kernel per unit cross section.
At each $r$, the solid and dashed lines almost coincide, meaning that $\Gamma^{\rm sph}$ 
and $\Gamma^{\rm cyl}$ are nearly equal at all $St$ and $r$.   
This suggests that $\langle |w_{\rm r}|\rangle \simeq 0.5 \langle |{\bs w}| \rangle$ since 
$\Gamma^{\rm sph}/\Gamma^{\rm cyl} = 2\langle |w_{\rm r} | \rangle/\langle |{\bs w}| \rangle$. 
The two collision kernels have a noticeable difference only at $St = 0.1-0.2$, where $\Gamma^{\rm sph}$ 
is smaller than $\Gamma^{\rm cyl}$ by $\lsim 5\%$. This is consistent with the result of 
Wang et al.\ (2000). 
Wang et al.\ (2000) also showed that the spherical formulation provides an almost exact description for the particle collision rate. The 
near equality of $\Gamma^{\rm sph}$ and $\Gamma^{\rm cyl}$ at all $St$ and $r$ suggests that 
one can apply either formulation to evaluate the collision rate.

Fig.\ \ref{product} shows that $2g(St, r) \langle |w_{\rm r}| \rangle$ and $g(St, r) \langle |{\bs w}| \rangle$ 
are independent of $r$ for $St \ge 1$. Apparently, this is because, at $St \gsim 1$,  the $r$-dependences of 
the RDF $g(St, r)$ and the absolute average $\langle |w_{\rm r}| \rangle $ (or $\langle |{\bs w}| \rangle$) almost 
cancel out (see Figs.\ \ref{rdf} and \ref{absolute}).  A more interesting perspective is that the 
inverse scaling of $g(St, r)$ and $\langle |w_{\rm r}| \rangle$ is expected from the intuition that 
the normalized kernel approaches a finite constant, corresponding to a finite collision rate, at sufficiently small $r$. 
Therefore, the scalings of $g(St, r)$ and $\langle |w_{\rm r}| \rangle$ must cancel out once the kernel converges 
(see more discussions in \S 7.2). 
At $St =1.55$,  $g(St, r)$ increases with decreasing $r$ as $\propto r^{-0.55}$, while both 
$\langle |w_{\rm r}| \rangle $ and $\langle |{\bs w}| \rangle$ scale with $r$ as $\propto r^{0.57}$. 
Note that, if the $r^{0.57}$ scaling of the relative speed persists in the $r\to 0$ limit, it does mean the
collision energy is zero for nearly-point particles. This is because $\langle |w_{\rm r}| \rangle $ or $\langle |{\bs w}| \rangle$ does 
not represent the collision energy. The average collision energy per collision is expected to be finite even 
if $\langle |w_{\rm r}| \rangle $ or $\langle |{\bs w}| \rangle$ approaches zero at $r\to0$ (see \S7.2). 
The cancellation between the overall $g(St, r)$ and $\langle |w_{\rm r}| \rangle$ for $St \gsim 1$ particles 
needs to be interpreted with care. As discussed in \S 6.2, at a given $r$, there are two 
types of particle pairs, i.e., continuous and caustic pairs. The two types have 
different properties,  and the inverse scalings of the overall RDF and 
relative velocity could be an artifact of not properly splitting the two types of particle pairs\footnote{For example, if for $St \gsim 1$ particles 
one type of pairs dominates the contribution to the RDF, while the other provides a dominant and  $r$-independent 
contribution to the collision kernel, then the cancellation of the scaling exponents of the overall RDF and relative velocity would not be meaningful physically.}.
In \S 7.2, we will evaluate the contribution from each type of pairs.  

We find that the collision kernel per unit cross section shows an abrupt increase as $St$ 
increases toward 1 (see, e.g., Sundaram \& Collins 1997), increases only slightly for $St$ between $1$ 
and $St_{\rm m} \simeq 30$, 
and starts to decrease at $St \simeq 30$. It finally scales with $St$ as $St^{-1/2}$ at $\tau_{\rm p} \gg T_{\rm L}$. 
A physical discussion for the behaviors is given in \S 7.2.   

The $r-$independence of the kernel of $St \ge 1$ particles at $r\simeq \eta$ implies that one may apply our result to 
estimate the collision kernel of $St \gsim 1$ dust particles in protoplanetary turbulence, even though 
the dust particle diameter $d_{\rm p}$ is smaller than $\eta$ by orders of magnitude. On the other hand, 
at $St \lsim 1$, the measured collision kernel depends on $r$ in the range of $r$ considered, and is 
thus not directly applicable for dust particles with $St \lsim 1$. To achieve a general 
understanding of the $r \to 0$ limit, it would be useful if one could isolate an $r$-independent contribution. 
For this purpose, we make a preliminary attempt to separate the contributions 
from the continuous and caustic pairs. 

\subsection{Decomposing the Continuous and Caustic Contributions}

As discussed in \S 6.2, at a given distance, $r$, there are two types of particle pairs, 
corresponding to the continuous (inner) and caustic (tail) parts of the relative velocity PDF. 
For the radial relative speed  PDF of approaching particle pairs, the two parts can be 
roughly divided by a critical value, $w_{\rm r}^{\rm c} \simeq - r/\tau_{\rm p}$.  
In this section, we separate the contributions of the two types of pairs to the 
collision kernel in the spherical formulation. Using the relation $\rho(r, w_{\rm r}; St) = g(r, St) P(w_{\rm r}, St)$, 
and splitting the left wing of $P(w_{\rm r}, St)$ into two parts, 
the collision kernel, eq.\ (\ref{sphericalformal}), is written as, 
\begin{equation}
\Gamma^{\rm sph} =  \Gamma^{\rm con} +  \Gamma^{\rm cau},
\label{split}
\end{equation}
where $\Gamma^{\rm con} =  - 4\pi d_{\rm p}^2 g(r, St)  \int_{w_{\rm r}^{\rm c}}^{0}  w_{\rm r} P(w_{\rm r}, St)  d w_{\rm r} $ and $\Gamma^{\rm cau} =  - 4\pi d_{\rm p}^2 g(r, St) \int_{-\infty}^{w_{\rm r}^{\rm c}} w_{\rm r} P(w_{\rm r}, St)  d w_{\rm r}$.
The continuous and caustic contributions  can be expressed in more convenient forms,  
\begin{gather}
\Gamma^{\rm con} =   4\pi d_{\rm p}^2  g^{\rm con} \langle w_{\rm r}^{\rm con} \rangle \notag \\
\Gamma^{\rm cau} =   4\pi d_{\rm p}^2  g^{\rm cau} \langle w_{\rm r}^{\rm cau} \rangle
\label{sphericalformal3}
\end{gather} 
where $g^{\rm con}(r, St) \equiv g(r, St) \int_{w_{\rm r}^{\rm c}}^{0} P(w_{\rm r}, St) dw_{\rm r}$ 
and $g^{\rm cau}(r, St) \equiv g(r, St)  \int_{-\infty}^{w_{\rm r}^{\rm c}} P(w_{\rm r}, St) dw_{\rm r}$ 
are the RDFs of continuous and caustic pairs, corresponding to the probabilities of  
finding an approaching neighbor that makes a continuous and caustic pair, respectively. The 
sum of $g^{\rm con}$ and  $g^{\rm cau}$ is given by $g^{-} \equiv g(r, St) \int_{-\infty}^{0} P(w_{\rm r}, St) dw_{\rm r}$. 
Because there are more separating pairs than approaching ones, $g^{-}$ is smaller (by 10-20\%) than 
$0.5g(r, St)$ at $St \lsim10$. It becomes equal to $0.5 g$ at $St \gsim 10$. The average  radial  
relative speeds for the two types of pairs are defined as $\langle w_{\rm r}^{\rm con} \rangle = - \int_{w_{\rm r}^{\rm c}}^{0}  w_{\rm r} 
P(w_{\rm r}, St) dw_{\rm r}/  \int_{w_{\rm r}^{\rm c}}^{0} P(w_{\rm r}, St) dw_{\rm r}$, and $\langle w_{\rm r}^{\rm cau} \rangle = - \int_{-\infty}^{w_{\rm r}^{\rm c}} w_{\rm r} P(w_{\rm r}, St) dw_{\rm r}/ \int_{-\infty}^{w_{\rm r}^{\rm c}} P(w_{\rm r}, St) dw_{\rm r}$. 
The four quantities, $g^{\rm con}$, $g^{\rm cau}$, $\langle w_{\rm r}^{\rm con} \rangle$, and $\langle w_{\rm r}^{\rm cau} \rangle$, 
depend on $w_{\rm r}^{\rm c}$, whose exact value is uncertain. We will set 
$w_{\rm r}^{\rm c}=- r/\tau_{\rm p}$ here, and defer a study on the effect of the choice of $w_{\rm r}^{\rm c}$ to a later work. 

In the formulation of Falkovich et al.\ (2002) for the collision kernel of $St \lsim 1$ particles, 
the clustering effect enters only for the S-T (continuous) contribution. The RDF was 
theoretically modeled and evaluated by considering the continuous-type particles only.
For the continuous pairs, $\langle w_{\rm r}^{\rm con} \rangle$ scales linearly with $r$, as in the S-T prediction. The 
continuous contribution to the kernel is amplified by the clustering effect.  On the other hand, the 
Falkovich et al.\ formulation suggests that clustering plays no role for the sling (caustic) contribution. Both 
$\langle w_{\rm r}^{\rm cau} \rangle$, and the normalized kernel $\Gamma^{\rm cau}/(\pi d_{\rm p}^2)$ due to the slings 
are independent of the particle distance (Falkovich et al.\ 2002). These also imply that $g^{\rm cau}$ is $r$-independent. 
The model of Gustavsson \& Mehlig (2011) provides a similar formulation for the collision kernel. The
continuous contribution is essentially the same as that in Falkovich et al.\ (2002).  They also predicted an $r$-independent caustic 
contribution to the normalized kernel. However, unlike  Falkovich et al.\ (2002), their derivation accounts for the possibility that
 both $g^{\rm cau}$ and $\langle w_{\rm r}^{\rm cau} \rangle$ may depend on $r$. But their scaling exponents exactly cancel out, giving 
 an $r$-independent contribution to the kernel.   

\begin{figure*}[t]
\includegraphics[height=2.75in]{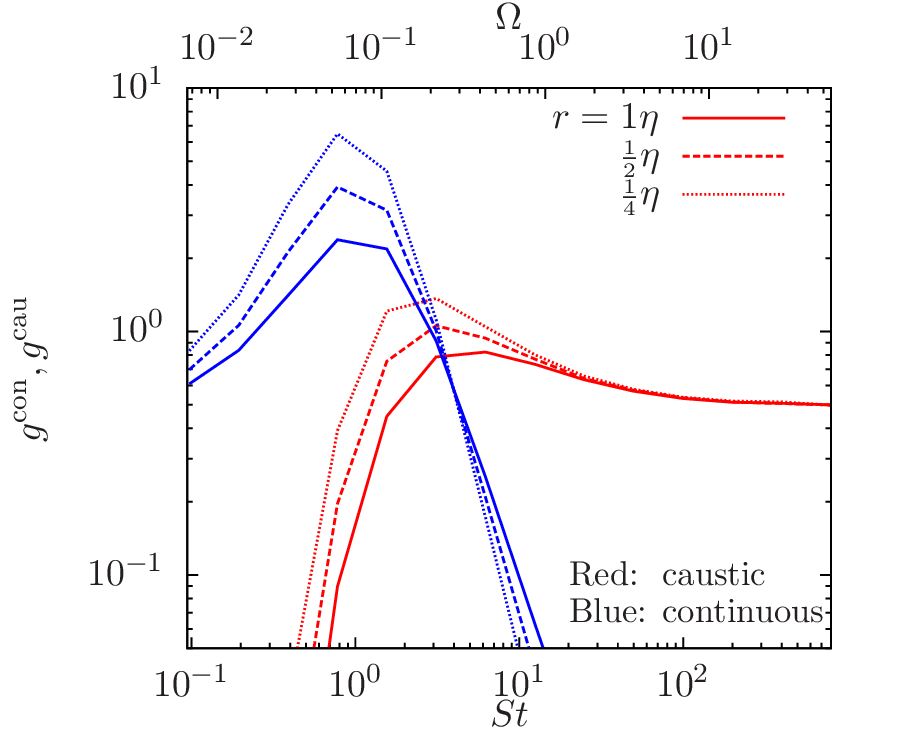}
\includegraphics[height=2.75in]{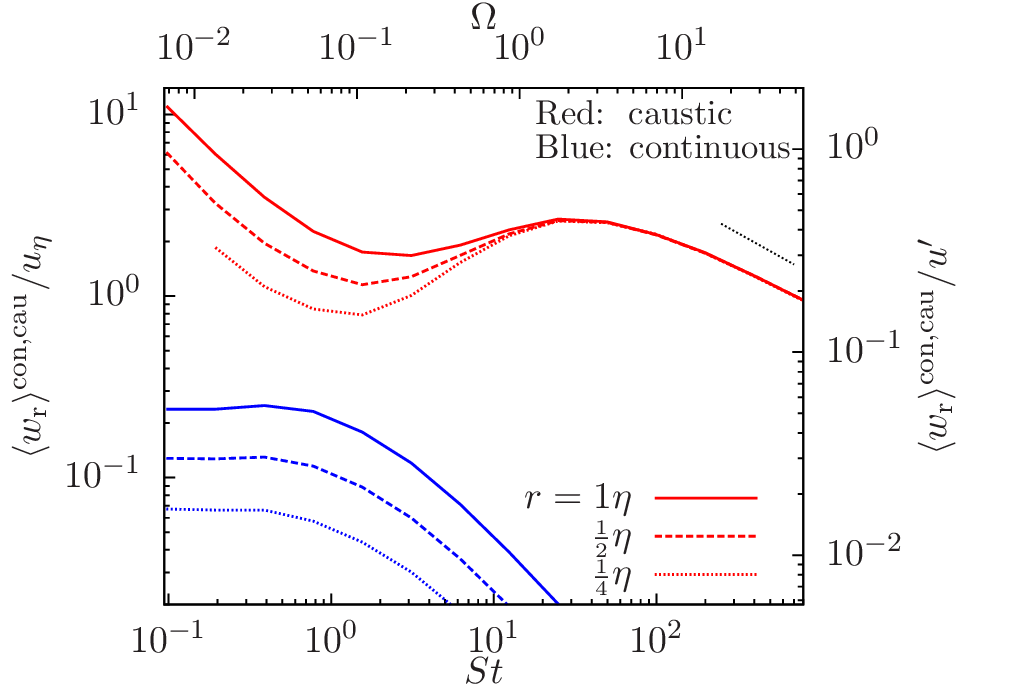}
\caption{Left panel: RDFs for continuous (blue) and caustic (red) particle pairs at $r=1\eta$ (solid), 
0.5 $\eta$ (dashed) and 0.25 $\eta$ (dotted), respectively. Right panel: Average velocities, 
$\langle w_{\rm r}^{\rm con} \rangle$ and $\langle w_{\rm r}^{\rm cau} \rangle$, of continuous and caustic pairs 
at the same three distances. The dotted line segment denotes a $St^{1/2}$ scaling. 
In both panels, the critical value, $w_{\rm r}^{\rm c}$, is set to $-r/\tau_{\rm p}$. 
The red dotted-curve in the right panel lacks a point at $St =0.1$ because no caustic pairs were found 
for $St =0.1$ particles at $r=\eta/4$ in the simulation. For the corresponding curve in the left panel, 
$g^{\rm cau}$ drops to zero at $St=0.1$, which is not visible due to the selected Y-range in the figure.}  
\label{splitrdfandv} 
\end{figure*}

In the left panel of Fig.\ \ref{splitrdfandv}, we plot $g^{\rm con}$ (blue) and $g^{\rm cau}$ (red) computed from 
our simulation data at three distances. At $St \lsim 1$, the number of caustic pairs is small, and most particle pairs belong to the continuous 
type. The continuous-type particles show significant clustering at $St \lsim 3$, and the RDF $g^{\rm con}$ 
increases toward smaller $r$ with similar scaling exponents as the overall RDF. These suggest 
that the origin of strong clustering at $St \lsim 3$ corresponds to the continuous-type particles 
(Falkovich et al.\ 2002). Similar to the overall RDF (Fig.\ \ref{rdf}), $g^{\rm con}$ peaks at $St \simeq 1$. 
At larger $St$, $g^{\rm con}$ decreases rapidly with $St$, and becomes 
comparable to $g^{\rm cau}$ at $St \simeq 3$. The fast drop of $g^{\rm con}$ at $St \gsim 3$ is because it 
becomes sensitive to the critical value, $w_{\rm r}^{\rm c}$, as the range $(w_{\rm r}^{\rm c}, 0)$ narrows
with increasing $St$. This range also becomes narrower with decreasing $r$, 
and this causes $g^{\rm con}$ to slightly decease with decreasing $r$ at $St \gsim 3$. 
At $St \simeq 6.2$,  $g^{\rm con}$ is significantly smaller than $g^{\rm cau}$, and essentially 
all particle pairs are of the caustic type.  
The caustic RDF $g^{\rm cau}$ increases with decreasing $r$ for particles with $St \lsim 10$, and 
is $r$-independent at larger $St$. 

Concerning the $r$-dependence of $g^{\rm cau}$ at $St \lsim 10$, there are 
several possibilities. The first is that this dependence is generic, and $g^{\rm cau} \to \infty$ as $r\to 0$, as 
considered by Gustavsson \& Mehlig (2011). Second, the $r$-dependence of $g^{\rm cau}$ may 
disappear at sufficiently small $r$, and $g^{\rm cau}$ finally converges to a finite constant, as assumed 
in Falkovich et al.\ (2002). In this case, one may in principle achieve a converged $g^{\rm cau}$ 
by using a larger number of $St\lsim 10$ particles in the simulation to resolve smaller scales. 
We checked $r= \eta/8$ in our simulation and found that the convergence is not reached yet. 
Third, the $r$-dependence of $g^{\rm cau}$ may be an artifact of our selected value for $w_{\rm r}^{\rm c}$. We attempted 
to vary $w_{\rm r}^{\rm c}$ around $-r/\tau_{\rm p}$ by a factor 4, but were not able to find a considerable decrease 
in the $r$-dependence of $g^{\rm cau}$. It is also possible that the transition from the continuous to the caustic type 
is gradual, and thus exactly splitting two types by a single critical value ($w_{\rm r}^{\rm c}$) is impossible at 
least at the scales explored here. 
We note that $g^{\rm cau}$ is larger than 0.5 at $St \simeq 3.11$, suggesting 
some degree of clustering for caustic-type particles.  This, however, could be an artifact that 
the two types of  pairs are not split precisely. 

The right panel of Fig.\ \ref{splitrdfandv} shows the average relative speeds, 
$\langle w_{\rm r}^{\rm con} \rangle$ and $\langle w_{\rm r}^{\rm cau} \rangle$, for continuous (blue) and 
caustic (red) pairs. At $St \ll 1$, $\langle w_{\rm r}^{\rm con} \rangle$ is constant with $St$ and 
decreases linearly with $r$, as in the S-T prediction. At $St \gsim 1$, $\langle w_{\rm r}^{\rm con} \rangle$ 
decreases with $St$. This is because the number of continuous pairs becomes small and the average relative 
speed starts to have a sensitive dependence on the lower limit $w_{\rm r}^{\rm c}$, which becomes closer to 
zero as $St$ increases. By definition,  the caustic pairs have larger relative velocity than continuous pairs. 
At $St\ll 1$, the sling events or caustic formation occur only in regions with very large velocity gradient, 
and this is responsible for the large relative velocity for caustic pairs at small $St$. 
As $St$ increases, the criterion for the sling events to occur becomes weaker, as 
reflected by the decrease of $|w_{\rm r}^{\rm c}|$. This results in a decrease 
in $\langle w_{\rm r}^{\rm cau} \rangle$ toward $St \simeq 1$. At $St \gsim 3$, most particles belong to 
the caustic type, and thus the trend of $\langle w_{\rm r}^{\rm cau} \rangle$ as a function $St$ becomes similar to the 
overall relative speed $\langle |w_{\rm r}|\rangle$ (see Fig.\ \ref{absolute}). It first increases with $St$ for $3 \lsim St \lsim St_{\rm m} \simeq 30$, 
and then decreases at $St \gsim St_{\rm m}$. At the three distances shown here, $\langle w_{\rm r}^{\rm cau} \rangle$ decreases with 
decreasing $r$ for  $St \lsim 10$ particles, as the condition to have a caustic pair is weaker at 
smaller $r$.  Again, the $r$-dependence of $\langle w_{\rm r}^{\rm cau} \rangle$ may disappear at much smaller $r$ and/or with a
 better method to split the two types of pairs. At $St \gsim 10$, both $g^{\rm cau}$ and $\langle w_{\rm r}^{\rm cau} \rangle$ are independent of $r$, indicating full resolution of 
 their statistics in our simulation.

\begin{figure}[t]
\centerline{\includegraphics[width=1.1\columnwidth]{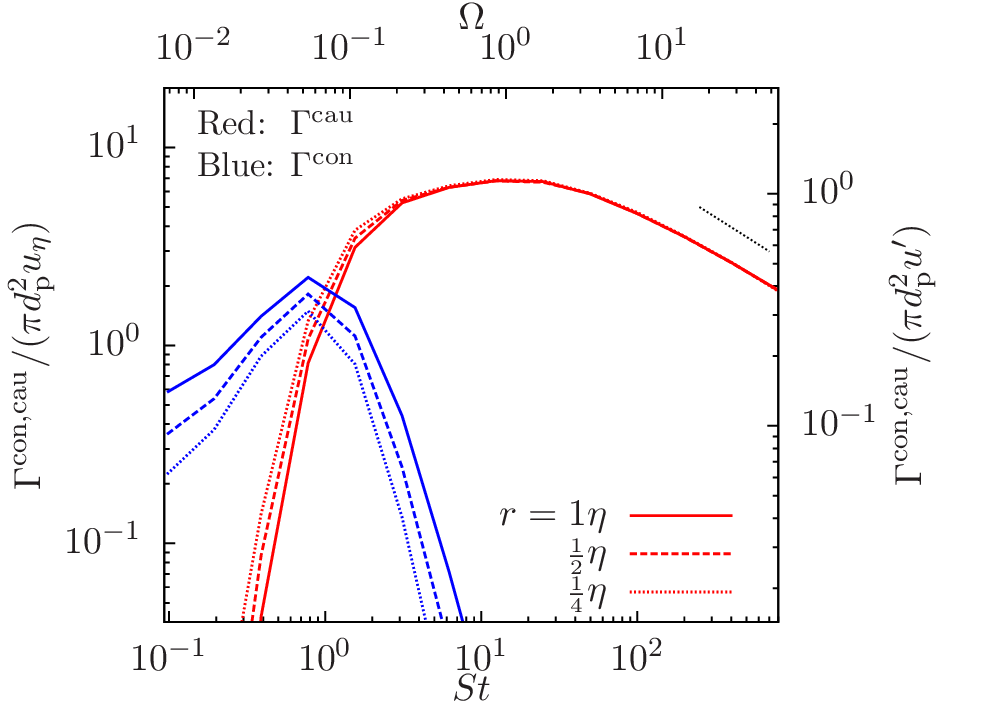}}
\caption{The continuous ($\Gamma^{\rm con}$; blue) and caustic ($\Gamma^{\rm cau}$; red) contributions to the 
kernel normalized to $\pi d_{\rm p}^2 u_\eta$ or $\pi d_{\rm p}^2 u'$ on the left and right Y-axis, respectively.  
The solid, dashed and dotted lines correspond to normalized kernels measured at $r=1$, $0.5$ and $0.25\eta$, respectively. 
The dotted line segment shows a $St^{-1/2}$ scaling.}    
\label{splitproduct} 
\end{figure}

In Fig.\ \ref{splitproduct}, we show the continuous (blue) and caustic (red) contributions to the 
collision kernel per unit cross section. In the $r$ range shown here, 
the continuous and caustic contributions dominate for $St \lsim 1$ and $St \gsim 1$, respectively. 
The continuous contribution decreases with $r$, corresponding to the $r$-dependence of the 
overall kernel, $\Gamma^{\rm sph}$, in Fig.\ \ref{product}. The decrease of the continuous contribution 
with $r$ is expected from the scaling exponents of $g^{\rm con}$ and $\langle w^{\rm con}\rangle$. 
With decreasing $r$, $\langle w^{\rm con}\rangle$ decreases linearly, which is faster than the 
increase of $g^{\rm con}$. In the limit $r \to 0$, the continuous contribution is expected to 
vanish for all particles. This means that, although amplified by clustering, the continuous-type particles 
do not significantly contribute to the collision rate of dust particles, as pointed out by Hubbard (2012).


The contribution of the caustic-type particles to the normalized kernel is found to 
be approximately $r$-independent. 
The differences in three red lines are small, and  one may claim achieving numerical 
convergence at $r=\eta/4$ 
for all particles. 
The $r$-independence of $\Gamma^{\rm cau}/(\pi r_{\rm p}^2)$ for $St \lsim 10$ 
particles at the distances explored here is related to the cancellation of the scalings 
of $g^{\rm cau}$ and $\langle w_{\rm r}^{\rm cau} \rangle$. Intuitively, one would expect 
that the collision rate of inertial particles converges to a finite constant in the  $r\to 0$ limit. 
Therefore, 
$g^{\rm cau}$ and $\langle w_{\rm r}^{\rm cau} \rangle$ must have inverse scalings with $r$, once the caustic kernel converges.  
On the other hand, one may prove the intuition of a finite constant caustic kernel at sufficiently small $r$ by showing that  
$g^{\rm cau}$ and $\langle w_{\rm r}^{\rm cau} \rangle$ scale inversely with $r$, which was actually 
predicted by the theoretical model of Gustavsson \& Mehlig (2011). 
As discussed earlier, both $g^{\rm cau}$ and $\langle w_{\rm r}^{\rm cau} \rangle$ may 
become independent of $r$ at much smaller $r$. 
However, it is also possible that, as $r\to 0$, $g^{\rm cau} \to \infty$ and $\langle w_{\rm r}^{\rm cau} \rangle \to 0$, 
especially for $St \lsim 1$ particles. Note that $\langle w_{\rm r}^{\rm cau} \rangle$ does not 
represent the collisional energy at all. Even if $\langle w_{\rm r}^{\rm cau} \rangle \to 0$, it does 
not suggest the average collisional energy is zero. In fact, a computation of the average collision 
energy per collision (i.e.,  the relative velocity variance weighted by the collision rate; see, e.g., Hubbard 2012) 
by caustic pairs for particles with $1\lsim St \lsim 10$  in our simulation shows that it already converges to a finite constant at $r=\eta/4$.  
The collision energy (velocity) of $St \lsim 10$ particles by caustic pairs in the 
$r\to 0$ limit will be examined systematically in a future study. 

The $r$-independence of the caustic contribution is responsible for that of the 
overall normalized kernel at $St \gsim 1$ observed in Fig.\ \ref{product}.  Because the 
continuous contribution vanishes as $r\to 0$, and due to the convergence of the 
caustic contribution, one may use our result for $\Gamma^{\rm cau}$ 
to estimate the collision rate of nearly point-like dust particles. 
One caveat, however, is that $\Gamma^{\rm cau}$ has a dependence on the 
critical value, $w_{\rm r}^{\rm c}$, for $St \lsim 1$ particles, an issue that will 
be refined in a future work.  

From the behaviors of  $\Gamma^{\rm con}$ and $\Gamma^{\rm cau}$ with $r$, 
there exists a  distance, $r_{\rm c}$, below which the caustic contribution dominates. 
The critical distance $r_{\rm c}$ is generally a function of $St$. 
As seen in Fig.\ \ref{splitproduct}, $r_{\rm c} \gsim 1 \eta $ for $St \gsim 1$.  At $St =0.78$, 
the two contributions become about equal at $r_{\rm c} \simeq \eta/4$.  Below $St =0.78$, $r_{\rm c}$ decreases very rapidly. For example, 
$r_{\rm c} \lsim 0.01\eta$ for $St =0.39$. Achieving sufficient statistics at such a small distance is computationally 
challenging, as it requires 
a huge number of particles. Directly resolving  $\Gamma^{\rm sph}$ for nearly point particles with 
$St \lsim 0.8$ would be extremely difficult, and the best approach is thus to properly isolate the caustic and continuous contributions.  

The normalized caustic kernel experiences a sharp rise as $St$ increase 
from 0.1 to 1, suggesting that the collision frequency greatly accelerates as the particle size 
grows toward $St \simeq 1$. The rise corresponds to the fast increase of $g^{\rm cau}$ 
(see the left panel of Fig.\ \ref{splitrdfandv}), which has been interpreted as due 
to the rapid caustic formation (Wilkinson et al.\ 2006). It has been claimed that the 
effect of caustic formation may be modeled as an activation process. If we fit the normalized 
caustic kernel by $\exp(-A/St)$ (Wilkinson et al.\ 2006, Falkovich \& Pumir 2007), the 
activation value, $A$, is found to be around $1.7$.  The sling events or caustic 
formation occur at places where the flow velocity gradient is larger than $\tau_{\rm p}^{-1}$ (Falkovich et al.\ 2002). 
At these locations, the flow velocity changes faster than the response 
of the particle. For small particles with $\tau_{\rm p} \ll \tau_{\eta}$, the probability of the 
sling events corresponds to the high PDF tail of the flow velocity gradient.  
With increasing $\tau_{\rm p}$, the probability becomes larger as it samples toward 
the inner parts of the flow velocity gradient PDF. Since the rms flow gradient is on 
the order of $\tau_{\eta}^{-1}$, the probability would increase rapidly
as $\tau_{\rm p}^{-1}$ approaches $\tau_{\eta}^{-1}$, leading to a sharp 
rise in the sling frequency and the caustic kernel at $St\simeq1$. 


For $St$ between $1$ and the peak Stokes number, $St_{\rm m} \simeq 30$, the 
normalized caustic kernel is almost constant, increasing only by 50\%. This is apparently 
due to the opposite trends of $g^{\rm cau}$ and $\langle w_{\rm r}^{\rm cau} \rangle$ with $St$ in this range. 
The kernel starts to decrease at $St \simeq 30$ due to the decrease of $\langle w_{\rm r}^{\rm cau} \rangle$. 
It finally scales with $St$ as $St^{-1/2}$ (dotted line segment in Fig.\ \ref{splitproduct}) for particles 
with $\tau_{\rm p} \gg T_{\rm L}$. 

We finally comment on the collision kernel commonly used in coagulation models for the dust particle 
growth in protoplanetary disks. In these models, the kernel is typically set to $\Gamma^{\rm coa} =\pi d_{\rm p}^2 \langle w^2\rangle^{1/2}$ 
(e.g., Dullemond and Dominik 2005). The models usually adopt the rms collision velocities 
based on the model of Volk et al.\ (1980) or its later developments. 
$\Gamma^{\rm coa}$ is of the cylindrical type discussed in \S 7.1. However, it would be more appropriate 
to use $\langle |{\bs w}| \rangle$, rather than the 3D rms, $\langle w^2\rangle^{1/2}$, in the 
cylindrical formulation (Wang et al.\ 2000; see the inset in the right panel of Fig.\ \ref{absolute} for the 
difference between $\langle |{\bs w}| \rangle$ and $\langle w^2\rangle^{1/2}$). We argue that $\Gamma^{\rm coa}$ 
cannot sufficiently capture the wealth of physics for the turbulence-induced collision rate, 
especially the interesting phenomena that occur at small $St$. The prediction of Volk et al.\ for  $\langle w^2\rangle^{1/2}$ 
should be viewed as for the $r \to 0$ limit, as it does not consider the $r$-dependence of the 
relative speed. Thus, we compare $\Gamma^{\rm coa}$ with our simulation result for 
$\Gamma^{\rm cau}$, which provides the dominant contribution in the $r \to 0$ limit. 
We point out several problems in the qualitative behavior of the commonly-used collision kernel. 

First,  Volk et al.'s model predicts that $\langle w^2\rangle^{1/2}$ drops exactly to zero for small 
particles with $\tau_{\rm p}$ below the turnover time of the smallest eddies in the assumed 
flow energy spectrum (e.g., Cuzzi \& Hoggan 2003). However, our simulation result shows that, 
for any finite $St \lsim1$, $\Gamma^{\rm cau}$ is nonzero, and it can be estimated from the 
activation formula, $\propto \exp(-A/St)$.  Clearly, the Volk et al.\ model does not capture the 
sling effect, and cannot accurately estimate the kernel at $St\lsim 1$. Second,  based 
on Fig.\ 5 of Cuzzi \& Hoggan (2003), $\Gamma^{\rm coa}$ would increase quite rapidly as $St$ 
increases from 1 to 2-3, and then connect to a $St^{1/2}$ scaling at $St \gsim 3$. This is in contrast to $\Gamma^{\rm cau}$ 
found in our simulation, which changes quite slowly in the range $1 \lsim St \lsim 10$. This suggests that 
the commonly-used $\Gamma^{\rm coa}$ is inaccurate for $St \lsim 10$. This range of $St$ may include particle 
sizes up to chondrules for typical disk parameters at a radius of 1 AU. 
It remains to be checked by simulations at higher resolutions whether and how $\Gamma^{\rm coa}$ switches to a $St^{1/2}$ 
scaling for inertial-range particles. Finally, the Volk et al.\ model overestimates the peak 
relative speed at $St_{\rm m}$ (\S 6.1.1), and would thus overestimate the collision rate by a factor of 2 or 
so for inertial-range particles and/or large particles with $St > St_{\rm m}$.

\section{Conclusions}

We investigated the turbulence-induced relative velocity 
of inertial particles using both theoretical modeling and numerical simulations. 
We conducted a 512$^3$ simulation of a weakly compressible turbulent 
flow with an rms Mach number of $\simeq 0.1$, and evolved 14 species of inertial particles with friction timescales, 
$\tau_{\rm p}$, covering the entire scale range of the flow. The Stokes number, $St$, of the particles spans 
about 4 order of magnitudes, ranging from 0.1 to 800. The friction time of the largest particles is about 41
times the large eddy turnover time, or 54 times the Lagrangian correlation time. We used the simulation 
to test the theoretical model for the rms relative velocity of inertial particles by Pan \& Padoan (2010), in the case 
of identical particles (equal friction times). 
We explored the probability distribution function (PDF) of the relative velocity. 
Consistent with previous studies, our simulation result for the 
PDF indicates two different types of particle pairs, named as the continuous and caustic pairs, respectively.  
In this work, the relative velocity statistics is measured at distances, $r$, in the 
range from $\eta/4$ to $\eta$. An accurate measurement at smaller scales 
would require a larger simulation with a larger number of particles. 
Our study for the relative velocity helps reveal the fundamental physics, and provides a theoretical preparation 
toward modeling the collision velocity  of nearly-point dust particles at $r \to 0$. Using the simulation data, 
we computed the particle collision kernel, and, in particular, we evaluated the contributions from the continuous
and caustic pairs. Distinguishing the two types of contributions is crucial for the estimate of 
the kernel at $r \to 0$.  
We summarize the main results of this work.   

\begin{enumerate}
\item We introduced the formulation for the particle relative velocity by Pan \& Padoan (2010), which reveals 
an insightful physical picture. The relative velocity of two nearby identical particles is determined by the memory of the 
flow velocity difference along their trajectories in the past, and hence depends on the separation behavior 
of particle pairs backward in time. We adopted a two-phase separation behavior consisting of a ballistic and a Richardson phase, 
and showed that the model predicts a $St^{1/2}$ scaling for inertial-range particles in turbulent flows with an 
extended inertial range. The model can correctly reproduce the expected behaviors in the extreme limits of small and large particles. 
The model prediction for the rms relative velocity is in good agreement with the simulation results 
for $ \eta/4 \lsim r \lsim \eta$. The physical picture also provides a successful explanation for the 
the relative velocity PDF as a function of $St$. 

\item To improve the understanding of the inertial particle statistics, we analyzed both Lagrangian 
and Eulerian temporal correlation functions in the simulated flow. While the flow velocity along the trajectory 
of a small particle is close to Lagrangian, the velocity seen by a very large particle may be better approximated 
by Eulerian. The Eulerian and Lagrangian correlation timescales, $T_{\rm E}$ and $T_{\rm L}$, were 
found to be close to each other, with $T_{\rm E}$ slightly larger (by 10\%). Our model predictions for
both 1-particle rms velocity and the rms relative speed between two identical particles depend 
mainly on the correlation timescale and are insensitive to the function form of the temporal correlation. 
These provides a validation for using the Lagrangian correlation 
function form for all particles.   

\item 
Our simulation data shows that, in the small particle limit ($St \ll 1$), the 3D rms relative velocity 
of particle pairs at a distance $r = \eta$ is constant, $\simeq u_\eta/\sqrt{3}$, consistent with the 
Saffman-Turner prediction. It starts to rise at $St \gsim 1$, and peaks for particles with $\tau_{\rm p} \simeq 2 T_{\rm L}$, 
corresponding to $St = St_{\rm m} \simeq 30$ in our simulated flow. As expected, the relative 
velocity scales with $St$ as $St^{-1/2}$ in the limit  $\tau_{\rm p} \gg T_{\rm L}$. The PP10 
model with reasonable parameters provides an excellent fit to the simulation result for the 3D rms. 
The maximum relative speed at $St_{\rm m}$ is twice smaller than the rms velocity of the 
turbulent flow, indicating a factor of $\simeq 2$ overestimate by the commonly-used 
model of Volk et al.\ (1980) and its later developments (e.g., Markiewicz et al.\ 1991; Cuzzi \& Hogan 2003; Ormel \& Cuzzi 2007). 

The rms relative speed of particles with $St \lsim 6$ shows a $r$-dependence in the range $\eta/4 \lsim r \lsim \eta$. 
The dependence for the smallest particles ($St =0.1-0.2$) in our simulated flow 
at $r< \eta$ was found to be slower than the linear scaling predicted by the Saffman-Turner formula,
suggesting considerable contributions from the sling events or caustic formation. 
For larger particles with $St \gsim 6$, the backward separation at a friction timescale ago becomes 
insensitive to the initial distance $r$, and the rms relative speed is independent of $r$. 
It remains to be examined whether and at which scale the rms relative velocity of $St\lsim 6$ 
particles would converge and become $r$-independent as $r\to 0$. 

The rms relative speeds in the radial and tangential directions are nearly equal for all $St \gsim 0.1$ particles. 
For $St \ll 1$ particles, this is in contrast to the Saffman-Turner formula, which predicts the tangential rms is larger than the 
radial rms by $\sqrt{2}$. This near equality is due to  the randomization of the relative velocity direction with 
respect to the particle separation $\bs{r}$, which is caused by the deviation of particle trajectories from 
the fluid elements and/or the stochastic backward separation of particle pairs.

In the distance range explored, we find an asymmetry in the relative speed of $St \lsim 6$ particles: 
approaching pairs that may lead to collisions have a larger relative speed than separating ones. 
The asymmetry is related to the phenomenon of turbulent clustering. The asymmetry is expected to 
decrease with decreasing $r$, but it remains to be tested whether it completely vanishes  as $r \to 0$.


\item The probability distribution function for the particle relative velocity is 
highly non-Gaussian, exhibiting extremely fat tails. 
For small particles with $St \lsim 1$, the effects of the particle memory and 
the backward separation lead to a self-amplification starting at the far tails, 
corresponding to the effect of slings or caustic formation. As $St$ increases, the amplification 
becomes stronger and proceeds toward the inner part of the PDF, causing an increase 
in the fatness of the overall PDF shape. On the other hand, as $St$ increases above $\simeq 1$, 
the PDF shape becomes less fat. For the larger particles, the relative velocity samples the PDF, $P_{\rm u}(\Delta u, \ell)$, 
of the flow velocity increment, $\Delta u$, at larger scales, $\ell$. As the fatness
of $P_{\rm u}$ decreases with increasing $\ell$,  the PDF of  $St\gsim1$ particles 
keeps thinning with increasing $St$. At a particle distance of $r\simeq 1\eta$, 
the PDF shape is fattest at $St \simeq 1$, with a kurtosis of $\simeq 30$. 

We identified two sources of non-Gaussianity for the particle relative velocity: the imprint of 
intermittency of the turbulent flow and an intrinsic contribution from the particle dynamics. 
We predicted a 4/3 stretch exponential PDF, $\propto \exp(-(|w|/\beta)^{4/3})$, for inertial-range particles 
in an exactly Gaussian velocity field with Kolmogorov scaling. This 4/3 stretched exponential is
observed in the PDF tails of particles with $St \simeq St_{\rm m}$ (or $\tau_{\rm p} \simeq T_{\rm L}$), confirming the validity of the 
physical picture of PP10. 
   
Based on the PP10 picture and our simulation result for the PDF, the particle pairs at a give distance 
can be split into two types, namely, the continuous type and the caustic type (Wilkinson et al.\ 2006). 
The two types correspond to the inner part of the PDF that follows the flow velocity difference, and the outer PDF part that 
is affected by the particle memory and the backward separation, respectively.    

The PDF of the particle collision velocity is expected to play a crucial role in the growth 
of dust particles in protoplanetary turbulence, as it determines the fractions of collisions 
resulting in sticking, bouncing and fragmentation. The relative velocity PDFs of $St \gsim12.4$ 
particles already converge at $r \simeq \eta/4$. On the other hand,  the PDFs of smaller particles
show $r$-dependence at $r\gsim \eta/4$, and an appropriate extrapolation is needed  for the
application to dust particle collisions at $r \to 0$. The shape of the measured PDFs provides 
interesting clues for the collisional energy. For example, the PDF of the 3D amplitude of the relative 
velocity suggests much higher probabilities of extremely small and large collision speeds 
than estimated from a Gaussian PDF. Only for very large particles with $\tau_{\rm p} \gsim 50 T_{\rm L}$ 
does the PDF approach Gaussian. The highly non-Gaussian nature of 
the relative/collision velocity needs to be incorporated into dust coagulation models 
for protoplanetary disks. Pushing the PDF toward smaller $r$ and finally to $r \to 0$ will be pursued in a future work.  

\item We computed the particle collision kernel from the simulation data 
using both spherical and cylindrical formulations. 
It was found that the two formulations give nearly equal predictions for all particles. 
Adopting the formulation of Falkovich et al.\ (2002) and Gustavsson and Mehlig (2011), 
we calculated the contributions to the collision kernel from continuous and
caustic particle pairs separately. 
We showed that, although amplified by the effect clustering, the collision kernel due to the continuous-type pairs 
decreases with $r$, and would vanish in the $r\to0$ limit.
Consistent with the theoretical prediction of Gustavsson and Mehlig (2011), the contribution to the normalized 
kernel by caustic pairs is found to be $r$-independent, and the convergence is reached at $r =\eta/4$. 
This caustic contribution is dominant at sufficiently small $r$, and can be used to estimate the collision rate of nearly point-like dust particles. 

The caustic contribution to the collision kernel per unit cross section shows an 
abrupt rise as $St$ increases toward 1, which may be viewed as an activation process 
corresponding to the rapid formation of caustics. As $St$ increases from 
$\simeq 1$ to $St_{\rm m}$, the normalized caustic kernel is roughly constant, increasing 
only slightly by 50\%. 
It finally decreases as $St ^{-1/2}$ for large particles with $\tau_{\rm p} \gg T_{\rm L}$. 
Coagulation models for dust particle growth need  to incorporate 
these important features. We will provide fitting functions for the caustic collision kernel 
and the collisional energy as a function of $St$ in a separate study. 

\end{enumerate}

In this work, we have focused on the monodisperse case with identical particles. 
A systematic analysis for the relative velocity of different particles will be conducted 
in a follow-up paper.  Future simulations at higher resolutions will further improve 
our understanding of the problem. For example, a $1024^3$ simulation would help 
verify the existence of the predicted $St^{1/2}$ scaling for inertial-range particles by various models, 
which is not yet confirmed numerically. We have only partially addressed the  $r\to0$ limit, necessary 
for the application to dust particle collisions. A computationally more demanding 
simulation with a larger number of inertial particles would allow us to examine the particle 
statistics at smaller scales. With such simulations and by isolating 
the caustic pairs from the continuous ones, we can systematically evaluate the collision rate and the 
PDF of the collisional energy at $r \to 0$.  
The results of these future studies will significantly improve the formulation of 
coagulation models to compute the  dust particle evolution in protoplanetary disks. 

\acknowledgements
We thank the referee, Alexander Hubbard, for an extensive report that helped us improve the paper, and Anders 
Johansen for useful discussions and for support with the Pencil code. PP is supported by the FP7-PEOPLE-2010-RG 
grant PIRG07-GA-2010-261359. The simulations were carried out on the NASA/Ames Pleiades supercomputer.

\appendix

\section{A: Separation of Tracer Particle Pairs} 

Our model for the relative velocity of inertial particles depends on the particle pair dispersion 
backward in time (\S 3.2.3). 
We adopted a two-phase behavior, consisting of a ballistic and a Richardson phase. 
To constrain the Richardson constant, $g$, in the latter phase,  
we study the dispersion of tracer particles in our simulated flow, and 
take the measured value of $g$ as a reference for inertial particles.     

\begin{figure}[b]
\centerline{\includegraphics[width=0.65\columnwidth]{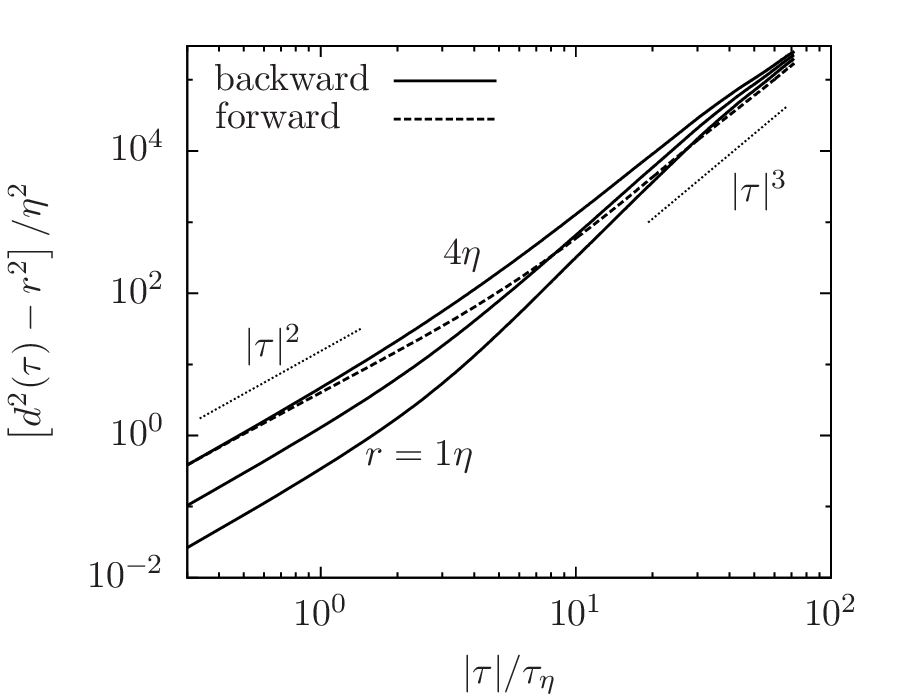}}
\caption{Separation of tracer particle pairs in our simulated flow. 
The time lag and the particle separation are normalized to 
Kolmogorov time and length scales. Solid lines from top to bottom correspond 
to the backward separation of particle pairs with ``initial" distance $r=1$, $2$ and $4 \eta$, 
respectively. The dashed line plots the forward 
separation with $r=4\eta$. The initial time is set to 0, and $\tau$ is negative (positive) 
for the backward (forward) separation. The separation shows a ballistic phase 
and a Richardson behavior at small and large $|\tau|$, respectively. 
} 
\label{separation} 
\end{figure}

In Fig.\ \ref{separation},  the three solid lines from bottom to top show the backward-in-time separation 
of tracer particle pairs with ``initial" distance $r =1 \eta$, $2\eta$ and $4\eta$, respectively. 
We subtracted $r^2$ from the separation variance, $d^2(\tau)$. The ``initial" time is set to 0, 
and $\tau$ is negative for the backward separation. As seen in Fig.\ \ref{separation}, the 
particle separation at small $|\tau|$ shows a ballistic behavior with $d^2(\tau) - r^2$ 
increasing as $\tau^2$. The physical origin of the ballistic behavior is that the velocity 
at which two tracer particles separate is determined by the flow 
velocity difference across the initial distance, $r$, and
remains roughly constant before the particle distance becomes 
significantly larger than $r$. This ballistic phase of the tracer particles is physically 
different from that of inertial particles discussed in \S 3.2.3. For 
inertial particles, the duration and the separation speed of the ballistic phase 
depend on the particles' memory timescale. But for tracer particles, 
the ballistic phase is determined purely by the initial
distance. 
For $r \lsim 4 \eta$, the ballistic phase lasts for a few Kolmogorov timescales. 

The Richardson separation behavior (the $|\tau|^3$ scaling) is observed at large 
$|\tau|$ after the particle separation enters the inertial range of the flow. In the 
bottom solid curve for $r=1\eta$, a $|\tau|^3$ scaling exists in a very limited range. 
A rough estimate of the Richardson constant in that range gives $g \simeq 0.5$.  
Consistent with previous studies (e.g., Sawford et al.\ 2008), the time range that exhibits 
the Richardson scaling becomes broader as $r$ increases to $4\eta$. 
This allows a more accurate estimate of $g$, and we find $g \simeq 1.2$ for $r=4\eta$, 
consistent with the experimental results of Berg et al.\ (2006).  Similar to Sawford et al.\ (2008), the measured 
$g$ has a dependence on $r$. If the inertial range of the flow is considerably broader and the 
Richardson behavior exists in a larger time range, the curves for different initial distances are expected to converge at sufficiently 
large time lags, with $g$ eventually approaching a universal value. 

The dashed line in Fig.\ \ref{separation} plots the forward-in-time pair dispersion 
of tracer particles at $r=4\eta$. A comparison with the top solid line shows that the 
forward separation is slower than the backward separation. For $r=4\eta$,  $g$ 
is estimated to be 0.5 in the forward separation, about twice smaller than the value 
(1.2) for the corresponding backward separation. This is consistent with the result 
of Berg et al.\ (2006). 
A physical explanation for the faster backward separation will be given in Appendix B.  In our model 
for the relative velocity of inertial particles, it is the backward separation that is relevant, 
and the purpose of showing an example for the forward separation in Fig.\ \ref{separation} 
is to illustrate the difference between the forward and backward separations. 


The Richardson constant, $g$, in the tracer-like phase of inertial particle 
separation may be different from tracers. However, it is reasonable to assume that $g$ for 
the backward separation of inertial particles lies in a similar range. Like tracers, the value of $g$ for 
inertial particles may also depend on the initial distance, $r$.  In \S 6.1, we adjusted the 
value of $g$ in our model to obtain best fits to the simulation results for the rms relative velocity at different $r$. 
%

\section{B: The PDFs of the Turbulent Velocity Field}

In this Appendix, we analyze the probability distribution functions (PDF) of the 
flow velocity increments. 
We measure the PDFs, $P_{\rm u}(\Delta u_{\rm r}, \ell)$ and $P_{\rm u} (\Delta u_{\rm t}, \ell)$, of the 
longitudinal and transverse increments as a function of the length scale, $\ell$. Similar to the 
computation of the structure functions in \S 4.2, the PDF measurement also uses the velocity differences 
along the three directions of the simulation grid.  
The variances of $P_{\rm u}(\Delta u_{\rm r}, \ell)$ and $P_{\rm u}(\Delta u_{\rm t}, \ell)$ 
correspond to the longitudinal ($S_{\rm ll}(\ell)$) and  transverse ($S_{\rm nn}(\ell)$) structure 
functions, which have been shown in Fig.\ \ref{structureandpower} (see \S 4.2). 
Clearly, the PDFs are wider at larger scales. 
Also, $P_{\rm u}(\Delta u_{\rm t}, \ell)$ is wider than $P_{\rm u}(\Delta u_{\rm r}, \ell)$ because $S_{\rm nn}(\ell) \ge S_{\rm ll}(\ell)$ at all $\ell$ (see Fig.\ \ref{structureandpower}). 

\begin{figure}[b]
\includegraphics[height=2.6in]{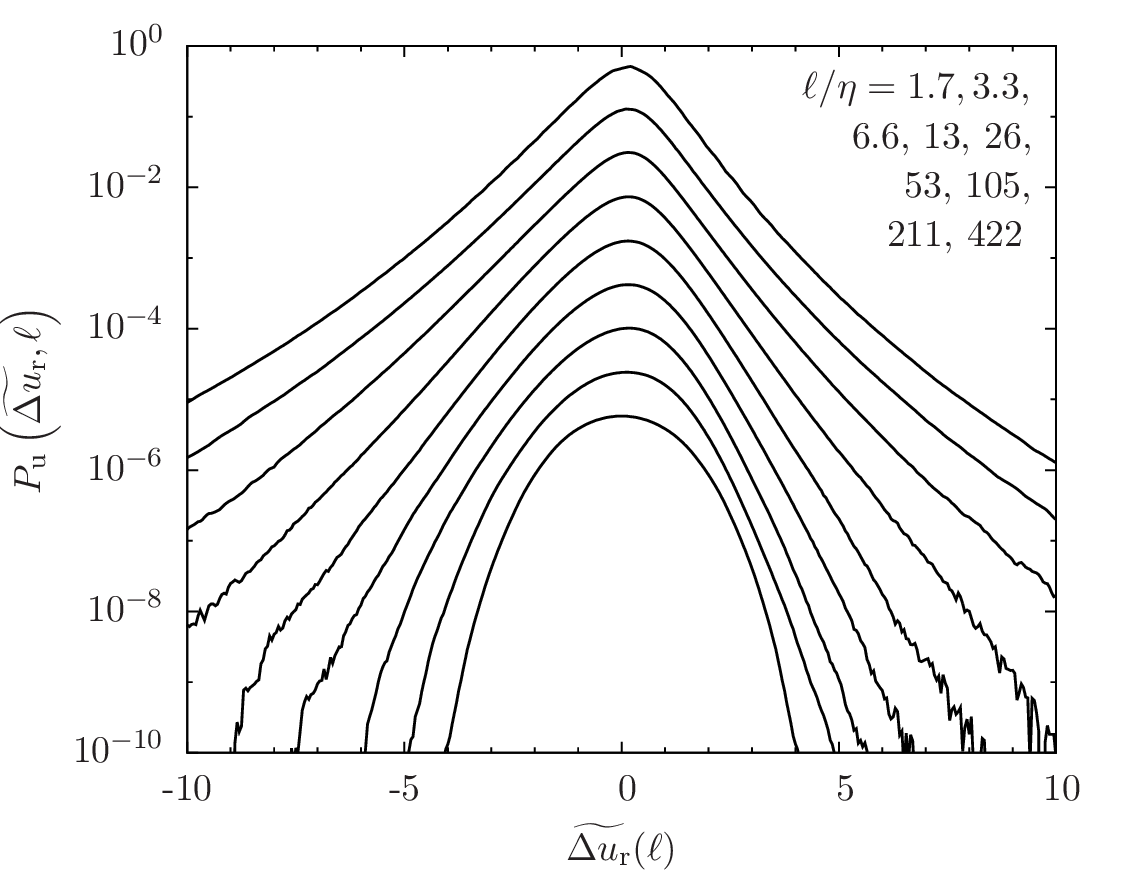}
\includegraphics[height=2.6in]{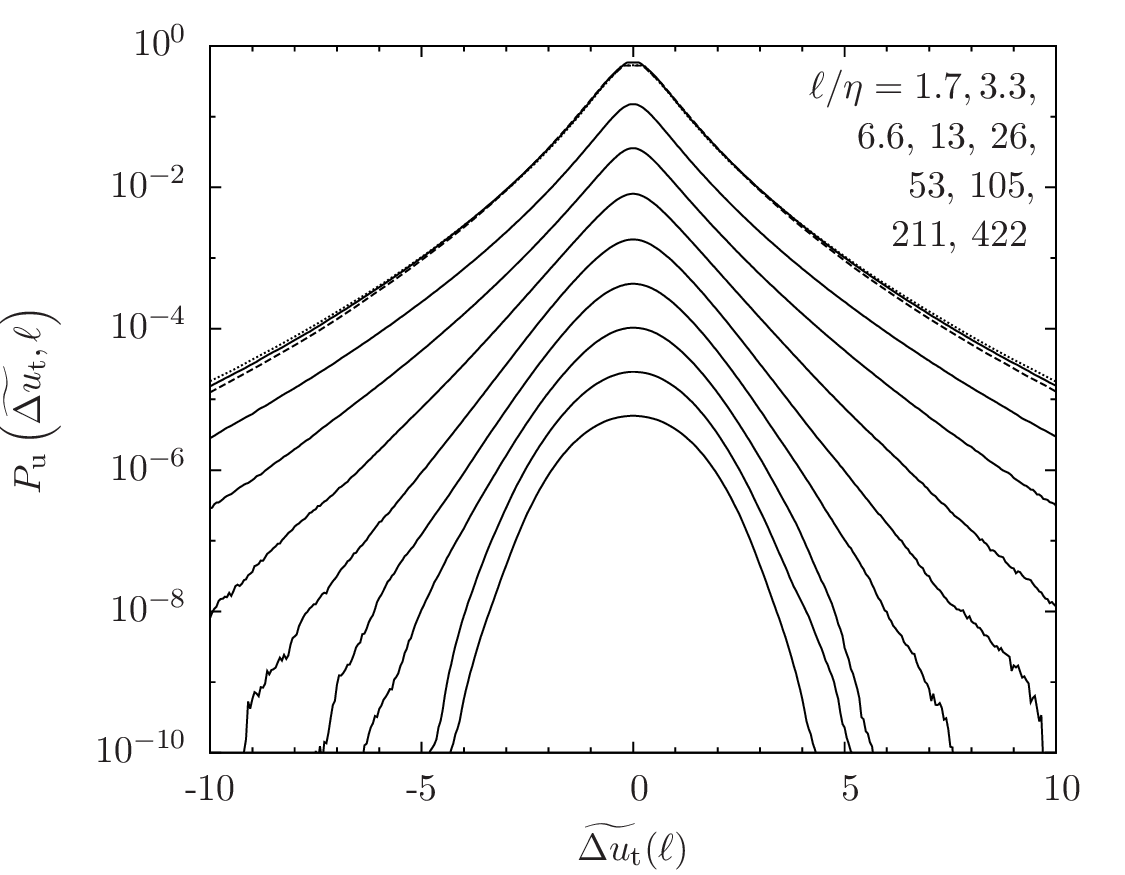}
\caption{The normalized PDFs of the flow velocity increment in the radial (left panel) and transverse (right panel) 
directions at different length scales, $\ell$. The normalized velocity 
increments are defined as $ \widetilde{\Delta u_{\rm r}}(\ell)  \equiv {\Delta u_{\rm r}}(\ell)/S_{\rm ll}^{1/2}(\ell)$ 
and $\widetilde{\Delta u_{\rm t}}(\ell) \equiv {\Delta u_{\rm t}}(\ell)/S_{\rm nn}^{1/2}(\ell)$. 
The top line in each panel plots the exact PDF at $\ell = 1.7 \eta$ (the cell size), 
and, for clarity, the PDF at each larger $\ell$ is shifted downward by a factor of 4.   
Except at the largest scales, the PDF, $P_{\rm u}(\widetilde{\Delta u_{\rm r}}, \ell)$,
of the radial increment has a negative skewness, whereas $P_{\rm u}(\widetilde{\Delta u_{\rm t}}, \ell)$ for 
the transverse increment is symmetric at all scales. The PDF tails are highly non-Gaussian at small scales. 
With increasing $\ell$, the PDFs become less fat and finally approach Gaussian. In the right panel, 
the dashed and dotted lines for $\ell = 1.7 \eta$ are the normalized PDFs of 
the transverse increment conditioned on $\Delta u_{\rm r} <0$ and $\Delta u_{\rm r }>0$, 
respectively. One may change the normalization of $\ell$ to the integral scale, $L$, using 
$L= 135\eta$.} 
\label{flowpdf} 
\end{figure}

To better see the PDF {\it shape} as a function of $\ell$, we normalized the PDFs at each 
scale to have unit variance. The radial and transverse velocity increments are normalized to 
their rms values, 
i.e., $\widetilde{\Delta u_{\rm r}}(\ell) \equiv {\Delta u_{\rm r}}(\ell)/S_{\rm ll}^{1/2}(\ell)$ 
and $\widetilde{\Delta u_{\rm t}}(\ell) \equiv {\Delta u_{\rm t}}(\ell)/S_{\rm nn}^{1/2}(\ell)$.  
The left panel of Fig.\ \ref{flowpdf} shows the normalized PDF, $P_{\rm u}(\widetilde{\Delta u_{\rm r}}, \ell)$, of the radial 
increment.  Except at the largest scales, the PDF is negatively  skewed. For inertial-range scales, 
this can be understood from Kolmogorov's 4/5 law, $\langle  \Delta u_{\rm r} (\ell)^3\rangle = -\frac{4}{5} \bar{\epsilon} \ell$, 
which indicates a negative skewness for the PDF of $\Delta u_{\rm r}$. The connection 
of the 3rd order moment of  $\Delta u_{\rm r}$ to the energy dissipation rate suggests 
that the skewness originates from the dissipative nature of turbulence. 
The skewness of the PDF of $\Delta u_{\rm r}$ also provides an explanation for 
the faster backward separation found in Appendix A. The left and right tails of 
$P_{\rm u}(\widetilde{\Delta u_{\rm r}}, \ell)$ correspond to tracer pairs receding from each other backward and forward in time, respectively. 
The broader left tail of the PDF thus suggests that the backward separation of tracer particles is faster 
than the forward case. 
Unlike $P_{\rm u}(\widetilde{\Delta u_{\rm r}}, \ell)$, the PDF, $P_{\rm u}(\widetilde{\Delta u_{\rm t}}, \ell)$, 
of the transverse increment in the right panel is symmetric 
at all $\ell$, as expected from statistical isotropy. 

Both $P_{\rm u}(\widetilde{\Delta u_{\rm r}}, \ell)$ and $P_{\rm u}(\widetilde{\Delta u_{\rm t}}, \ell)$ 
are close to Gaussian at the largest scales, $\ell = 211 \eta$ (1/4 box size, or 1.6 times the 
integral scale, $L$) and $422\eta$ (1/2 box size, or $3.1L$), of the simulated flow. 
This is consistent with the Gaussian 1-point statistics in fully developed turbulence. At smaller $\ell$, the 
PDFs become non-Gaussian, and the tails keep getting fatter
with decreasing $\ell$, a phenomenon known as intermittency 
in turbulence theory (Frisch 1995).  As mentioned in the text,  we use the word ``fat" (or ``thin")  
specifically for the shape of the PDF, while ``broad" (or  ``narrow") refers to the 
extension or width of the PDF.  The smallest scale, $1.7\eta$, in the figure corresponds to 
the size, $\Delta x$, of the computational cell. The shape of the normalized PDF is expected to remain 
unchanged once $\ell$ becomes smaller than $\sim \eta$. 
Physically, the viscosity acts to smooth the velocity field and makes it differentiable 
in the dissipation range, and thus the velocity increment at any scale $\ell \lsim \eta$ is 
proportional to the local velocity gradient, whose PDF is fixed.  In our simulation,  the velocity field 
inside a computation cell is obtained by interpolation, and thus the PDF of the velocity 
difference below the cell size is controlled by the velocity gradient PDF at $\Delta x$.
In \S 6.2, we showed that the trend of the PDF shape of the flow velocity difference 
with $\ell$ has interesting effects on the PDF of the relative velocity of 
inertial particles as a function of the particle inertia. 


The tails of $P_{\rm u}(\widetilde{\Delta u_{\rm t}}, \ell)$ for the transverse increment can 
be approximately described by stretched exponentials, $P_{\rm se}$ (see eq.\ (\ref{se}) 
in \S 6.2). At largest scales, $P_{\rm u}(\widetilde{\Delta u_{\rm t}}, \ell)$ are nearly Gaussian, 
and $\alpha =2$. With decreasing $\ell$, $\alpha$ decreases, corresponding to fatter tails. 
For example, the best-fit  $\alpha$ for $P_{\rm u}(\widetilde{\Delta u_{\rm t}})$ at $\ell = 26 \eta$ is 
$\simeq 1$, and it further decreases to $\simeq 0.72$ at $\ell = 1.7 \eta $. 
Due to the asymmetry of the radial PDF, $P_{\rm u}(\widetilde{\Delta u_{\rm r}}, \ell)$, 
one needs to obtain the fits separately for the left and right wings. Comparing 
the left wing of $P(\widetilde{\Delta u_{\rm r}}, \ell)$ with $P_{\rm u}(\widetilde{\Delta u_{\rm t}}, \ell)$, 
we see that their shape has a similar level of fatness at the same scale $\ell$. In fact, 
the best-fit $\alpha$ for the left tail of $P_{\rm u}(\widetilde{\Delta u_{\rm r}}, \ell)$ is 
very close to that for $P_{\rm u}(\widetilde{\Delta u_{\rm t}}, \ell)$. 
We also find that the best-fit values of $\alpha$ for the left and 
right tails of $P_{\rm u}(\widetilde{\Delta u_{\rm r}}, \ell)$ are close, indicating that the two tails have a similar 
shape and differ only in the  fluctuation amplitude.

To quantify the fluctuation amplitudes in the left and right wings of $P_{\rm u}(\Delta u_{\rm r}, \ell)$, 
we define the variances in the two wings as $\langle \Delta u_{\rm r}^2 \rangle_- =  \int_{-\infty}^0\Delta u_{\rm r}^2 P_{\rm u}(\Delta u_{\rm r}, \ell) 
d \Delta u_{\rm r}/\int_{-\infty}^0 P_{\rm u}(\Delta u_{\rm r}, \ell) d \Delta u_{\rm r}$ and  $\langle \Delta u_{\rm r}^2 \rangle_+ =  \int_{0}^{\infty}\Delta u_{\rm r}^2 P_{\rm u}(\Delta u_{\rm r}, \ell) d \Delta u_{\rm r}/\int_{0}^{\infty} P_{\rm u}(\Delta u_{\rm r}, \ell) d \Delta u_{\rm r}$.
The definition of  $\langle \Delta u_{\rm r}^2 \rangle_{\mp}$ is similar to $\langle w_{\rm r}^2 \rangle_{\mp}$ for the 
relative velocity of inertial particles (\S 6.1.3).
We find that  the ratio of $\langle \Delta u_{\rm r}^2 \rangle_- $ to $\langle \Delta u_{\rm r}^2 \rangle_+ $  is 
$\simeq 1.47$ at $\ell =1.7 \eta$. This ratio decreases with increasing $\ell$, and reaches unity at the largest 
scales. The variances of the left and right wings of $P_{\rm u}(\Delta u_{\rm r}, \ell)$ was used in  the discussion on the 
relative velocity of approaching and separating particle pairs in the $St \ll 1$ limit (see \S 6.1.3). 
We also considered the PDF of $\Delta u_{\rm t}$ conditioned on 
the sign of $\Delta u_{\rm r}$. We denote two conditional PDFs as $P_{\rm u}(\Delta u_{\rm t}|\Delta u_{\rm r} \lessgtr 0, \ell)$ 
and their variances as $\langle \Delta u_{\rm t}^2 \rangle_{\mp} \equiv \int_{-\infty}^{\infty} \Delta u_{\rm t}^2 
P_{\rm u}(\Delta u_{\rm t}|\Delta u_{\rm r} \lessgtr 0, \ell)d \Delta u_{\rm t}$.   
At $\ell =1.7 \eta$, $\langle \Delta u_{\rm t}^2 \rangle_-$ is found to be larger than $\langle \Delta u_{\rm t}^2 \rangle_+$ by 28\%. 
In the right panel of Fig.\ \ref{flowpdf}, the dashed and dotted lines show the normalized conditional PDFs, 
$P_{\rm u}(\widetilde{\Delta u_{\rm t}}|\Delta u_{\rm r} < 0, \ell)$ and 
$P_{\rm u}(\widetilde{\Delta u_{\rm t}}|\Delta u_{\rm r} > 0, \ell)$, at $\ell =1.7 \eta$.  
The conditional variances and PDFs of $\Delta u_{\rm t}$ are useful to understand 
the tangential relative velocity of approaching and separating particle pairs with $St \ll 1$ (\S 6.1.3 and 6.2.4).

\end{document}